\documentclass[oneside,11pt]{article}

\usepackage[utf8]{inputenc}
\usepackage[T1]{fontenc}

\usepackage{amsthm}
\usepackage{amsmath}
\usepackage{amsfonts}
\usepackage{amssymb}

\usepackage{graphicx} 
\usepackage{float}
\usepackage{booktabs}
\usepackage{multirow}
\usepackage{rotating}
\usepackage{array}
\usepackage{xcolor}

\usepackage{breakcites}

\usepackage{nowidow}

\usepackage{enumitem}

\usepackage[top=1.5cm,bottom=2cm,right=2.5cm,left=2.5cm]{geometry}
\usepackage{titling}
\usepackage{natbib}

\usepackage{ifplatform}

\usepackage{textcomp}

\ifwindows
\fi

\ifwindows
	\usepackage{bbm}
\else
	\iflinux
		\usepackage{bbm}
	\else
		\usepackage{bbm}
	\fi
\fi

\newcommand{\R}{\mathbb{R}}

\DeclareFontFamily{OT1}{pzc}{}
\DeclareFontShape{OT1}{pzc}{m}{it}{<-> s * [1.10] pzcmi7t}{}
\DeclareMathAlphabet{\mathpzc}{OT1}{pzc}{m}{it}

\newcommand{\n}{^{(n)}}

\newcommand{\ny}{n\rightarrow\infty}

\newcommand{\sirc}{{\scriptscriptstyle
\circ }}

\newcommand{\Xb}{\mathbf{X}}
\newcommand{\Sb}{\mathbf{S}}

\newcommand{\Zb}{\mathbf{Z}}

\newcommand{\ub}{\ensuremath{\mathbf{u}}}

\newcommand{\xb}{\ensuremath{\mathbf{x}}}

\newcommand{\Ab}{\ensuremath{\mathbf{A}}}
\newcommand{\Bb}{\ensuremath{\mathbf{B}}}

\newcommand{\Ob}{\ensuremath{\mathbf{O}}}

\newcommand{\tb}{\ensuremath{\mathbf{t}}}
\newcommand{\Ub}{\ensuremath{\mathbf{U}}}

\newcommand{\Mb}{\ensuremath{\mathbf{M}}}

\newcommand{\Yb}{\ensuremath{\mathbf{Y}}}

\newcommand{\thetab}{{\pmb \theta}}

\newcommand{\Lamb}{{\pmb \Lambda}}
\newcommand{\Sigb}{{\pmb \Sigma}}
\newcommand{\deltab}{{\pmb \delta}}
\newcommand{\Deltab}{{\pmb \Delta}}
\newcommand{\taub}{{\pmb \tau}}

\newcommand{\Gamb}{{\pmb \Gamma}}

\newcommand{\vecho}{\ensuremath{{\rm vech}\,}}
\newcommand{\vechrond}{{\rm vech}^{^{\!\!\!\!\!\!\!\!\!\!\sirc}}\,\,\,\,}
\newcommand{\veco}{\ensuremath{{\rm vec}\,}}

\newcommand{\inlaw}{\stackrel{\mathcal{D}}{\rightsquigarrow}}

\usepackage[pdftex,final,bookmarksnumbered,bookmarksopen=false,breaklinks,colorlinks]{hyperref}
\hypersetup{
	final,
    bookmarksnumbered=true,
    bookmarksopen=true,
    bookmarksopenlevel=0,
    unicode=false,          
    pdftoolbar=true,        
    pdfmenubar=true,        
    pdffitwindow=false,     
    pdftitle={On optimal tests for rotational symmetry against new classes of hyperspherical distributions},    
	pdfdisplaydoctitle=true, 
	pdftoolbar=true,
	pdfmenubar=true,
	pdflang={English},
    pdfauthor={Eduardo Garcia-Portugues, Davy Paindaveine, Thomas Verdebout},     
    pdfsubject={arXiv paper},   
    pdfcreator={Eduardo Garcia-Portugues},   
    pdfproducer={Eduardo Garcia-Portugues}, 
    pdfkeywords={Directional data}{Local asymptotic normality}{Locally asymptotically maximin tests}{Rotational symmetry}, 
    pdfnewwindow=true,      
    breaklinks=true,
    hidelinks,       
    linkcolor=black,          
    citecolor=black,        
    filecolor=magenta,      
    urlcolor=cyan           
}

\newtheorem{defin}{Definition}
\newtheorem{theo}{Theorem}
\newtheorem{coro}{Corollary}

\newtheorem{prop}{Proposition}
\newtheorem{lem}{Lemma}

\allowdisplaybreaks

\begin{document}

\title{On optimal tests for rotational symmetry against new classes of hyperspherical distributions}
\setlength{\droptitle}{-1cm}
\predate{}%
\postdate{}%

\date{}

\author{Eduardo Garc\'ia-Portugu\'es$^{1,2,6}$, Davy Paindaveine$^{3,4,5}$, and Thomas Verdebout$^{3,4}$}

\footnotetext[1]{
	Department of Statistics, Carlos III University of Madrid (Spain).}
\footnotetext[2]{
	UC3M-Santander Big Data Institute, Carlos III University of Madrid (Spain).}
\footnotetext[3]{
	D\'{e}partement de Math\'{e}matique, Universit\'{e} libre de Bruxelles (Belgium).}
\footnotetext[4]{
	ECARES, Universit\'{e} libre de Bruxelles (Belgium).}
\footnotetext[5]{
	Toulouse School of Economics, Universit\'{e} Toulouse Capitole (France).}
\footnotetext[6]{Corresponding author. e-mail: \href{mailto:edgarcia@est-econ.uc3m.es}{edgarcia@est-econ.uc3m.es}.}

\maketitle


\begin{abstract}
Motivated by the central role played by rotationally symmetric distributions in directional statistics, we consider the problem of testing rotational symmetry on the hypersphere. We adopt a semiparametric approach and tackle problems where the location of the symmetry axis is either specified or unspecified. For each problem, we define two tests and study their asymptotic properties under very mild conditions. We introduce two new classes of directional distributions that extend the rotationally symmetric class and are of independent interest. We prove that each test is locally asymptotically maximin, in the Le Cam sense, for one kind of the alternatives given by the new classes of distributions, both for specified and unspecified symmetry axis. The tests, aimed to detect location-like and scatter-like alternatives, are combined into convenient hybrid tests that are consistent against both alternatives. We perform Monte Carlo experiments that illustrate the finite-sample performances of the proposed tests and their agreement with the asymptotic results. Finally, the practical relevance of our tests is illustrated on a real data application from astronomy. The R package \texttt{rotasym} implements the proposed tests and allows practitioners to reproduce the data application.
\end{abstract}

\begin{flushleft}
\end{flushleft}

\section{Introduction}

\subsection{Motivation}

Directional statistics deals with data belonging to the unit hypersphere ${\cal S}^{p-1}:=\{ \xb \in \R^p: \|\xb\|^2=\xb^{T} \xb=1 \}$ of $\R^p$. The most popular parametric model in directional statistics, which can be traced back to the beginning of the 20th century, is the von Mises--Fisher (vMF) model characterized by the density $\xb \mapsto c^{\cal M}_{p, \kappa} \exp( \kappa\, \xb^{T} \thetab)$ (densities on~${\cal S}^{p-1}$ throughout are with respect to the surface area measure~$\sigma_{p-1}$ on~${\cal S}^{p-1}$), where $\thetab \in {\cal S}^{p-1}$ is a location parameter (it is the modal location on the sphere), $\kappa>0$ is a concentration parameter (the larger the value of~$\kappa$, the more the probability mass is concentrated about~$\thetab$), and~$c^{\cal M}_{p, \kappa}$ is a normalizing constant. The vMF model belongs to a much broader model characterized by \emph{rotationally symmetric} densities of the form $\xb \mapsto c_{p,g} \, g(\xb^{T} \thetab)$, where~$g$ is a function from~$[-1,1]$ to~$[0,\infty)$ and where~$c_{p,g}$ is a normalizing constant. The rotationally symmetric model is indexed by the finite-dimensional parameter~$\thetab$ and the infinite-dimensional parameter~$g$, hence is of a semiparametric nature. Clearly, the (parametric) vMF submodel is obtained with $g(t)= \exp(\kappa t)$, $\kappa>0$. Note that for axial distributions ($g(-t)=g(t)$ for any~$t$), only the pair~$\{\pm\thetab\}$ is identified, whereas non-axial distributions allow identifying~$\thetab$ under mild conditions  (identifiability of~$\thetab$ is discussed below). \\

Rotationally symmetric distributions are often regarded as the most natural non-uniform distributions on the sphere and tend to have more tractable normalizing constants than non-rotationally symmetric models. As the following examples show, rotational symmetry plays a central role in applied directional statistics. Rotational symmetry is a common assumption in the analysis of fibre textures, as the distribution of crystallographic orientations typically has a favored direction that may be taken as the mode of a rotationally symmetric distribution, see, e.g., Chapter~5 of \cite{Bu15}. \cite{GuaSmi2017} used a rotationally symmetric distribution as a noise in their probabilistic camera model. Mixtures of vMF distributions are used in \cite{Las10} in their analysis of Functional Magnetic Resonance Imaging (FMRI) data. They used mixtures of vMF to model their concept of \emph{selectivity profiles}, each component of the mixture corresponding to a functional system concentrated about a favored direction on the sphere. Both rotationally and non-rotationally symmetric models are used in \cite{XJ15} in the study of neuronal population volumes. The objective was to fit the distribution of neuronal spike activity around one or more active electrodes. Finally, the real data application provided in Section~\ref{sec:realdata} indicates that rotational symmetry may also be relevant in astronomy.\\

Rotational symmetry has also been extensively adopted in the literature as a core assumption for performing inference with directional data. A (far from exhaustive) list of references illustrating this follows: \cite{Ri89}, \cite{KoCha1993}, and \cite{ChaRi01} considered regression and M-estimation under rotationally symmetric assumptions; \cite{LC87} considered Bayes procedures for rotationally symmetric models; \cite{La02} considered vMF likelihood ratios; \cite{TS07}, \cite{Leyetal2013}, and \cite{PaiVer2013} proposed rank tests and estimators for the mode of a rotationally symmetric distribution; \cite{Ley2014} proposed a concept of quantiles for rotationally symmetric distributions; \cite{PaiVer2017} considered inference for the direction of weak rotationally symmetric signals, whereas \cite{PaiVer2019} tackled high-dimensional hypothesis testing in the same framework. \\

Rotationally symmetric distributions, however, impose a rather stringent symmetry on the hypersphere, which makes it crucial to test for rotational symmetry prior to basing data analysis or inference on rotational symmetry assumptions. The problem of testing rotational symmetry has mainly been considered in the circular case ($p=2$), where rotational symmetry is referred to as \emph{reflective symmetry}. Tests of reflective symmetry about a specified location~$\thetab$ have been considered in \cite{Schach1969}, using a linear rank test, and in \cite{MJ00}, using sign-based statistics, whereas \cite{P02} introduced a test based on second-order trigonometric moments for unspecified $\thetab$. \cite{LV14} and \cite{MeiVer2018} developed specified-$\thetab$ tests that are locally asymptotically optimal against specific alternatives. For~$p\geq 3$, however, the problem is much more difficult, which explains that the corresponding literature is much sparser: to the best of our knowledge, for~$p\geq 3$, only \cite{Jupp1983} and \cite{LVD16} addressed the problem of testing rotational symmetry in a \textit{semiparametric} way (i.e., without specifying the function~$g$). The former considered a test for symmetry in dimension~$p\geq2$ using the Sobolev tests machinery from \cite{Gine1975}, whereas the latter established the efficiency of the \cite{Wat1983} test against a new type of non-rotationally symmetric alternatives. Both papers restricted to the specified-$\thetab$ problem, so that no tests for rotational symmetry about an unspecified location~$\thetab$ are available in dimension~$p\geq 3$. Goodness-of-fit tests within the directional framework (i.e., tests for checking that the distribution on ${\cal S}^{p-1}$ belongs to a given parametric class of distributions) have received comparatively more attention in the literature. For instance, \cite{BD97} proposed goodness-of-fit tests based on spherical harmonics for a specific class of rotationally symmetric distributions. More recently, \cite{Fig12} considered goodness-of-fit tests for vMF distributions, while \cite{Bo14} introduced goodness-of-fit tests based on kernel density estimation for any (possibly non-rotationally symmetric) distribution.

\subsection{Summary of the main contributions}

In this paper, we consider the problem of testing rotational symmetry on the {unit hypersphere~${\cal S}^{p-1}$ in any dimension~$p\geq 3$}. The methodological contributions are threefold. Firstly, we tackle the specified-$\thetab$ problem and propose two tests that aim to detect \textit{scatter}-like and \textit{location}-like departures from the null hypothesis. Secondly, we introduce two new classes ${\cal C}_1$ and  ${\cal C}_2$ of distributions on~${\cal S}^{p-1}$ that are of independent interest and that may serve as natural alternatives to rotational symmetry. In particular, the class ${\cal C}_1$ is an ``elliptical'' extension of the class of rotationally symmetric distributions based on the angular Gaussian distributions from \cite{tyler1987}. We prove that, for the specified-$\thetab$ problem,  the proposed scatter and location tests are locally asymptotically maximin against alternatives in ${\cal C}_1$ and ${\cal C}_2$, respectively. Thirdly, we tackle the more challenging unspecified-$\thetab$ problem. We prove that the scatter test is unaffected asymptotically by the estimation of~$\thetab$ under the null (and therefore also under contiguous alternatives) but that the location test shows a much more involved asymptotic behavior affected by the estimation of~$\thetab$. We therefore propose corrected versions of the location test that keep strong optimality properties against alternatives in~${\cal C}_2$. Finally, using the asymptotic independence (under the null) between the location and scatter test statistics, we introduce{, both for the specified and unspecified-$\thetab$ problems}, hybrid tests that asymptotically show power both against alternatives in ${\cal C}_1$ and alternatives in~${\cal C}_2$.\\

The outline of the paper is as follows. In Section~\ref{sec:specloc}, we consider the problem of testing rotational symmetry about a specified location~$\thetab$. The asymptotic distributions of two tests proposed for that aim are given in Section~\ref{sec:thetests}. Section~\ref{sec:distribs} introduces two non-rotationally symmetric extensions of the class of rotationally symmetric distributions, which are used in Sections~\ref{sec:optim1}--\ref{sec:optim2} to investigate the non-null asymptotic behavior of the proposed tests. In Section~\ref{sec:unspecloc}, our tests are extended to the unspecified-$\thetab$ problem. Hybrid tests are introduced in Section~\ref{sec:hybrid}. Monte Carlo experiments that illustrate the finite-sample performances of the proposed tests and their agreement with the asymptotic results are given in Section~\ref{sec:simus}. We present a real data application in Section~\ref{sec:realdata} and discuss perspectives for future research in Section~\ref{sec:discuss}. Supplementary materials collect the proofs of the main results and detail the delicate construction of optimal location tests in the unspecified-$\thetab$ problem. The R package \texttt{rotasym} allows practitioners to perform the proposed tests and reproduce the data application.

\section{\texorpdfstring{Testing rotational symmetry about a specified~$\thetab$}{Testing rotational symmetry about a specified theta}}
\label{sec:specloc}

A random vector~$\Xb$ with values in~$\mathcal{S}^{p-1}$ is said to be \emph{rotationally symmetric} about $\thetab\in\mathcal{S}^{p-1}$ if and only if~$\Ob\Xb\stackrel{\mathcal{D}}{=}\Xb$ for any $p\times p$ orthogonal  matrix~$\Ob$ satisfying $\Ob\thetab=\thetab$ (throughout, $\stackrel{\mathcal{D}}{=}$ denotes equality in distribution). For any~$\xb\in\mathcal{S}^{p-1}$, write
\begin{align}
\label{vandsignU}
v_{\thetab}(\xb)
:=
\xb^{T}\thetab
\qquad
\textrm{and}\qquad
\ub_{\thetab}(\xb)
:=
\frac{\Gamb_{\thetab}^{T} \xb}{\|\Gamb_{\thetab}^{T} \xb\|}
=
\frac{\Gamb_{\thetab}^{T} \xb}{(1-v^2_{\thetab}(\xb))^{1/2}}\,,
\end{align}
where $\Gamb_{\thetab}$ denotes an arbitrary  $p \times (p-1)$ matrix whose columns form an orthonormal basis of the orthogonal complement to~$\thetab$ (so that~$\Gamb_{\thetab}^{T}\Gamb_{\thetab}={\bf I}_{p-1}$ and $\Gamb_{\thetab}\Gamb_{\thetab}^{T}={\bf I}_p - \thetab \thetab^{T}$). This allows considering the \emph{tangent-normal} decomposition
\begin{align}
\xb = v_{\thetab}(\xb) \thetab + ({\bf I}_p- \thetab \thetab^{T}) \xb =v_{\thetab}(\xb) \thetab +
(1-v^2_{\thetab}(\xb))^{1/2} \, \Gamb_{\thetab} \ub_{\thetab}(\xb).
\label{tann}
\end{align}
If $\Xb$ is rotationally symmetric about $\thetab$, then the distribution of the random $(p-1)$-vector~$\Gamb_{\thetab}^{T} \Xb$ is spherically symmetric about the origin of~$\R^{p-1}$, so that the \emph{multivariate sign}~$\ub_{\thetab}(\Xb)$ is uniformly distributed over~${\cal S}^{p-2}$, hence it satisfies the moment conditions
\begin{align}
\label{mom}
{\rm E}[\ub_{\thetab}(\Xb)]={\bf 0}\quad\text{and}\quad
{\rm E}[\ub_{\thetab}(\Xb) \ub^{T}_{\thetab}(\Xb)]= \frac{1}{p-1}\, {\bf I}_{p-1}.
\end{align}
Note also that~$\ub_{\thetab}(\Xb)$ and the cosine~$v_{\thetab}(\Xb)$ are then mutually independent. This multivariate sign is therefore a quantity that is more appealing than the ``projection''~$\Gamb_{\thetab}^{T} \Xb$, that is neither distribution-free nor independent of~$v_{\thetab}(\Xb)$. If~$\Xb$ admits a density, then this density is of the form
\begin{align}
\label{densitys}
{\bf x}
\mapsto
f_{\thetab,g}({\bf x})
=
c_{p,g} \, g({\bf x}^{T} \thetab),
\end{align}
where $c_{p,g}(>0)$ is a normalizing constant and~$g: [-1,1] \longrightarrow [0,\infty)$ is henceforth referred to as an \emph{angular function}. Then,~$v_{\thetab}(\Xb)$ is absolutely continuous with respect to the Lebesgue measure on~$[-1,1]$ and the corresponding density is
\begin{align} \label{coss}
v \mapsto \tilde{g}_p(v):=\omega_{p-1} c_{p,g}(1-v^2)^{(p-3)/2} g(v),
\end{align}
where $\omega_{p-1}:=2\pi^\frac{p-1}{2}/\Gamma(\frac{p-1}{2})$ is the surface area of~${\cal S}^{p-2}$. Application of \eqref{coss} to the vMF with location~$\thetab$ and concentration $\kappa$ (notation: ${\cal M}_{p}(\thetab,\kappa)$){,  that is, to~$g(t)=\exp(\kappa t)$,} gives $c_{p,g}=c^{\cal M}_{p, \kappa}=\kappa^{\frac{p-2}{2}}/((2\pi)^{\frac{p}{2}} I_{\frac{p-2}{2}}(\kappa))$, where $I_\nu$ is the order-$\nu$ modified Bessel function of the first kind.

\subsection{The proposed tests}
\label{sec:thetests}

In view of the above considerations, it is natural to test the null of rotational symmetry about~$\thetab$ by testing that~$\ub_{\thetab}(\Xb)$ is uniformly distributed over~$\mathcal{S}^{p-2}$. Since there are extremely diverse alternatives to uniformity on~$\mathcal{S}^{p-2}$, one may first want to consider \textit{location} alternatives and \textit{scatter} alternatives, the ones associated with violations of the expectation and the covariance conditions in \eqref{mom}, respectively. The tests we propose in this paper are designed to detect such alternatives.\\

Let~$\Xb_1,\ldots,\Xb_n$ be a random sample from a distribution on~$\mathcal{S}^{p-1}$ and consider the problem of testing the null~$\mathcal{H}_{0,\thetab}:$~``$\Xb_1$ is rotationally symmetric about~$\thetab$''. Writing~$\Ub_{i,\thetab}:=\ub_{\thetab}(\Xb_i)$, $i=1,\ldots,n$, the first test we propose rejects the null hypothesis for large values of
$$
Q^{{\rm loc}}_{\thetab}
:=
\frac{p-1}{n} \sum_{i,j=1}^n \Ub_{i,\thetab}^{T} \Ub_{j,\thetab}=n(p-1)\|\bar{\Ub}_{\thetab}\|^2,
$$
with $\bar{\Ub}_{\thetab}:=\frac{1}{n}\sum_{i=1}^n\Ub_{i,\thetab}$. This test statistic coincides with the celebrated \cite{Ray1919} test statistic computed from the $\Ub_{i,\thetab}$'s. Alternatively, if it is assumed that the~$\Xb_i$'s are sampled from a rotationally symmetric distribution (about an unspecified location), then the test also coincides with the \cite{PaiVer2013} \emph{sign} test for the null that the unknown location is equal to~$\thetab$. Since, under the null~$\mathcal{H}_{0,\thetab}$, the $\Ub_{i,\thetab}$'s form a random sample from the uniform distribution over~$\mathcal{S}^{p-2}$, the Central Limit Theorem (CLT) readily entails that $\sqrt{n}\bar{\Ub}_{\thetab}\inlaw\mathcal{N}({\bf0},\frac{1}{p-1}{\bf I}_{p-1})$, and hence that~$Q^{{\rm loc}}_{\thetab}\inlaw\chi^2_{p-1}$ under~$\mathcal{H}_{0,\thetab}$, where $\inlaw$ denotes convergence in distribution.  The resulting test, $\phi_\thetab^{\rm loc}$ say, then rejects the null hypothesis $\mathcal{H}_{0,\thetab}$ at asymptotic level~$\alpha$ whenever
$
Q^{{\rm loc}}_{\thetab} > \chi_{p-1,1- \alpha}^2,
$
where~$\chi_{\ell, 1- \alpha}^2$ denotes the $\alpha$-upper quantile of the chi-squared distribution with $\ell$ degrees of freedom. As we will show, this test typically detects the location alternatives violating the expectation condition in~\eqref{mom}.\\

In contrast, the second test we propose is designed to show power against the scatter alternatives that violate the isotropic covariance condition in~\eqref{mom}. This second test rejects~$\mathcal{H}_{0,\thetab}$ for large values \nolinebreak[4]of
$$
Q^{{\rm sc}}_{\thetab}
:=
\frac{p^2-1}{2n} \sum_{i,j=1}^n \left( (\Ub_{i,\thetab}^{T}\Ub_{j,\thetab})^2-\frac{1}{p-1}\right)
=
\frac{n(p^2-1)}{2}
\,
\bigg(
{\rm tr}\big[\Sb_{\thetab}^2\big]-\frac{1}{p-1}
\bigg),
$$
where we let
$
\Sb_{\thetab}
:=
\frac{1}{n} \sum_{i=1}^n \Ub_{i,\thetab}\Ub_{i,\thetab}^{T}.
$
Using again the fact that, under~$\mathcal{H}_{0,\thetab}$, the $\Ub_{i,\thetab}$'s form a random sample from the uniform distribution over~$\mathcal{S}^{p-2}$, it readily follows from \cite{HalPai2006} that~$Q^{{\rm sc}}_{\thetab}\inlaw\chi^2_{(p-2)(p+1)/2}$ under~$\mathcal{H}_{0,\thetab}$. The resulting test, $\phi_\thetab^{\rm sc}$ say, then rejects the null hypothesis $\mathcal{H}_{0,\thetab}$ at asymptotic level~$\alpha$ whenever~$Q^{{\rm sc}}_{\thetab} > \chi_{(p-2)(p+1)/2,1- \alpha}^2$.\\

For each test, thus, the asymptotic distribution of the test statistic, under the null of rotational symmetry about a \emph{specified} location~$\thetab$, follows from results available in the literature. Note that these are results under the usual fixed-$p$ large-$n$ asymptotic framework. In a high-dimensional asymptotic framework where~$p=p_n$ goes to infinity with~$n$ (at an arbitrary rate), it can be showed that, under the null, 
\begin{align}
\label{highW}
\frac{Q_{\thetab}^{\rm loc}-(p_n-1)}{\sqrt{2 (p_n-1)}}
\inlaw
\mathcal{N}(0,1)
\quad\textrm{and}
\quad
\frac{Q_{\thetab}^{\rm sc}-\frac{(p_n-2)(p_n+1)}{2}}{\sqrt{(p_n-2)(p_n+1)}}
\inlaw
\mathcal{N}(0,1),
\end{align} 
which allows also to apply the tests in high dimensions. The proofs of these high-dimensional limiting results are actually based on \cite{PaiVer2016}. For instance, the asymptotic normality result for~$Q_{\thetab}^{\rm loc}$ in~(\ref{highW}) is Theorem~2.2 from that paper, whereas the one for~$Q_{\thetab}^{\rm sc}$ follows rather directly from Theorem~2.5 \textit{ibid}. \\

Yet, two important questions remain open at this stage: (\textit{i}) do~$\phi_\thetab^{\rm loc}$ and~$\phi_\thetab^{\rm sc}$ behave well under non-null distributions? In particular, are there alternatives to~$\mathcal{H}_{0,\thetab}$ against which these tests would enjoy some power optimality? (\textit{ii}) To test rotational symmetry about an unspecified location~$\thetab$, can one use the tests~$\phi_{\hat{\thetab}}^{\rm loc}$ and~$\phi_{\hat{\thetab}}^{\rm sc}$ obtained by replacing~$\thetab$ with an appropriate estimator~$\hat{\thetab}$ in~$\phi_\thetab^{\rm loc}$ and~$\phi_\thetab^{\rm sc}$? We address~(\textit{i}) in Sections~\ref{sec:optim1} and~\ref{sec:optim2} (for appropriate scatter and location alternatives, respectively), and~(\textit{ii}) in Section~\ref{sec:unspecloc}. 

\subsection{Non-rotationally symmetric tangent distributions}
\label{sec:distribs}

As explained in the previous section, if $\Xb$ is rotationally symmetric about~$\thetab$, then the sign~$\Ub:=\ub_{\thetab}(\Xb)$ (see~\eqref{vandsignU}) is uniformly distributed over~$\mathcal{S}^{p-2}$ and is independent of the cosine~$V:=v_{\thetab}(\Xb)$. Vice versa, it directly follows from the tangent-normal decomposition in~\eqref{tann} that any rotational distribution on~$\mathcal{S}^{p-1}$ can be obtained as the distribution of
\begin{align}
\label{ff}
V \thetab +
\sqrt{1-V^2}
\
\Gamb_{\thetab}
\Ub,
\end{align}
where~$\Ub$ is a random vector that is uniformly distributed over~$\mathcal{S}^{p-2}$ and where the random variable~$V$ with values in~$[-1,1]$ is independent of~$\Ub$. In this section, we introduce natural alternatives to rotational symmetry by relaxing some of the distributional constraints on~$\Ub$ in~\eqref{ff}. Rather than assuming that~$\Ub$ is uniformly distributed over~$\mathcal{S}^{p-2}$, we construct two families of non-rotationally symmetric distributions for which $\Ub$ follows an \textit{angular central Gaussian distribution} (\citealp{tyler1987}) and a vMF distribution.\\

For the first family, recall that the random $(p-1)$-vector~$\Ub$ has an angular central Gaussian distribution on~$\mathcal{S}^{p-2}$ with shape parameter~$\Lamb$ (notation: $\Ub\sim \mathcal{A}_{p-1}(\Lamb)$) if it admits the density $\ub \mapsto c^{\cal A}_{p-1,\Lamb}  (\ub^{T} \Lamb^{-1} \ub)^{-(p-1)/2}$ with respect to the surface area measure~$\sigma_{p-2}$ on~$\mathcal{S}^{p-2}$, where~$c^{\cal A}_{p-1,\Lamb}:=\big(\omega_{p-1}\allowbreak({\rm det}\,\Lamb)^{1/2}\big)^{-1}$ is a normalizing constant. Here, the scatter parameter~$\Lamb$ is a $(p-1) \times (p-1)$ symmetric and positive-definite matrix that is normalized into a shape matrix in the sense that~${\rm tr}[\Lamb]=p-1$ (without this normalization,~$\Lamb$ would be identified up to a positive scalar factor only). Letting~$\mathcal{G}$ be the set of all cumulative distribution functions~$G$ over~$[-1,1]$, and~$\mathcal{L}_{p-1}$ be the collection of shape matrices~$\Lamb$, we then introduce the family of \textit{tangent elliptical distributions}.

\begin{defin}
	\label{def1}
	Let~$\thetab\in\mathcal{S}^{p-1}$,~$G\in\mathcal{G}$, and $\Lamb\in\mathcal{L}_{p-1}$. Then the random vector~$\Xb$ has a \emph{tangent elliptical distribution} on~$\mathcal{S}^{p-1}$ with location~$\thetab$, angular distribution function~$G$, and shape~$\Lamb$ if and only if~$
	\Xb
	\stackrel{\mathcal{D}}{=}
	V \thetab +
	\sqrt{1-V^2}
	\
	\Gamb_{\thetab}
	\Ub,
	$
	where~$V\sim G$ and~$\Ub\sim \mathcal{A}_{p-1}(\Lamb)$ are mutually independent. If~$V\!$ admits the density~\eqref{coss} involving the angular function~$g$, then we will write~$\Xb\sim \mathcal{TE}_p(\thetab,g,\Lamb)$.
\end{defin}

Clearly, rotationally symmetric distributions are obtained for~$\Lamb=\mathbf{I}_{p-1}$. Since the distribution~$\mathcal{A}_{p-1}(\Lamb)$ can be obtained by projecting radially on~$\mathcal{S}^{p-2}$ a $(p-1)$-dimensional  elliptical distribution with location~$\mathbf{0}$ and scatter~$\Lamb$, the distributions in Definition~\ref{def1} form an \textit{elliptical extension} of the class of the (by nature, spherical) rotationally symmetric distributions, which justifies the terminology. In the absolutely continuous case, the following result provides the density of a tangent elliptical distribution.

\begin{theo}
	\label{pdfellipt}
	If~$\Xb\sim \mathcal{TE}_p(\thetab, g,\Lamb)$, then~$\Xb$ is absolutely continuous  and the corresponding density is
	$
	\xb
	\mapsto
	f^{\cal TE}_{\thetab, g,\Lamb}
	(\xb)
	=
	\omega_{p-1} c_{p,g}c^{\cal A}_{p-1,\Lamb} g(v_{\thetab}(\xb)) ({\bf u}_{\thetab}^{T}(\xb) {\Lamb}^{-1} {\bf u}_{\thetab}(\xb))^{-(p-1)/2}
	$.
\end{theo}

As mentioned above, tangent elliptical distributions provide an elliptical extension of the class of rotationally symmetric distributions, hence in particular of vMF distributions. Another elliptical extension of vMF distributions is the \cite{Kent1982} class of Fisher--Bingham distributions. The tangent elliptical distributions show several advantages with respect to the latter: (\textit{i}) they form a semiparametric class of distributions that contains \textit{all} rotationally symmetric distributions; (\textit{ii}) the densities of tangent elliptical distributions involve normalizing constants that are simple to compute (see, e.g., \cite{KuWo05} for the delicate problem of approximating normalizing constants in the Fisher--Bingham model); (\textit{iii}) simulation is straightforward. \\

\begin{figure}[!h]
	\vspace*{-0.5cm}
	\centering
	\includegraphics[width=0.9\textwidth,clip,trim={0cm 0cm 0cm 1.5cm}]{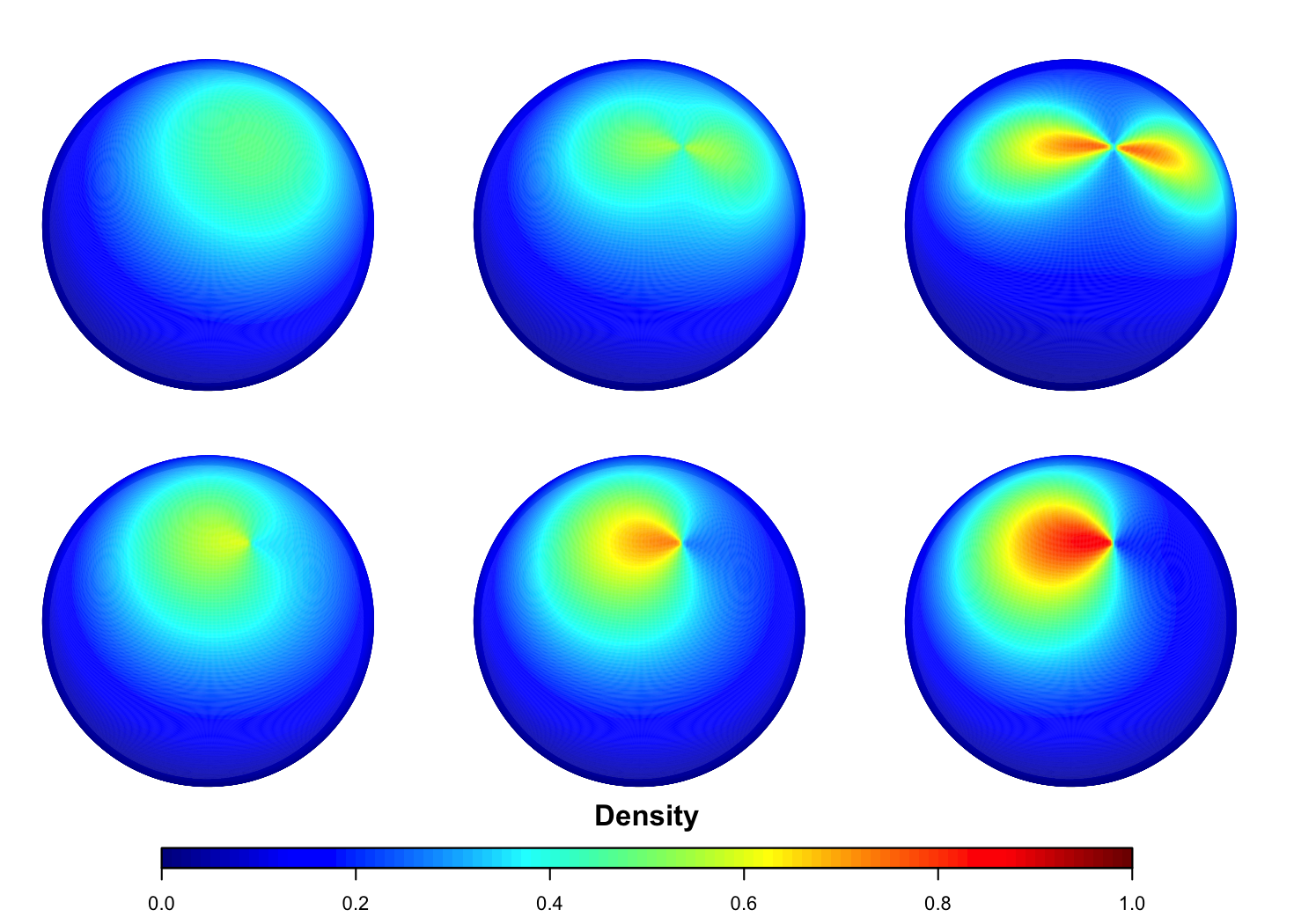}
	\vspace*{-0.25cm}
	\caption{\small Contour plots of tangent elliptical and tangent vMF densities, both with~$g(z)=\exp(3z)$. Top row: from left to right, tangent elliptical with shape matrices~$\Lamb=\binom{
			\,1+a\quad0\,}{
			\,0\quad1-a\,}$, $a=0$ (rotationally symmetric), $a=0.15$, and $a=0.45$. Bottom row: from left to right, tangent vMF densities with skewness intensities~$\kappa=0.25,0.50$, and $0.75$. \label{fig:1}}
\end{figure}

The second class of distributions we introduce, namely the class of \emph{tangent vMF distributions}, is obtained by assuming that~$\Ub\sim\mathcal{M}_{p-1}({\pmb \mu},\kappa)$. Unlike the tangent elliptical distributions, under which~$\Ub$ assumes an axial distribution on~$\mathcal{S}^{p-2}$, the unimodality of $\mathcal{M}_{p-1}({\pmb \mu},\kappa)$ in the tangent space provides a skewed distribution for~$\Xb$ about $\thetab$ (see the bottom row of Figure \ref{fig:1}). Theorem~\ref{pdf2} provides the density of the tangent vMF distributions in the absolutely continuous case. Its proof is along the same lines as the proof of Theorem~\ref{pdfellipt}, hence is omitted.

\begin{defin}
	\label{def2}
	Let~$\thetab \in\mathcal{S}^{p-1}$, $G\in\mathcal{G}$, ${\pmb \mu} \in {\cal S}^{p-2}$, and $\kappa\geq0$. Then the random vector~$\Xb$ has a \emph{tangent vMF distribution} on~$\mathcal{S}^{p-1}$ with location~$\thetab$, angular distribution function~$G$, skewness direction~${\pmb \mu}$, and skewness intensity~$\kappa$ if and only if~$
	\Xb
	\stackrel{\mathcal{D}}{=}
	V \thetab +
	\sqrt{1-V^2}
	\
	\Gamb_{\thetab}
	\Ub,
	$
	where~$V\sim G$ and~$\Ub\sim {\cal M}_{p-1}({\pmb \mu}, \kappa)$ are mutually independent. If~$V\!$ admits the density~\eqref{coss} involving the angular function~$g$, then we will write~$\Xb\sim \mathcal{TM}_p(\thetab, g, {\pmb \mu}, \kappa)$.
\end{defin}

\begin{theo}
	\label{pdf2}
	If~$\Xb\sim \mathcal{TM}_p(\thetab, g, {\pmb \mu}, \kappa)$, then~$\Xb$ is absolutely continuous  and the corresponding density is
	$
	\xb
	\mapsto
	f^{\cal TM}_{\thetab, g, {\pmb \mu}, \kappa}(\xb)= \omega_{p-1} c_{p,g} c^{\cal M}_{p-1, \kappa}  g(v_{\thetab}(\xb)) \exp(\kappa {\pmb \mu}^{T}{\bf u}_{\thetab}(\xb))
	$.
\end{theo}

Note that, albeit our framework is $p\geq3$, the distributions of Definitions \ref{def1} and \ref{def2} are also properly defined for $p=2$. In that case, the sign~${\bf U}$ takes values in~$\mathcal{S}^{0}=\{-1,1\}$, $\omega_{1}=2$, and the angular central Gaussian and the vMF densities become probability mass functions over~$\mathcal{S}^{0}$. The former, since it is an axial distribution, puts equal mass in $\pm1$. Since $I_{-\frac{1}{2}}(\kappa)=\sqrt{2/(\pi\kappa)}\cosh(\kappa)$, the vMF associated with~$\mu\in\mathcal{S}^{0}$ assigns probabilities $\exp(\pm\mu\kappa)/(\exp(-\mu\kappa)+\exp(\mu\kappa))$ to $\pm1$, respectively. Therefore, only the tangent vMF distributions provide alternatives to rotational symmetry when $p=2$. This is coherent with the fact that $Q_{\thetab}^{\rm sc}$ is a non-random test statistic when $p=2$ and therefore does not provide any reasonable test. To deal with non-degenerate tests, we restrict to~$p\geq3$ in\nolinebreak[4] the\nolinebreak[4] sequel. 

\subsection{Non-null results for tangent elliptical alternatives}
\label{sec:optim1}

In this section, we investigate the performances of the tests~$\phi^{\rm loc}_\thetab$ and~$\phi^{\rm sc}_\thetab$ under the tangent elliptical alternatives to rotational symmetry introduced above. To do so, we need the following notation: $\vecho({\bf A})$ for the $(p(p+1)/2)$-dimensional vector stacking the upper-triangular entries of a $p \times p$ symmetric matrix ${\bf A}=(A_{ij})$; $\vechrond({\bf A})$ for $\vecho({\bf A})$ with the first entry ($A_{11}$) excluded; ${\bf M}_p$ for the matrix satisfying~${\bf M}_p^{T} \vechrond({\bf A})= \veco({\bf A})$ for any $p \times p$ symmetric matrix ${\bf A}$ with ${\rm tr}[{\bf A}]=0$; ${\bf J}_p:= \sum_{i,j =1}^{p}({\bf e}_{p,i}{\bf e}_{p,j}^{T})\otimes ({\bf e}_{p,i}{\bf e}_{p,j}^{T}) = ({\rm vec}\,{\bf I}_p) ({\rm vec}\,{\bf I}_p)^{T}$;
${\bf K}_{p}:= \sum_{i,j =1}^{p}({\bf e}_{p,i}{\bf e}_{p,j}^{T})\otimes ({\bf e}_{p,j}{\bf e}_{p,i}^{T})$ for the {\em commutation matrix}, where~${\bf e}_{p,\ell}$ denotes the $\ell$-th vector of the canonical basis of~$\R^p$. Since the shape matrix~$\Lamb$ of a tangent elliptical distribution is symmetric and satisfies~${\rm tr}[\Lamb]=p-1$, it is completely characterized by~$\vechrond(\Lamb)$.  Throughout, we let~$V_{i,\thetab}:=v_{\thetab}(\Xb_i)=\Xb_i^{T}\thetab$  and we denote as $\chi^2_\nu(\lambda)$ the chi-squared distribution with $\nu$ degrees of freedom and $\lambda$ being the non-centrality parameter (so that $\chi^2_\nu(0)\stackrel{\mathcal{D}}{=}\chi^2_\nu$). \\

In order to examine log-likelihood ratios involving the angular functions $g$, we need to assume some regularity conditions on~$g$. More precisely, we will restrict to the collection~$\mathcal{G}_a$ of non-constant angular functions~$g:[-1,1]\longrightarrow(0,\infty)$ that are absolutely continuous and for which~$\mathcal{J}_p({g}):= \int_{-1}^{1} \varphi_{g}^2(t) (1-t^2) \tilde{g}_p(t) \, dt$ is finite, where~$\varphi_{g}:=\dot{g}/g$ involves the almost everywhere derivative~$\dot{g}$ of~$g$.\\

Consider then the semiparametric model $\big\{ {\rm P}_{\thetab, g,\Lamb}^{{\cal TE}(n)} : \thetab \in {\cal S}^{p-1}, g\in {\cal G}_a, \Lamb \in {\cal L}^{p-1}  \big\}$, where~${\rm P}_{\thetab, g,\Lamb}^{{\cal TE}(n)}$ denotes the probability measure associated with observations $\Xb_1, \ldots, \Xb_n$ that are randomly sampled from the tangent elliptical distribution~$\mathcal{TE}_p(\thetab, g,\Lamb)$. In the rotationally symmetric case, that is, for~$\Lamb=\mathbf{I}_{p-1}$, we will simply write~${\rm P}_{\thetab,g}^{(n)}$ instead of~${\rm P}_{\thetab,g,\mathbf{I}_{p-1}}^{{\cal TE}(n)}$. Investigating the optimality of tests of rotational symmetry against tangent elliptical alternatives requires studying the asymptotic behavior of tangent elliptical log-likelihood ratios associated with local deviations from~$\Lamb=\mathbf{I}_{p-1}$. This leads to the following \emph{Local Asymptotic Normality} (LAN) result.

\begin{theo}
	\label{LANell}
	Fix~$\thetab\in\mathcal{S}^{p-1}$ and $g\in\mathcal{G}_a$.
	Let~${\taub_n}:=({\bf t}_n^{T}, \vechrond(\mathbf{L}_n)^{T})^{T}$, where~$({\bf t}_n)$ is a bounded sequence in~$\R^p$ such that~$\thetab_n:=\thetab+n^{-1/2} {\bf t}_n\in\mathcal{S}^{p-1}$ for any~$n$, and where~$({\mathbf{L}}_n)$ is a bounded sequence of $(p-1)\times (p-1)$ matrices such that~$\Lamb_n:={\bf I}_{p-1}+n^{-1/2} {\mathbf{L}}_n\in\mathcal{L}_{p-1}$ for any~$n$.
	Then, the tangent elliptical log-likelihood ratio associated with local deviations~$\Lamb_n$ from $\Lamb={\bf I}_{p-1}$ satisfies
	\begin{align} \label{LAQell}
	\log
	\frac{ d{\rm P}^{{\cal TE}(n)}_{\thetab_n,g,\Lamb_n}}{ d{\rm P}\n_{\thetab,g}}
	=
	\taub_n^{T}\Deltab_{\thetab,g}^{{\cal TE}(n)}
	- \frac{1}{2} \, \taub_n^{T}\Gamb_{\thetab, g}^{\cal TE} {\taub_n}+o_{\rm P}(1)
	\end{align}
	as $\ny$ under $ {\rm P}_{\thetab,g}\n$, where the central sequence
	$$
	\Deltab_{\thetab,g}^{{\cal TE}(n)}
	:=
	\bigg(
	\begin{array}{c}
	\Deltab_{\thetab,g;1}\n
	\\
	\Deltab_{\thetab;2}^{{\cal TE}(n)}
	\end{array}
	\bigg)
	:=
	\Bigg(
	\begin{array}{c}
	\frac{1}{\sqrt{n}} \sum_{i=1}^n \varphi_g(V_{i,\thetab}) (1-V_{i,\thetab}^2)^{1/2} \, \Gamb_\thetab {\bf U}_{i,\thetab}
	\\
	\frac{p-1}{2\sqrt{n}} \, {\bf M}_{p} \sum_{i=1}^n \veco \big({\bf U}_{i,\thetab}{\bf U}_{i,\thetab}^{T} - \frac{1}{p-1}\,{\bf I}_{p-1} \big)
	\end{array}
	\Bigg)
	$$
	is, still under $ {\rm P}_{\thetab,g}\n$, asymptotically normal with mean zero and covariance matrix
	$$
	\Gamb_{\thetab,g}^{{\cal TE}}
	:=
	\bigg(
	\begin{array}{cc}
	\Gamb_{\thetab,g;11}
	& \mathbf{0} \\
	\mathbf{0} & \Gamb_{\thetab;22}^{\cal TE}
	\end{array}
	\bigg)
	:=
	\bigg(
	\begin{array}{cc}
	\frac{{\cal J}_p(g)}{p-1} ({\bf I}_p- \thetab\thetab^{T}) & \mathbf{0} \\
	\mathbf{0} & \frac{p-1}{4(p+1)}{\bf M}_p \big( {\bf I}_{(p-1)^2}+ {\bf K}_{p-1}\big) {\bf M}_p^{T}
	\end{array}
	\bigg).
	$$
\end{theo}

The restriction that~$g\in\mathcal{G}_a$ guarantees that the $g$-parametric submodel of the tangent elliptical model has a finite Fisher information for~$\thetab$ in the vicinity of rotational symmetry. Any LAN result requires a finite Fisher information condition of this sort (along with a smoothness condition that allows defining Fisher information). The LAN result in Theorem~\ref{LANell} easily provides the following corollary.

\begin{coro}
	\label{powerE}
	Fix~$\thetab\in\mathcal{S}^{p-1}$ and $g\in \mathcal{G}_a$.
	Let~$(\Lamb_n)$ and $({\mathbf{L}}_n)$ be as in Theorem \ref{LANell}, now with ${\mathbf{L}}_n \to  {\mathbf{L}}\neq {\bf 0}$. Then, under~${\rm P}^{{\cal TE}(n)}_{\thetab,g,\Lamb_n}$: (\textit{i})~$Q^{{\rm loc}}_{\thetab}\inlaw\chi^2_{p-1}$; (\textit{ii})~$Q^{{\rm sc}}_{\thetab}\inlaw\chi^2_{(p-2)(p+1)/2}(\lambda)$, with $\lambda=(p-1){\rm tr}[{\mathbf{L}}^2]/(2(p+~1))$.
\end{coro}

First note that (\textit{i}) implies that, for the local alternatives considered, the null and non-null asymptotic distributions of~$Q^{{\rm loc}}_{\thetab}$ do coincide, so that the test~$\phi^{\rm loc}_\thetab$ has asymptotic power~$\alpha$ against such alternatives. On the contrary, (\textit{ii}) shows that the test~$\phi^{\rm sc}_\thetab$ exhibits non-trivial asymptotic powers against any alternatives associated with~$\Lamb_n={\bf I}_{p-1}+n^{-1/2} {\mathbf{L}}_n$, ${\mathbf{L}}_n \to  {\mathbf{L}}\neq {\bf 0}$ (note indeed that~${\rm tr}[{\mathbf{L}}^2]$ is the squared Frobenius norm of~${\mathbf{L}}$). Note also that, since~${\mathbf{L}}$ has trace zero by construction, the non-centrality parameter~$(p-1){\rm tr}[{\mathbf{L}}^2]/(2(p+1))$ above is proportional to the variance of the eigenvalues of~${\mathbf{L}}$, which is in line with the fact that~$\phi^{\rm sc}_{\thetab}$ has the nature of a sphericity test.\\

While Corollary~\ref{powerE} shows that the test~$\phi^{\rm sc}_{\thetab}$ can detect local alternatives of a tangent elliptical nature, it does not provide information on the possible optimality of this test. General results on the Le Cam's asymptotic theory of statistical experiments (see, e.g., Chapter 5 of \citealp{LeyVerbook}) together with Theorem~\ref{LANell} directly entail that, when testing~$\big\{ {\rm P}\n_{\thetab,g} \big\}$ against~$\bigcup_{\Lamb\in \mathcal{L}_{p-1}\setminus\{\mathbf{I}_p\}} \big\{{\rm P}^{{\cal TE}(n)}_{\thetab, g,\Lamb}\big\}$, a locally asymptotically maximin test at asymptotic level~$\alpha$  rejects the null hypothesis whenever
\begin{align}
\label{optell}
Q^{\rm sc}_{\thetab}
=\big(\Deltab_{\thetab;2}^{{\cal TE}(n)}\big)^{T}
\big(\Gamb_{\thetab;22}^{\cal TE}\big)^{-1}
\Deltab_{\thetab;2}^{{\cal TE}(n)}
>
\chi_{(p-2)(p+1)/2,1- \alpha}^2.
\end{align}
Now, using the closed form for the inverse of~$\Gamb_{\thetab;22}^{\cal TE}$ in Lemma~5.2 from \cite{HPDec}, it is easy to show that the test statistic in~\eqref{optell} coincides with~$Q^{{\rm sc}}_{\thetab}$. Since this holds at \emph{any} angular function~$g$ in~$\mathcal{G}_a$, we proved the following result.

\begin{coro}
	\label{optimE}
	When testing~$\bigcup_{g\in \mathcal{G}_a}\big\{ {\rm P}\n_{\thetab,g}\big\}$ against~$\bigcup_{g\in \mathcal{G}_a} \bigcup_{\Lamb\in \mathcal{L}_{p-1}\setminus\{\mathbf{I}_p\}} \big\{{\rm P}^{{\cal TE}(n)}_{\thetab, g,\Lamb}\big\}$, the test~$\phi^{{\rm sc}}_{\thetab}$ is locally asymptotically maximin at asymptotic level~$\alpha$.
\end{coro}

We conclude that, when testing rotational symmetry about a specified location~$\thetab$ against contiguous tangent elliptical alternatives, the location test~$Q^{{\rm loc}}_{\thetab}$ \textit{does not show any power}, while the scatter test~$Q^{{\rm sc}}_{\thetab}$ is \textit{optimal} in the Le Cam sense, at any angular function~$g\in \mathcal{G}_a$.

\subsection{Non-null results under tangent vMF alternatives}
\label{sec:optim2}

To investigate the non-null behavior of the proposed tests under tangent vMF alternatives, we consider the semiparametric model~$\big\{ {\rm P}_{\thetab, g,{\pmb \mu},\kappa}^{{\cal TM}(n)} : \thetab\in {\cal S}^{p-1}, g\in {\cal G}_a  ,{\pmb \mu} \in {\cal S}^{p-2},\kappa\geq 0\big\}$, where~${\rm P}_{\thetab, g,{\pmb \mu},\kappa}^{{\cal TM}(n)}$ denotes the probability measure associated with observations $\Xb_1, \ldots, \Xb_n$ that are randomly sampled from the tangent vMF distribution~$\mathcal{TM}_p(\thetab, g, {\pmb \mu}, \kappa)$; for~$\kappa=0$, ${\rm P}_{\thetab, g,{\pmb \mu},\kappa}^{{\cal TM}(n)}$ is defined as the rotationally symmetric hypothesis~${\rm P}_{\thetab,g}^{(n)}$ (see the notation introduced in Section~\ref{sec:optim1}). To investigate the optimality properties of tests of rotational symmetry against such alternatives, it is convenient to parametrize this model with~$\thetab$, $g$, and $\deltab$, where we let~$\deltab:=\kappa {\pmb \mu}$; obviously, we will then use the notation~${\rm P}_{\thetab, g,\deltab}^{{\cal TM}(n)}$. In this new parametrization, the null hypothesis of rotational symmetry about~$\thetab$ coincides with ${\cal H}_{0,\thetab}: \deltab={\bf 0}$. The main advantage of the parametrization in $\deltab \in \R^{p-1}$ over the original one in $(\kappa, {\pmb \mu}) \in [0,\infty) \times {\cal S}^{p-2}$ is that the $\deltab$-parameter space is standard (it is the Euclidean space~$\R^{p-1}$), while the $(\kappa, {\pmb \mu})$-one is curved.\\

As for tangent elliptical distributions, our investigation of optimality issues is based on a LAN result.

\begin{theo}
	\label{LANvM}
	Fix~$\thetab\in\mathcal{S}^{p-1}$ and $g\in\mathcal{G}_a$.
	Let~${\taub_n}:=({\bf t}_n^{T}, {\bf d}_n^{T})^{T}$, where~$({\bf t}_n)$ is a bounded sequence in~$\R^p$ such that~$\thetab_n:=\thetab+n^{-1/2} {\bf t}_n\in\mathcal{S}^{p-1}$ for any~$n$, and~$\deltab_n:=n^{-1/2} {\bf d}_n$ with~$({\bf d}_n)$ a bounded sequence in~$\R^{p-1}$.
	Then, the tangent vMF log-likelihood ratio associated with local deviations~$\deltab_n$ from $\deltab={\bf0}$~is
	\begin{align*}
	\log
	\frac{ d{\rm P}^{{\cal TM}(n)}_{\thetab_n,g,\deltab_n}}{ d{\rm P}\n_{\thetab,g}}
	=
	\taub_n^{T} \Deltab_{\thetab,g}^{{\cal TM}(n)}
	- \frac{1}{2} \, \taub_n^{T} \Gamb_{\thetab, g}^{\cal TM} {\taub_n}+o_{\rm P}(1),
	\end{align*}
	as $\ny$ under $ {\rm P}_{\thetab,g}\n$, where the central sequence
	$$
	\Deltab_{\thetab,g}^{{\cal TM}(n)}
	:=
	\bigg(
	\begin{array}{c}
	\Deltab_{\thetab,g;1}\n
	\\
	\Deltab_{\thetab;2}^{{\cal TM}(n)}
	\end{array}
	\bigg)
	:=
	\Bigg(
	\begin{array}{c}
	\frac{1}{\sqrt{n}} \sum_{i=1}^n \varphi_g(V_{i,\thetab}) (1-V_{i,\thetab}^2)^{1/2} \, \Gamb_\thetab {\bf U}_{i,\thetab}
	\\
	\frac{1}{\sqrt{n}} \sum_{i=1}^n  {\bf U}_{i,\thetab}
	\end{array}
	\Bigg)
	$$
	is, still under $ {\rm P}_{\thetab,g}\n$,  asymptotically normal with mean zero and covariance matrix
	$$
	\Gamb_{\thetab,g}^{{\cal TM}}
	:=
	\bigg(
	\begin{array}{cc}
	\Gamb_{\thetab,g;11}
	& \Gamb_{\thetab,g;12}^{\cal TM} \\
	\Gamb_{\thetab,g;21}^{\cal TM} & \Gamb_{22}^{\cal TM}
	\end{array}
	\bigg)
	:=
	\bigg(
	\begin{array}{cc}
	\frac{{\cal J}_p(g)}{p-1} ({\bf I}_p- \thetab\thetab^{T}) & \frac{\mathcal{I}_p(g)}{p-1}\, \Gamb_\thetab \\[.5mm]
	\frac{\mathcal{I}_p(g)}{p-1}\, \Gamb_\thetab^{T} & \frac{1}{p-1}\, {\bf I}_{p-1}
	\end{array}
	\bigg),
	$$
	with~$\mathcal{I}_p(g):=\int_{-1}^{1} \varphi_{g}(t) \sqrt{1-t^2}\, \tilde{g}_p(t) \, dt$.
\end{theo}

Unlike for the LAN property in Theorem~\ref{LANell}, the Fisher information matrix associated with the LAN property above, namely~$\Gamb_{\thetab,g}^{{\cal TM}}$, is not block-diagonal. Also, Jensen's inequality ensures that~$\mathcal{J}_p(g)\geq \mathcal{I}^2_p(g)$, which confirms the finiteness of $\mathcal{I}_p(g)$ and the positive semidefiniteness of~$\Gamb_{\thetab,g}^{{\cal TM}}$. It is also easy to check that the \textit{only} angular functions for which Jensen's inequality is actually an equality (hence, for which~$\Gamb_{\thetab,g}^{{\cal TM}}$ is singular) are of the form~$g(t)=C \exp(\kappa \arcsin(t))$ for some real constants~$C$ and~$\kappa$.

\begin{coro}
	\label{powerF}
	Fix~$\thetab\in\mathcal{S}^{p-1}$ and $g\in \mathcal{G}_a$.
	Let~$({\pmb \mu}_n)$ be a sequence in~$\mathcal{S}^{p-2}$ that converges to~${\pmb \mu}$. Let~$\kappa_n:=n^{-1/2} k_n$, where~$(k_n)$ is a sequence in~$(0,\infty)$ that converges to~$k>0$. Then, under~${\rm P}^{{\cal TM}(n)}_{\thetab,g,{\pmb\mu}_n,\kappa_n}$: (\textit{i})~$Q^{{\rm loc}}_{\thetab}\inlaw\chi^2_{p-1}(\lambda)$, with $\lambda=k^2/(p-1)$; (\textit{ii})~$Q^{{\rm sc}}_{\thetab}\inlaw\chi^2_{(p-2)(p+1)/2}$.
\end{coro}

The location test~$\phi^{\rm loc}_{\thetab}$ and the scatter test~$\phi^{\rm sc}_{\thetab}$ therefore exhibit opposite non-null behaviors under tangent vMF alternatives, compared to what occurs under tangent elliptical alternatives in Section~\ref{sec:optim1}: under contiguous tangent vMF alternatives,~$\phi^{\rm sc}_{\thetab}$ has asymptotic power equal to the nominal level~$\alpha$, whereas~$\phi^{\rm loc}_{\thetab}$ shows non-trivial asymptotic powers. Since the latter test is the test rejecting the null hypothesis of rotational symmetry about~$\thetab$ whenever
\begin{align*}
Q^{\rm loc}_{\thetab}
=
\big(\Deltab_{\thetab;2}^{{\cal TM}(n)}\big)^{T}
\big(\Gamb_{22}^{\cal TM}\big)^{-1}
\Deltab_{\thetab;2}^{{\cal TM}(n)} > \chi_{p-1,1- \alpha}^2,
\end{align*}
it is actually locally asymptotically maximin at asymptotic level~$\alpha$ when testing~$\big\{ {\rm P}\n_{\thetab,g} \big\}$ against $\bigcup_{{\pmb\mu}\in\mathcal{S}^{p-2}}\bigcup_{\kappa> 0}  \big\{{\rm P}^{{\cal TM}(n)}_{\thetab,g,{\pmb\mu},\kappa}\big\}$. Moreover, since~$Q^{\rm loc}_{\thetab}$ does not depend on~$g$, we have the following result.

\begin{coro}
	\label{optimF}
	When testing~$\bigcup_{g\in \mathcal{G}_a}\big\{{\rm P}\n_{\thetab,g}\big\}$ against~$\bigcup_{g\in \mathcal{G}_a} \bigcup_{{\pmb\mu}\in\mathcal{S}^{p-2}}\bigcup_{\kappa> 0}  \big\{{\rm P}^{{\cal TM}(n)}_{\thetab,g,{\pmb\mu},\kappa}\big\}$, the test~$Q^{{\rm loc}}_{\thetab}$ is locally asymptotically maximin at asymptotic level~$\alpha$.
\end{coro}

We conclude that the location test~$\phi^{\rm loc}_{\thetab}$ and the scatter test~$\phi^{\rm sc}_{\thetab}$ are optimal in the Le Cam sense, for any angular function in~$g\in \mathcal{G}_a$, against tangent vMF alternatives and tangent elliptical alternatives, respectively.

\section{\texorpdfstring{Testing rotational symmetry about an unspecified~$\thetab$}{Testing rotational symmetry about an unspecified theta}}
\label{sec:unspecloc}

The tests $\phi_{\thetab}^{\rm loc}$ and $\phi_{\thetab}^{\rm sc}$ studied above allow testing rotational symmetry about a given location~$\thetab$. Often, however, it is desirable to rather test for rotational symmetry about an unspecified~$\thetab$. Natural tests for this unspecified-$\thetab$ problem are obtained by substituting an estimator~$\hat{\thetab}$ for~$\thetab$ in~$\phi_{\thetab}^{\rm loc}$ and $\phi_{\thetab}^{\rm sc}$. In the present ULAN framework (it is easy to strengthen the LAN results in Theorems~\ref{LANell}--\ref{LANvM} into ULAN---\emph{Uniformly Locally Asymptotically Normal}---ones), this estimator~$\hat{\thetab}$ should actually satisfy the following assumption:

\begin{enumerate}[label=A$_{\mathcal{G}'}$,ref=A$_{\mathcal{G}'}$]
	\item The estimator~$\hat\thetab$ (with values in~${\mathcal S}^{p-1}$) is part of a sequence that is: (\textit{i}) \emph{root-$n$ consistent} under any~$g\in\mathcal{G}'$, i.e., $\sqrt{n}(\hat{\thetab}- \thetab)= O_{\rm P}(1)$ under $\bigcup_{g \in \mathcal{G}'} \big\{{\rm P}\n_{\thetab, g}\big\}$; (\textit{ii}) \emph{locally asymptotically discrete}, i.e., for all $\thetab$ and for all~$C>0$, there exists a positive integer~$M=M(C)$ such that the number of possible values of~$\hat{\thetab}$ in  $\{\tb \in \mathcal{S}^{p-1}: \sqrt{n}\, \|\tb-\thetab\|\leq C \}$ is bounded by~$M$, uniformly as $n\to\infty$.\label{assump:2}
\end{enumerate}

Part~(\textit{i}) of Assumption \ref{assump:2} requires that~$\hat{\thetab}$ is root-$n$ consistent under the null hypothesis for a broad range $\mathcal{G}'$ of angular functions~$g$. The restriction to such a~$\mathcal{G}'$ is explained by the fact that classical estimators of~$\thetab$ typically address either monotone rotationally symmetric distributions ($g$ is monotone increasing) or axial ones ($g(-t)=g(t)$ for any~$t$), but cannot deal with both types. Practitioners are thus expected to take~$\mathcal{G}'$ as the collection of monotone or symmetric angular functions, depending on the types of directional data (unimodal or axial data) they are facing. In the unimodal case, the classical choice is the \emph{spherical mean}~$\hat{\thetab}=\bar{\Xb}/\|\bar{\Xb}\|$, with~$\bar{\Xb}:=\frac{1}{n}\sum_{i=1}^n \Xb_i$, whereas, in the axial case, estimators of~$\thetab$ are typically based on the eigenvectors of the covariance matrix~$\Sb:=\frac{1}{n}\sum_{i=1}^n (\Xb_i-\bar{\Xb})(\Xb_i-\bar{\Xb})^T$. As for part~(\textit{ii}), it is a purely technical requirement with little practical implications in the sense that, for fixed~$n$, any estimate can be considered part of a locally asymptotically discrete sequence of estimators; see, e.g., page 2467 in \cite{IlmPai2011} for a discussion.\\

Now, the impact of plugging an estimator~$\hat\thetab$ satisfying Assumption \ref{assump:2} in our specified-$\thetab$ tests crucially depends on the tests considered. We first focus on scatter tests. 

\subsection{Scatter tests}
\label{sec:scattertestunsp}

The block-diagonality of the Fisher information matrix in Theorem~\ref{LANell} entails that the replacement in $Q_{\thetab}^{\rm sc}$ of~$\thetab$ with an estimator $\hat{\thetab}$ satisfying \ref{assump:2} has no asymptotic impact under the null hypothesis. More precisely, we have the following result.
\begin{prop}
	\label{asymplin}
	Let~$\hat{\thetab}$ satisfy \ref{assump:2}. Then, for any~$\thetab\in\mathcal{S}^{p-1}$ and any~$g\in \mathcal{G}_a\cap \mathcal{G}'$,
	$
	Q_{\hat \thetab}^{\rm sc}-Q_{\thetab}^{\rm sc}
	=
	o_{\rm P}(1)
	$
	as $\ny$ under ${\rm P}_{\thetab,g}\n$.
\end{prop}

From contiguity, the null asymptotic equivalence in this proposition extends to local alternatives of the form~${\rm P}^{{\cal TE}(n)}_{\thetab,g,\Lamb_n}$, with~$\Lamb_n={\bf I}_{p-1}+n^{-1/2} {\mathbf{L}}_n$ as in Theorem \ref{LANell}. Therefore, the test, $\phi_\dagger^{\rm sc}$ say, that rejects the null of rotational symmetry about an unspecified location when $Q_{\hat \thetab}^{\rm sc}> \chi_{(p-2)(p+1)/2, 1- \alpha}^2$ remains optimal in the Le Cam sense against the tangent elliptical alternatives introduced in Section~\ref{sec:distribs}. More precisely, this test is locally asymptotically maximin at asymptotic level~$\alpha$ when testing~$\bigcup_{\thetab\in\mathcal{S}^{p-1}}\bigcup_{g\in \mathcal{G}_a\cap \mathcal{G}'}\{ {\rm P}\n_{\thetab,g} \}$ against~$\bigcup_{\thetab\in\mathcal{S}^{p-1}}\bigcup_{g\in \mathcal{G}_a\cap \mathcal{G}'} \bigcup_{\Lamb\in \mathcal{L}_{p-1}\setminus\{\mathbf{I}_p\}} \big\{{\rm P}^{{\cal TE}(n)}_{\thetab, g,\Lamb}\big\}$. Of course, the same contiguity argument also implies that~$\phi_\dagger^{\rm sc}$ has asymptotic power~$\alpha$ against the local tangent vMF alternatives  considered in Corollary~\ref{powerF}.

\subsection{Location tests}
\label{sec:locationtestunsp}

Since, on the contrary, the Fisher information matrix is not block-diagonal in Theorem~\ref{LANvM}, the story is very different for location tests. The ULAN extension of this theorem yields that, if~$\hat \thetab$ satisfies Assumption~\ref{assump:2}, then
\begin{align}
\label{linlin}
\Deltab_{\hat{\thetab};2}^{{\cal TM}(n)}
-
\Deltab_{\thetab;2}^{{\cal TM}(n)}
=
-
\Gamb_{\thetab,g;21}^{\cal TM}
\sqrt{n}(\hat{\thetab}- \thetab)
+o_{\rm P}(1)
\end{align}
as $\ny$ under~${\rm P}\n_{\thetab,g}$, with~$g\in \mathcal{G}_a\cap \mathcal{G}'$, so that~$Q_{\hat{\thetab}}^{\rm loc}$ is no more asymptotically chi-squared distributed under the same sequence of (null) hypotheses. Unlike for~$Q_{\thetab}^{\rm sc}$, thus, the substitution of $\hat{\thetab}$ for~$\thetab$ in~$Q_{\thetab}^{\rm loc}$ unfortunately has a non-negligible asymptotic impact. The non-block-diagonality of the Fisher information matrix in Theorem~\ref{LANvM} also implies that the optimal unspecified-$\thetab$ location tests will show asymptotic powers under contiguous alternatives that are strictly smaller than those of the corresponding specified-$\thetab$ tests. \\

The construction of these optimal unspecified-$\thetab$ location tests relies on \emph{efficient central sequences} and is very technical, hence is deferred to the supplementary material. Here, we only describe the vMF version of the resulting tests and state its main properties. This vMF test, $\phi^{\rm loc}_{\rm vMF}$ say, rejects the null hypothesis of rotational symmetry about an unspecified~$\thetab$ whenever
\begin{align*}
Q^{\rm loc}_{\rm vMF}
:=
\widehat{\Deltab}^{T}
\widehat{\Gamb}^{-1}
\widehat{\Deltab}
>
\chi^2_{p-1,1-\alpha},
\end{align*}
where
$$
\widehat{\Deltab}
:=
\frac{1}{\sqrt{n}} \sum_{i=1}^n
\Big(
1-\hat{D}_{p}\,
(1-V_{i,\hat{\thetab}}^2)^{1/2}
\Big)
\Ub_{i, \hat{\thetab}}
\quad
\textrm{and}
\quad
\widehat{\Gamb}
:=
\frac{1}{p-1}
\,
\Big(
1-2 \hat{D}_{p} \hat{E}_{p} + \hat{D}_{p}^2  (1-\hat{F}_{p}) \Big)
\mathbf{I}_{p-1}
$$
involve the quantities
$$
\hat{D}_{p}:=\frac{(p-2)\,\sum_{i=1}^n
	V_{i,\hat{\thetab}} (1-V_{i,\hat{\thetab}}^2)^{-1/2}}{(p-1)\, \sum_{i=1}^n
	V_{i,\hat{\thetab}}},
\quad
\hat{E}_{p}
:=
\frac{1}{n} \sum_{i=1}^n
\, (1-V_{i,\hat{\thetab}}^2)^{1/2},
\quad
\textrm{and}
\quad
\hat{F}_{p}
:=
\frac{1}{n} \sum_{i=1}^n
\, V_{i,\hat{\thetab}}^2.
$$
Denoting as~$\mathcal{G}_b$ the collection of angular functions~$g$ for which
$
\int_{-1}^{1} t (1-t^2)^{-1/2}\, \tilde{g}_p(t)\, dt
<\infty
$
(note that, for any~$p\geq 3$,~$\mathcal{G}_b$ contains all angular functions~$g$ that are bounded in a neighborhood of~${\pm} 1$), we show in the supplementary material that~$\phi^{\rm loc}_{\rm vMF}$ is optimal in the Le Cam sense (more precisely, it is locally asymptotically maximin) at asymptotic level~$\alpha$ when testing
$$
\bigcup_{\thetab\in\mathcal{S}^{p-1}} \bigcup_{g\in \mathcal{G}_a\cap\mathcal{G}_b\cap\mathcal{G}'}\big\{ {\rm P}\n_{\thetab,g} \big\}
\quad\textrm{ against } \quad
\bigcup_{\thetab\in\mathcal{S}^{p-1}} \bigcup_{\eta>0} \bigcup_{{\pmb\mu}\in\mathcal{S}^{p-2}}\bigcup_{\kappa> 0}  \big\{{\rm P}^{{\cal TM}(n)}_{\thetab,g_\eta,{\pmb\mu},\kappa}\big\},
$$
with~$g_\eta(t):=\exp(\eta t)$. This vMF test therefore meets the asymptotic level constraint under all rotational symmetric densities and achieves Le Cam optimality against any tangent vMF alternatives involving a vMF angular density~$g$. It is easy to show, however, that this test still has asymptotic power~$\alpha$ against the local tangent elliptical alternatives considered in Corollary~\ref{powerE}.

\pagebreak

\section{Hybrid tests}
\label{sec:hybrid}

The location and scatter tests, either in the specified-$\thetab$ or unspecified-$\thetab$ situations, are based on the empirical checking of the moment conditions in \eqref{mom}. Both are necessary conditions for the uniformity of $\ub_{\thetab}(\Xb)$ over~$\mathcal{S}^{p-2}$, hence for rotational symmetry. For the families of alternatives introduced in Section \ref{sec:distribs}, the tests present rather extreme behaviors: either they are optimal (in the Le Cam sense), or they are blind to the contiguous alternatives. While this antithesis is desirable for testing against a specific kind of alternative, it is also a double-edged sword, since knowing the alternative on which rotational symmetry might be violated can be challenging in practice, specially for high-dimensional settings. As we explain below, a possible way out is to construct \emph{hybrid} tests that show non-trivial asymptotic powers against both types of alternatives considered.\\

Consider first the problem of testing rotational symmetry about a specified location~$\thetab$. Since ${\bf U}_{i,\thetab}$ and $\veco({\bf U}_{i,\thetab} {\bf U}_{i,\thetab}^T)$ are uncorrelated, $\Deltab_{\thetab;2}^{{\cal TM}(n)}$ and $\Deltab_{\thetab;2}^{{\cal TE}(n)}$ are uncorrelated, too. The CLT then readily entails that, under ${\rm P}_{\thetab,g}\n$,
\begin{align*}
\bigg(
\begin{array}{c}
\Deltab_{\thetab;2}^{{\cal TM}(n)}
\\[.5mm]
\Deltab_{\thetab;2}^{{\cal TE}(n)}
\end{array}
\bigg)
\inlaw\mathcal{N}\bigg(\bigg(
\begin{array}{c}
\mathbf{0} \\[.0mm]
\mathbf{0}
\end{array}
\bigg),\bigg(
\begin{array}{cc}
\Gamb_{22}^{\cal TM} & \mathbf{0} \\[.0mm]
\mathbf{0}  & \Gamb_{\thetab;22}^{\cal TE} 
\end{array}
\bigg)\bigg),
\end{align*}
which implies that, under $\mathcal{H}_{0,\thetab}$,
\begin{align*}
Q_{\thetab}^{\rm hyb}
:=
Q_{\thetab}^{\rm loc}+Q_{\thetab}^{\rm sc}
&=   
\big(\Deltab_{\thetab;2}^{{\cal TM}(n)}\big)^{T}
\big(\Gamb_{22}^{\cal TM}\big)^{-1}
\Deltab_{\thetab;2}^{{\cal TM}(n)}
+
\big(\Deltab_{\thetab;2}^{{\cal TE}(n)}\big)^{T}
\big(\Gamb_{\thetab;22}^{\cal TE}\big)^{-1}
\Deltab_{\thetab;2}^{{\cal TE}(n)}
\\[2mm] &\inlaw 
\chi^2_{(p-1)+(p-2)(p+1)/2}.
\end{align*}
The resulting hybrid test, $\phi_{\thetab}^{\rm hyb}$ say, then rejects the null hypothesis at asymptotic level $\alpha$ whenever $Q_{\thetab}^{\rm hyb}>\chi^2_{(p-1)+(p-2)(p+1)/2,1-\alpha}$. In a high-dimensional asymptotic framework where~$p=p_n$ goes to infinity with~$n$ (still at an arbitrary rate), we have that, under the null hypothesis of rotational symmetry about~$\thetab$,
\begin{align*}
\frac{Q_{\thetab}^{\rm hyb}-\frac{p_n(p_n+1)-4}{2}}{\sqrt{p_n{(p_n+1)}-4}}
\inlaw
\mathcal{N}(0,1).
\end{align*} 
This can be obtained by using the same CLT for martingale differences that was used in \cite{PaiVer2016}; see Section \ref{sec:thetests} for a discussion. Coming back to the low-dimensional setup, this test, as announced, can detect both contiguous tangent elliptical and tangent vMF alternatives.

\begin{coro}
	\label{powerH}
	Fix~$\thetab\in\mathcal{S}^{p-1}$ and $g\in \mathcal{G}_a$.  Let~$(\Lamb_n)$, $({\mathbf{L}}_n)$, $(\kappa_n)$, and $(k_n)$ be as in Corollaries \ref{powerE} and~\ref{powerF}. Then: (\textit{i}) under~${\rm P}^{{\cal TE}(n)}_{\thetab,g,\Lamb_n}$,~$Q^{{\rm hyb}}_{\thetab}\inlaw\chi^2_{(p-1)+(p-2)(p+1)/2}(\lambda)$, with $\lambda=(p-1){\rm tr}[{\mathbf{L}}^2]/(2(p+1))$; 
	(\textit{ii}) under~${\rm P}^{{\cal TM}(n)}_{\thetab,g,{\pmb\mu}_n,\kappa_n}$,~$Q^{{\rm hyb}}_{\thetab}\inlaw\chi^2_{(p-1)+(p-2)(p+1)/2}(\lambda)$, with $\lambda=k^2/(p-1)$.
\end{coro}

The same construction applies in the unspecified-$\thetab$ problem, where, under any (null) rotationally symmetric distribution~${\rm P}_{\thetab,g}\n$, the hybrid test statistic~$Q_{\rm vMF}^{\rm hyb}:=Q_{\rm vMF}^{\rm loc}+Q_{\hat\thetab}^{\rm sc}$ will be asymptotically~$\chi^2_{(p-1)+(p-2)(p+1)/2}$. The resulting test, that rejects the null hypothesis at asymptotic level $\alpha$ when $Q_{\rm vMF}^{\rm hyb}>\chi^2_{(p-1)+(p-2)(p+1)/2,1-\alpha}$, will be denoted as $\phi_{\rm vMF}^{\rm hyb}$ in the sequel. It is easy to check that, like the specified-$\thetab$ test~$\phi_{\thetab}^{\rm hyb}$, the test~$\phi_{\rm vMF}^{\rm hyb}$ can detect both types of alternatives considered. \\

The aggregation of the test statistics carried out in the hybrid statistic $Q_{\thetab}^{\rm hyb}$ can of course be performed in other ways. For instance, one could balance equally the contribution of the location and scatter test statistics by adopting the well-known Fisher's method \citep{Fisher1925} for combining independent test statistics. This approach yields, under $\mathcal{H}_{0,\thetab}$,
\begin{align}
\tilde{Q}_{\thetab}^{\rm hyb} := -2 \log (1 - F_{p-1}(Q_{\thetab}^{\rm loc}))-2\log (1 - F_{(p-2)(p+1)/2}(Q_{\thetab}^{\rm sc}))\inlaw\chi^2_4,\label{eq:Qtilde}
\end{align}
where $F_\nu$ is the cumulative distribution function of a $\chi^2_\nu$. The resulting test, $\tilde{\phi}_{\thetab}^{\rm hyb}$ say, rejects $\mathcal{H}_{0,\thetab}$ at asymptotic level $\alpha$ whenever 
$\tilde{Q}_{\thetab}^{\rm hyb}>\chi^2_{4,1-\alpha}$. For the unspecified-$\thetab$ case, obvious modifications give $\tilde{\phi}_{\thetab}^{\rm hyb}$, the test that rejects at asymptotic level $\alpha$ whenever $\tilde{Q}_{\rm vMF}^{\rm hyb}>\chi^2_{4,1-\alpha}$, where $\tilde{Q}_{\rm vMF}^{\rm hyb}$ is obtained by replacing $Q_{\thetab}^{\rm loc}$ and $Q_{\thetab}^{\rm sc}$ by $Q_{\rm vMF}^{\rm loc}$ and $Q_{\hat\thetab}^{\rm sc}$, respectively, in \eqref{eq:Qtilde}. As with the original hybrid tests, these new tests can detect both types of alternatives considered, either in the specified or unspecified-$\thetab$ cases. Note that, if $p=3$, then $\tilde{Q}_{\thetab}^{\rm hyb}=Q_{\thetab}^{\rm hyb}$ and $\tilde{Q}_{\rm vMF}^{\rm hyb}=Q_{\rm vMF}^{\rm hyb}$.

\section{Simulations}
\label{sec:simus}

In this section, we investigate the finite-sample performances of the proposed tests through two Monte Carlo exercises; see the supplementary material for two additional simulation exercises. In the specified-$\thetab$ problem, we will consider the tests~$\phi_\thetab^{\rm loc}$ and~$\phi_\thetab^{\rm sc}$ from Section~\ref{sec:thetests}, as well as the hybrid tests~$\phi_\thetab^{\rm hyb}$ and $\tilde{\phi}_\thetab^{\rm hyb}$ from Section~\ref{sec:hybrid} (the latter only performed for $\mathcal{S}^3$, as it equals $\phi_\thetab^{\rm hyb}$ for $\mathcal{S}^2$). As explained in Section~\ref{sec:thetests}, these tests look for possible departures from rotational symmetry about~$\thetab$ by checking whether or not the sign vector is uniformly distributed over~$\mathcal{S}^{p-2}$. Clearly, competing tests for rotational symmetry about~$\thetab$ can be obtained by applying other tests of uniformity over~${\cal S}^{p-2}$, such as (for~$p=3$) the well-known Kuiper test or (for~$p > 3$) the Gin\'e ($F_n$) test; see pages 99 and 209 of \cite{MJ00}, respectively. This generates a Kuiper test~$\phi_\thetab^{\rm KU}$ of rotational symmetry on~$\mathcal{S}^2$ and a Gin\'e test~$\phi_\thetab^{\rm GI}$ of rotational symmetry on~$\mathcal{S}^{p-1}$ with $p>3$, both about a specified~$\thetab$. Since they are based on omnibus tests of uniformity over~$\mathcal{S}^{p-2}$, both~$\phi_\thetab^{\rm KU}$ and~$\phi_\thetab^{\rm GI}$ are expected to show some power against both tangent vMF and tangent elliptical alternatives. Still for the specified-$\thetab$ problem, we will also consider the semiparametric test from \cite{LVD16}, denoted as $\phi_\thetab^{\rm LV}$. Now, for the unspecified-$\thetab$ problem, we will restrict to the proposed semiparametric tests~$\phi_{\dagger}^{\rm sc}$, $\phi_{\rm vMF}^{\rm loc}$, and~$\phi_{\rm vMF}^{\rm hyb}$, from Sections~\ref{sec:scattertestunsp}, \ref{sec:locationtestunsp}, and~\ref{sec:hybrid}, respectively. To the best of our knowledge, indeed, these unspecified-$\thetab$ tests have no competitors in the literature. In particular, it is unclear how to turn the omnibus specified-$\thetab$ tests~$\phi_\thetab^{\rm KU}$ and $\phi_\thetab^{\rm GI}$ into unspecified-$\thetab$ ones. 

\subsection{\texorpdfstring{The unspecified-$\thetab$ problem on~$\mathcal{S}^2$}{The unspecified-theta problem on S2}}
\label{sec:simuunspec}

The first simulation exercise focuses on the unspecified-$\thetab$ problem and intends to show, in particular, that using specified-$\thetab$ tests with a misspecified value of~$\thetab$ leads to severe violations of the nominal level constraint. For two sample sizes ($n=100,200$) and two types of alternatives to rotational symmetry~($r=1,2$), we generated
$N = 5,\!000$ mutually independent random samples of the form
$$
{\Xb}_{i;\ell}^{(r)}, \quad
i=1,\ldots, n,  \quad
\ell=0, \ldots, 5,  \quad
r=1, 2,
$$
with values in~$\mathcal{S}^2$. The ${\Xb}_{i;\ell}^{(1)}$'s follow a $\mathcal{TE}_3(\thetab_0,g_1,{\Lamb}_{\ell})$ with location~$\thetab_0:=(1/\sqrt{2}, -1/\sqrt{2}, 0)^T$, angular function~$t\mapsto g_1(t):=\exp(2t)$, and shape~${\Lamb}_{\ell}:=2{\rm diag}(1+\ell/2,1)/(2+\ell/2)$. The~${\Xb}_{i;\ell}^{(2)}$'s follow a $\mathcal{TM}_3(\thetab_0,g_1,{\pmb \mu},\kappa_{\ell})$ with skewness direction~${\pmb \mu}:=(1,0)^T$ and skewness intensity $\kappa_{\ell}:=\ell$. In both cases, $\ell=0$ corresponds to the null hypothesis of rotational symmetry, whereas~$\ell=1, \ldots, 5$ provide increasingly severe alternatives. For each replication, we performed, at asymptotic level~$\alpha=5\%$,  the specified-$\thetab$ tests~$\phi_\thetab^{\rm loc}$, $\phi_\thetab^{\rm sc}$, $\phi_\thetab^{\rm hyb}$, $\phi_\thetab^{\rm LV}$, and~$\phi_\thetab^{\rm KU}$, all based on the \emph{misspecified} location value~$\thetab:=(1,0,0)^T$, and the unspecified-$\thetab$ tests $\phi_\dagger^{\rm sc}$, $\phi_{\rm vMF}^{\rm loc}$, and $\phi_{\rm vMF}^{\rm hyb}$, all computed with the spherical mean to estimate~$\thetab$.\\

\begin{figure}[!h]
	\centering
	\includegraphics[width=.9\textwidth]{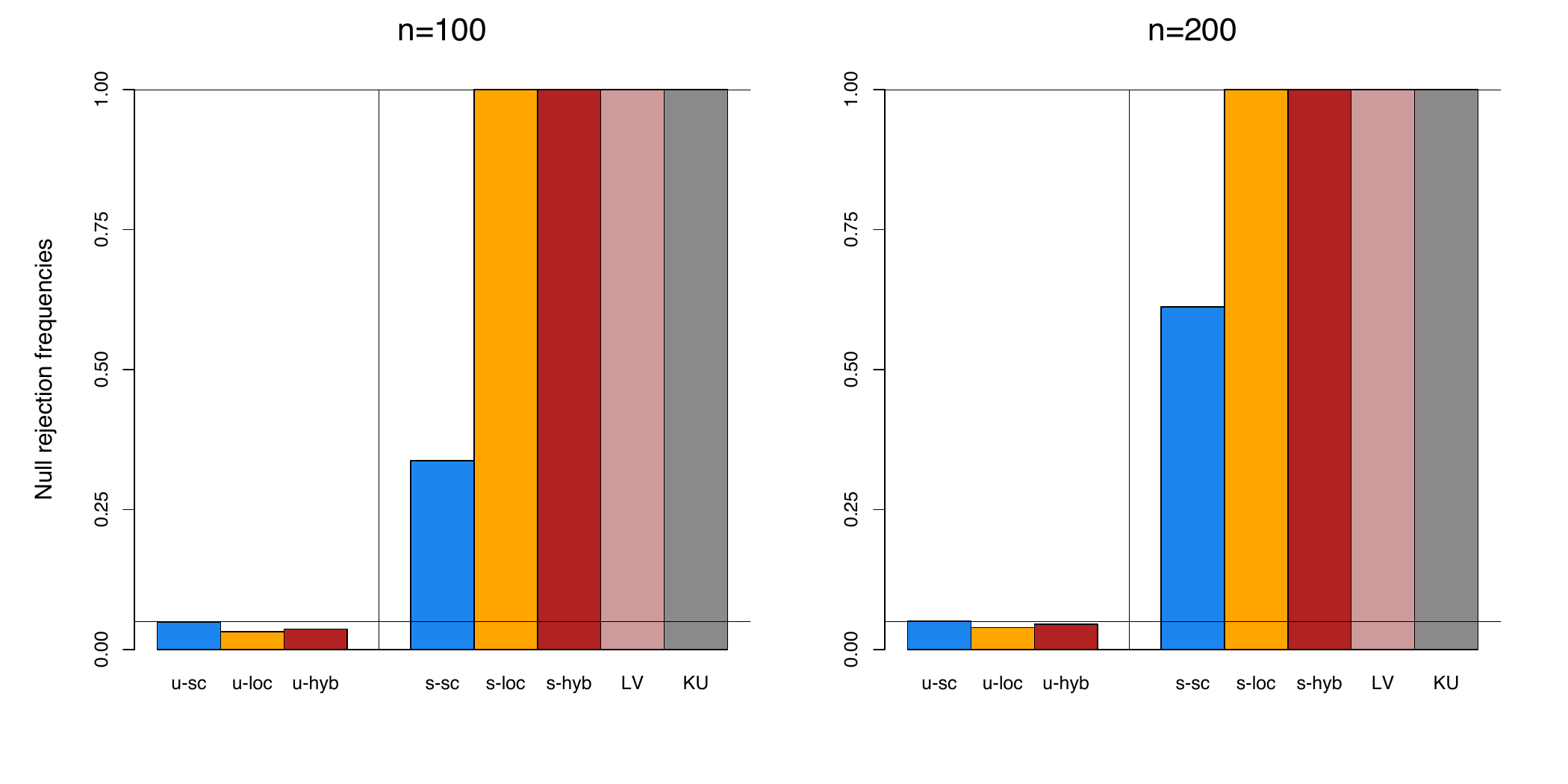}\vspace*{-0.75cm}\caption{\small Null rejection frequencies, for sample sizes $n=100$ and $n=200$, of the unspecified-$\thetab$ tests~$\phi_\dagger^{\rm sc}$ (u-sc), $\phi_{\rm vMF}^{\rm loc}$ (u-loc), and $\phi_{\rm vMF}^{\rm hyb}$ (u-hyb), as well as the (mis)specified-$\thetab$ tests~$\phi_\thetab^{\rm sc}$ (s-sc), $\phi_\thetab^{\rm loc}$ (s-loc), $\phi_\thetab^{\rm hyb}$ (s-hyb), $\phi_\thetab^{\rm LV}$ (LV), and $\phi_\thetab^{\rm KU}$ (KU). All tests are performed at asymptotic level~$5\%$; see  Section~\ref{sec:simuunspec} for details.
		\label{Levelplot}}
\end{figure}

\begin{figure}[!htpb]
	\centering
	\includegraphics[width=.9\textwidth]{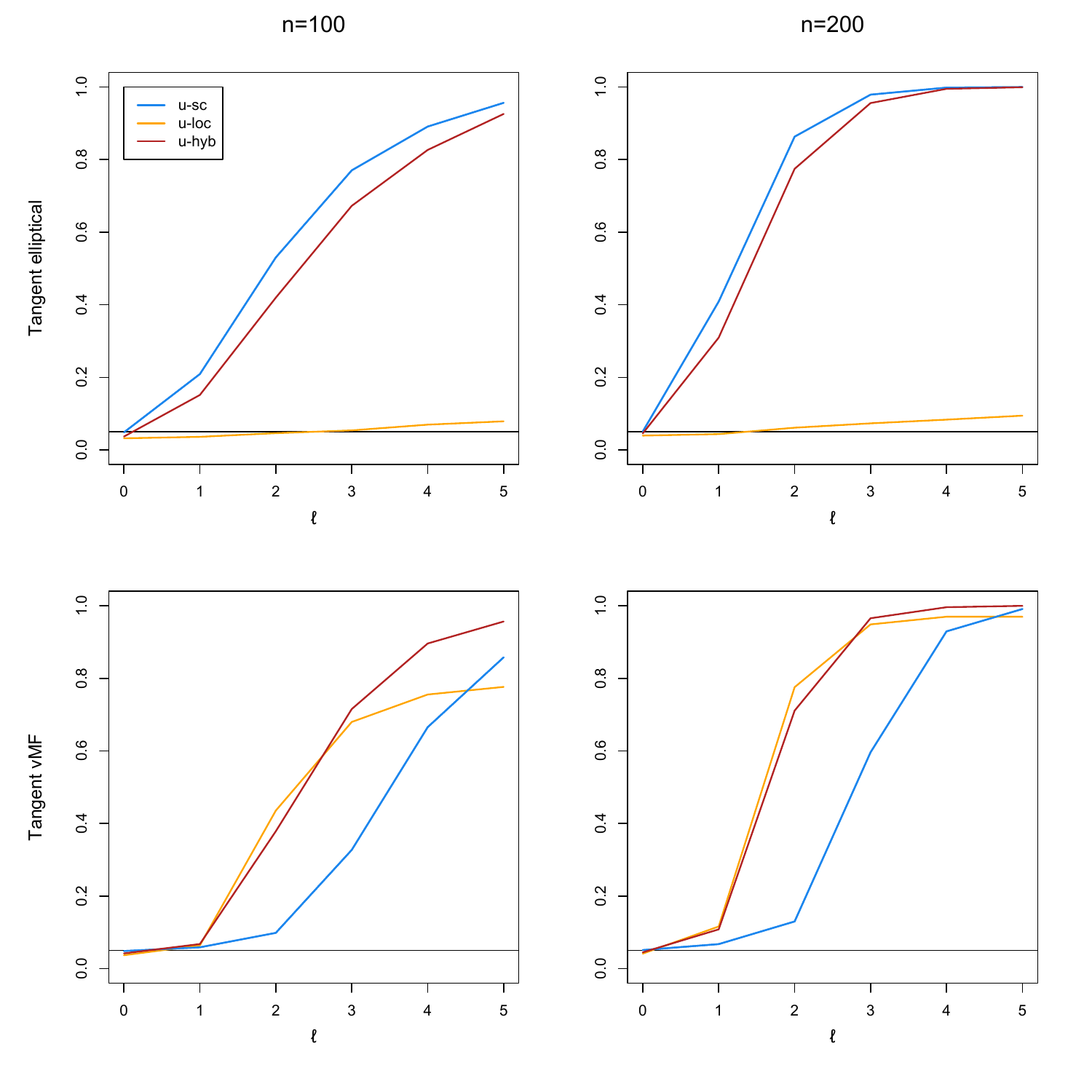}\vspace*{-0.75cm}
	\caption{\small Rejection frequencies, under tangent elliptical alternatives (top row) and tangent vMF ones (bottom row), of the unspecified-$\thetab$ tests~$\phi_\dagger^{\rm sc}$, $\phi_{\rm vMF}^{\rm loc}$, and~$\phi_{\rm vMF}^{\rm hyb}$ for $n=100$ (left column) and $n=200$ (right column). Both tests are performed at asymptotic level~$5\%$; see Section~\ref{sec:simuunspec} for details. \label{powerU}}
\end{figure}

Due to misspecification, it is expected that only the unspecified-$\thetab$ tests will exhibit null rejection frequencies close to~$5\%$. This is confirmed in Figure~\ref{Levelplot}, that shows that all (mis)specified-$\thetab$ tests are severely liberal. For the two samples sizes and the two types of alternatives considered, Figure~\ref{powerU} plots the empirical powers of the three \mbox{unspecified-$\thetab$} tests (a power comparison involving the specified-$\thetab$ tests would be meaningless since these tests do not meet the level constraint). Inspection of Figure~\ref{powerU} reveals that (\textit{i}) as expected, $\phi_\dagger^{\rm sc}$ dominates $\phi_{\rm vMF}^{\rm loc}$ under tangent elliptical alternatives while the opposite occurs under tangent vMF alternatives; (\textit{ii}) the hybrid test detects both types of alternatives and performs particularly well against tangent vMF ones.

\subsection{\texorpdfstring{The specified-$\thetab$ problem on~$\mathcal{S}^2$}{The specified-theta problem on S2}}
\label{sec:simuspecp3}

The second simulation exercise focuses on the specified-$\thetab$ problem on~${\cal S}^2$, {with~$\thetab:=(1,0,0)^T$}. We generated $N = 5,\!000$ mutually independent random samples of the form
$
{\Xb}_{i;\ell}^{(r)}$, 
$i=1,\ldots, n$, 
$\ell=0, \ldots, 5$, 
$r=1,2,3$,
with values in~$\mathcal{S}^2$. The ${\Xb}_{i;\ell}^{(1)}$'s follow a $\mathcal{TE}_3(\thetab,g_2, {\Lamb}_{\ell})$ with angular function~$t\mapsto g_2(t):=\exp(5t)$ and with ${\Lamb}_{\ell}$ as in Section \ref{sec:simuunspec}, whereas the~${\Xb}_{i;\ell}^{(2)}$'s follow a $\mathcal{TM}_3(\thetab,g_2, {\pmb \mu}, \kappa_{\ell})$ with {skewness direction~${\pmb \mu}:=(1,0)^T$ and}  skewness intensity $\kappa_{\ell}:=\ell/6$. The ${\Xb}_{i;\ell}^{(3)}$'s have a Fisher--Bingham distribution with location~$\thetab$, concentration~$2$, and shape matrix~${\bf A}_{\ell}:={\rm diag}(0, \ell/2, -\ell/2)$; we refer to \cite{MJ00} for details on Fisher--Bingham distributions, which, for the zero shape matrix, reduce to a vMF distribution. For~$r=1,2,3$, thus, the value $\ell=0$ corresponds to the null hypothesis of rotational symmetry about~$\thetab$, whereas~$\ell=1, \ldots, 5$ provide increasingly severe alternatives. For each replication, we performed, at asymptotic level~$\alpha=5\%$, the specified-$\thetab$ tests~$\phi_\thetab^{\rm loc}$, $\phi_\thetab^{\rm sc}$,  $\phi_\thetab^{\rm hyb}$, $\phi_\thetab^{\rm LV}$, and~$\phi_\thetab^{\rm KU}$ (based on the true value of~$\thetab$). For the sake of comparison, we also considered the unspecified-$\thetab$ tests~$\phi_\dagger^{\rm sc}$, $\phi_{\rm vMF}^{\rm loc}$, and~$\phi_{\rm vMF}^{\rm hyb}$, based on the spherical mean.\\

\begin{figure}[!htpb]
	\centering
	\includegraphics[width=.8\textwidth]{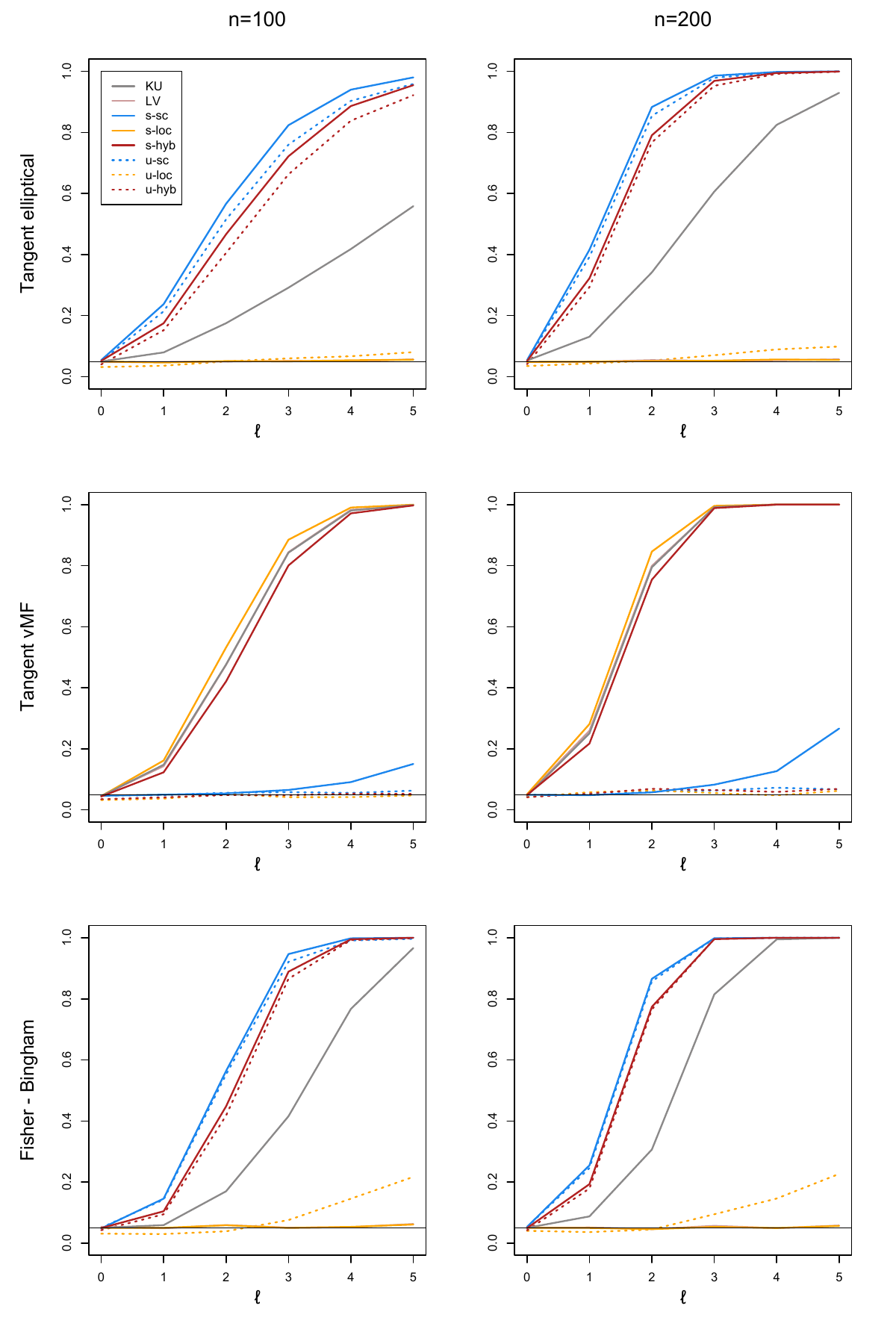}\vspace*{-0.75cm}
	\caption{\small Rejection frequencies, under tangent elliptical alternatives (top row), tangent vMF alternatives (middle row), and Fisher--Bingham alternatives (bottom row),  of the specified-$\thetab$ tests $\phi_\thetab^{\rm sc}$ (s-sc), $\phi_\thetab^{\rm loc}$ (s-loc), $\phi_\thetab^{\rm hyb}$ (s-hyb), $\phi_\thetab^{\rm LV}$ (LV), and $\phi_\thetab^{\rm KU}$ (KU), as well as the unspecified-$\thetab$ tests~$\phi_\dagger^{\rm sc}$ (u-sc), $\phi_{\rm vMF}^{\rm loc}$ (u-loc), and $\phi_{\rm vMF}^{\rm hyb}$ (u-hyb). Sample sizes are $n=100$ (left column) and $n=200$ (right column). All tests are performed at asymptotic level~$5\%$; see Section~\ref{sec:simuspecp3} for details.
		\label{power3d}}
\end{figure}

Figure~\ref{power3d} plots the resulting empirical power curves for sample sizes $n=100$ and $n=200$. Inspection of the figure confirms the theoretical results:
(\textit{i}) $\phi_\thetab^{\rm sc}$ dominates the other tests under tangent elliptical alternatives, whereas~$\phi_\thetab^{\rm loc}$ dominates the other tests under tangent vMF alternatives (even if the latter dominance is less prominent);
(\textit{ii}) $\phi_\thetab^{\rm sc}$ and  $\phi_\dagger^{\rm sc}$ exhibit extremely similar performances, as expected from Proposition~\ref{asymplin} and the comments below that result; 
(\textit{iii}) $\phi_\thetab^{\rm hyb}$ and $\phi_\thetab^{\rm KU}$ show non-trivial powers against each type of alternatives but are always dominated by some other test.
Moreover, it should be noted that $\phi_\thetab^{\rm sc}$ and $\phi_\thetab^{\rm hyb}$ perform well under Fisher--Bingham alternatives, which was expected since, parallel to tangent elliptical alternatives, Fisher--Bingham alternatives are of an elliptical~nature.

\begin{figure}[!htpb]
	\vspace*{-0.5cm}
	\centering
	\includegraphics[width=.5\textwidth]{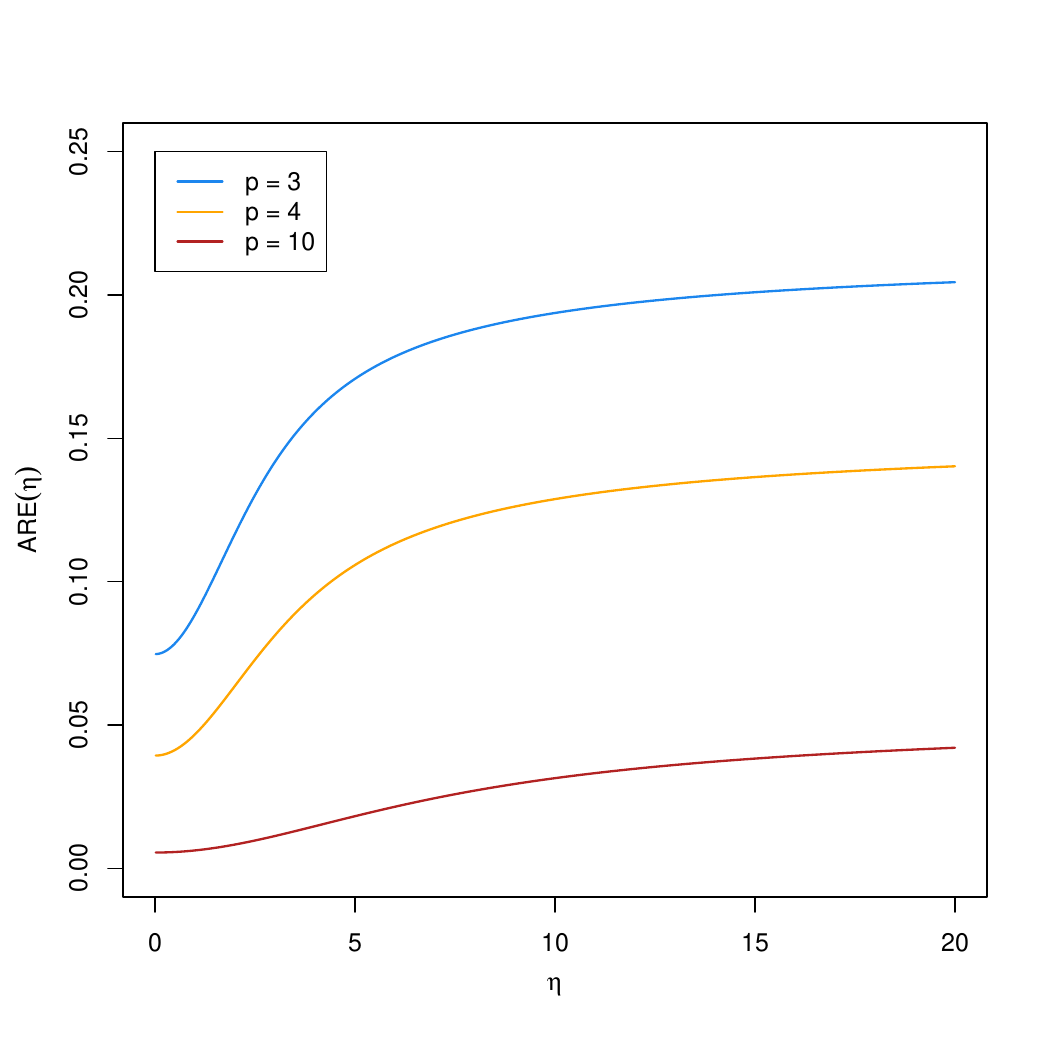}
	\vspace*{-0.75cm}
	\caption{\small Plots, for several dimensions~$p$, of the asymptotic relative efficiency, as a function of~$\eta$, of the unspecified-$\thetab$ test~$\phi_{\rm vMF}^{\rm loc}$ with respect to the specified-$\thetab$ test~$\phi_\thetab^{\rm loc}$ under the sequence of alternatives considered in Corollary~\ref{powerF} at the vMF with concentration~$\eta$. 
		\label{AREfig}}
\end{figure}

It may be surprising at first that, under tangent vMF alternatives, the (optimal) unspecified-$\thetab$ test~$\phi_{\rm vMF}^{\rm loc}$ shows little power compared to the specified-$\thetab$ test~$\phi_\thetab^{\rm loc}$. This, however, only reflects the fact that the cost of the unspecification of~$\thetab$ is high for the (vMF) angular function considered. Actually, the results of the previous sections allow us to quantify this cost theoretically. Under the sequence of alternatives considered in Corollary~\ref{powerF}, the Asymptotic Relative Efficiency (ARE) of the unspecified-$\thetab$ test~$\phi_{\rm vMF}^{\rm loc}$ with respect to the specified-$\thetab$ test~$\phi_\thetab^{\rm loc}$ is obtained as the usual ratio of the corresponding non-centrality parameters in the asymptotic non-null chi-squared distributions of the corresponding statistics. It follows from~\eqref{ncpsemi} in the supplementary material and from Corollary~\ref{powerF} that, at the vMF with concentration~$\eta$, $\mathrm{ARE}(\eta)=1-\mathcal{I}^2_p(g_\eta)/\mathcal{J}_p(g_\eta)=1-\big(2 \Gamma \left(\frac{p}{2}\right)^2 I_{\frac{p-1}{2}}(\eta){}^2\big)\big/\big((p-~1) \Gamma \big(\frac{p-1}{2}\big)^2 I_{\frac{p-2}{2}}(\eta) I_{\frac{p}{2}}(\eta)\big)$, still with~$g_\eta(r)=\exp(\eta r)$. Figure~\ref{AREfig} provides plots of this ARE as a function of~$\eta$, for various values of~$p$. For the tangent vMF alternatives considered in the present simulation exercise (for which~$\eta=5$ and~$p=3$), the ARE is~$0.171$, which explains the poor performance of~$\phi_{\rm vMF}^{\rm loc}$ compared to~$\phi_\thetab^{\rm loc}$. This, of course, is not incompatible with the optimality of~$\phi_{\rm vMF}^{\rm loc}$ in the unspecified-$\thetab$ problem.

\section{Real data application}
\label{sec:realdata}

We illustrate in this section the practical relevance of the proposed tests with a novel case study. The data we analyze is based on the Debrecen Photoheliographic Data (DPD) sunspot catalogue \citep{Baranyi2016,Gyoeri2016}. It contains observations of sunspots locations since 1974 and is a continuation of the Greenwich Photoheliographic Results (GPR) catalogue, which spanned 1872--1976.\\

Sunspots are darker, cooler regions on the Sun's photosphere that correspond to solar magnetic field concentrations. They are temporary phenomena that experience continuous change, lasting for hours to days, and with their shapes and areas varying notably along their lifespans. Prototypical sunspots come in pairs with opposite magnetic polarity, forming the so-called Bipolar Magnetic Regions (BMR), and they are usually clustered in groups that evolve with time. Sunspots are widely used to study and measure solar activity, whose effects, among others, may affect Earth's long-term climate (see, e.g., \citealp{Haigh2007}).\\

\begin{figure}[!htbp]
	\centering
	\includegraphics[width=0.49\textwidth]{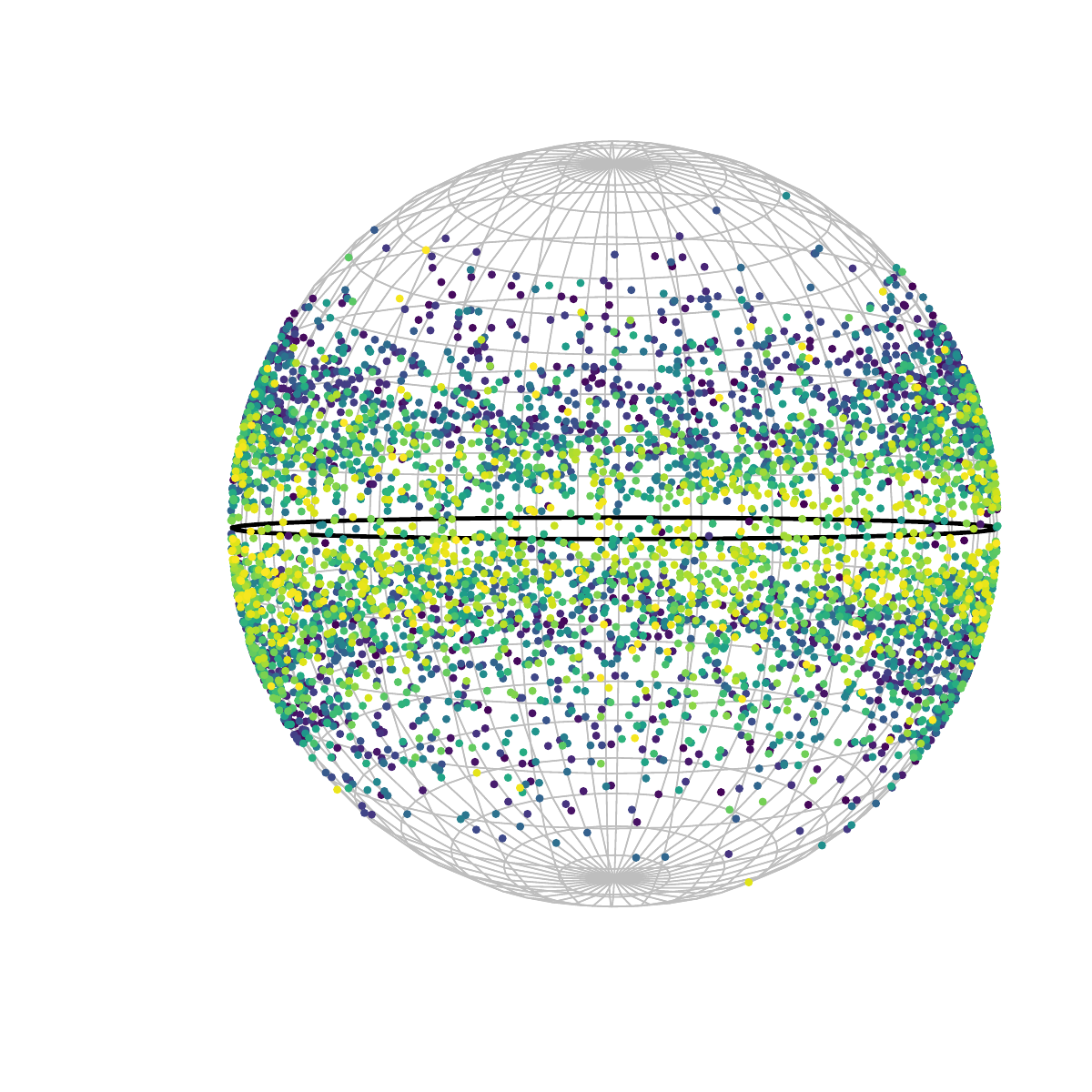}
	\includegraphics[width=0.49\textwidth]{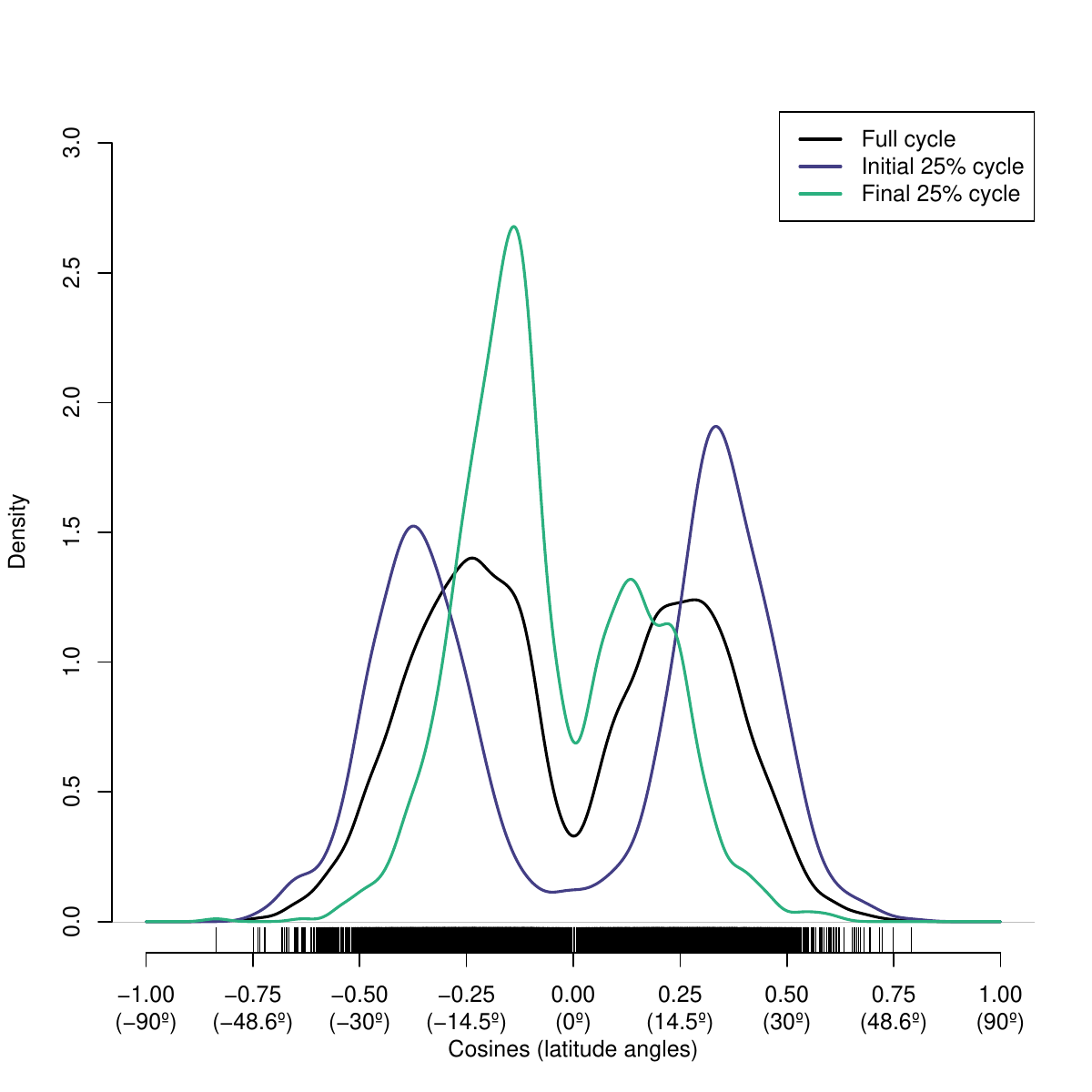}\vspace*{-0.05cm}
	\caption{\small Left: emerging locations of sunspot groups during the 23rd solar cycle. The locations are colored with a blue-yellow gradient according to the relative position of the sunspot appearance date within the solar cycle in order to visualize the Sp\"orer's law. Right: the kernel density estimator of the cosines $v_{\thetab_0}(\xb_i)$, $i=1,\ldots,n$, for the full cycle and for the data in the initial and final $25\%$ cycle duration. Recall how the hemisphere asymmetry of the initial and final regimes balances out in the full cycle density.}
	\label{fig:2}
\end{figure}

As it can be seen in the left panel of Figure \ref{fig:2}, sunspots originate following a nearly rotationally symmetric pattern. An explanation for this phenomenon is given by the \cite{Babcock1961} model for solar dynamics. It describes how the force lines of an \textit{initial} rotationally symmetric dipolar field are twisted by the Sun's differential rotation (solar plasma rotates slower near the poles than at the equator) to produce a spiral wrapping of the magnetic field, with an amplification of the field strength that depends on the \textit{latitude}. The magnetic buoyancy of locally intense concentrations of the field pushes inner magnetic flux tubes up to the solar surface, forming BMR and producing sunspots as the result of the intersection of BMR with the solar surface. The (magnetic) field intensity required for producing BMR is reached at progressively lower latitudes as the twisting of the initial field advances, a phenomenon known as the \textit{Sp\"orer's law}. After about 11 years, the magnetic field behaves as a dipolar field of reversed polarity. This period constitutes a \textit{solar cycle}. The process repeats itself, attaining the initial conditions after a 22-year cycle. Further details on sunspots and their origin can be consulted in \cite{Babcock1961} and \cite{Solanki2006}.

\newpage

Even though the main driving force generating sunspots is of a rotational symmetric nature (intense magnetic field concentrations at equal latitudes due to the Sun's differential rotation), \citet[pages 574 and 581]{Babcock1961} points out that sunspots tend to emerge in ``preferred zones of occurrence'' associated to longitudes where there has been prior activity. Hence, this phenomenon may trigger non-rotational symmetric patterns, of unclear significance, in the emergence of sunspots. We offer a quantification of this significance via the proposed tests. To that aim, we analyze the rotational symmetry of the appeared sunspots during the 23rd solar cycle (August 1996 -- December 2008), currently the last fully-observed cycle with curated measurements. The corresponding data~$\Xb_1, \ldots, \Xb_n$, shown in the left panel of Figure~\ref{fig:2}, consists of $n=5,\!373$ central positions of groups of sunspots, understood as the first-ever observations of each group. The DPD catalogue delimits the central position of a group of sunspots through an area-weighted average of the sunspots within the group. Note that, within this setting, both the temporal (observations belong to a single cycle that aggregates the different latitude-appearance regimes) and spatial (clusters of related sunspots are treated as a single observation) dependency of the data is mitigated, therefore better accommodating the independent and identically distributed framework considered in the paper.

Visual inspection reveals that rotational symmetry about the north pole~$\thetab_0 = (0, 0, 1)^T$ (or equivalently, about the south pole~$-\thetab_0$) may be suspected. The tests $\phi_{\thetab_0}^{\rm sc}$ (s-sc), $\phi_{\thetab_0}^{\rm loc}$ (s-loc), and $\phi_{\thetab_0}^{\rm hyb}$ (s-hyb) for rotational symmetry about~$\thetab_0$ yield $p$-values equal to $0.1656$, $0.4571$, and $0.2711$, respectively. Therefore, rotational symmetry about~$\thetab_0$ is not rejected at any usual significance level and a rotationally symmetric model is suitable for the data at hand. As explained in Section~\ref{sec:specloc}, this greatly simplifies inference. In particular, to estimate the density~$f=f_{\thetab_0,g}$ of the sunspot groups on $\mathcal{S}^2$, one only needs to estimate the common, \emph{univariate}, density~$\tilde{g}_p$ of the cosines~$v_{\thetab_0}(\Xb_i)= \Xb_i^{T} \thetab_0$, $i=1,\ldots,n$; see~\eqref{densitys}--\eqref{coss}. The left panel in Figure \ref{fig:2} provides a plot of the kernel density estimator (kde)~$\hat{\tilde{g}}_p$ based on Gaussian kernel and the direct plug-in bandwidth selector \citep{Sheather1991}. Note that, since the sample stays away from the endpoints of~$[-1,1]$, it is not required considering a kde that would specifically address issues associated with boundary effects. Then, from~\eqref{densitys}--\eqref{coss} and since $p=3$, the final estimator for the density is then~$\hat{f}(\xb)= \omega_{p-1}^{-1}\hat{\tilde{g}}_p(\xb^{T} \thetab_0)$. Given the large sample size and the fact that only a univariate density was estimated nonparametrically, one may be confident that~$\hat{f}$ is a good estimate of $f$. The kde $\hat{\tilde{g}}_p$ can be applied for further quantitative description of the data (see \cite{Chacon2018} for a survey of applications). In particular, it provides an estimate of the shortest set of latitudes containing at least $90\%$ of the probability of an emergence of a sunspot group; in the present example, this shortest set, namely $(-29.48^\circ, -2.51^\circ)\cup (2.56^\circ, 28.10^\circ)$, contains the latitudes for which $\hat{\tilde{g}}_p$ exceeds $0.4812$. Analysis of $\hat{\tilde{g}}_p$ also reveals two modes at latitudes $-13.69^\circ$ and $16.49^\circ$. \\

Remarkably, however, when performed on the 22nd cycle data, the tests $\phi_{\thetab_0}^{\rm sc}$, $\phi_{\thetab_0}^{\rm loc}$, and $\phi_{\thetab_0}^{\rm hyb}$ for rotational symmetry about $\thetab_0$ provide $p$-values $0.1077$, $0.0125$, and $0.0103$, respectively. Consequently, the location and hybrid tests flag a departure from rotational symmetry about~$\thetab_0$, which evidences a significant non-rotational symmetric emergence pattern during that cycle and points to the need of a more complex modeling. The analysis for both cycles were performed for the unspecified-$\thetab$ case through the tests $\phi_\dagger^{\rm sc}$, $\phi_{\rm vMF}^{\rm loc}$, and $\phi_{\rm vMF}^{\rm loc}$, reaching exactly the same conclusions. 

\section{Perspective for future research}
\label{sec:discuss}

As explained in Section~\ref{sec:thetests}, the random vector~$\Xb$ with values on~$\mathcal{S}^{p-1}$ is rotationally symmetric about~$\thetab$ if and only if, using the notation introduced in~\eqref{vandsignU}, (\textit{i}) the random vector~$\ub_{\thetab}(\Xb)$ is uniformly distributed over~$\mathcal{S}^{p-2}$ and (\textit{ii})~$\ub_{\thetab}(\Xb)$ is independent of $v_{\thetab}(\Xb)$. The tests proposed in this paper are designed to detect deviations from rotational symmetry by testing that~(\textit{i}) holds. As a consequence, they will be blind to alternatives of rotational symmetry for which~(\textit{i}) holds but~(\textit{ii}) does not. This could be fixed by testing that the covariance between~$\ub_{\thetab}(\Xb)$ and $v_{\thetab}(\Xb)$ is zero, which can be based on a statistic like
$$
\Deltab_{\thetab}^{{\rm cov}(n)}:= \frac{1}{\sqrt{n}} \sum_{i=1}^{n} v_{\thetab}(\Xb_i) \ub_{\thetab}(\Xb_i).
$$
Since~$\Deltab_{\thetab}^{{\rm cov}(n)}$ is asymptotically normal with mean zero and covariance matrix~$(p-1)^{-1}\allowbreak{\rm E}_{\thetab,g}[v^2_{\thetab}(\Xb_1)]\allowbreak{\bf I}_{p-1}$ under~${\rm P}\n_{\thetab,g}$, the resulting test would,  at asymptotic level~$\alpha$, reject the null hypothesis of rotational symmetry about~$\thetab$ whenever
$$
\frac{p-1}{\sum_{i=1}^n v^2_{\thetab}(\Xb_i)}
\sum_{i,j=1}^n
v_{\thetab}(\Xb_i)
v_{\thetab}(\Xb_j)
\ub_{\thetab}^{T}(\Xb_i)\ub_{\thetab}(\Xb_j)
> \chi_{p-1,1-\alpha}^2.
$$
Such a test of course would detect violations of~(\textit{ii}) only and it is natural to design a test that would be able to detect deviations from both~(\textit{i}) and~(\textit{ii}) by considering test statistics that are quadratic forms in~$\big(\big(\Deltab_{\thetab}^{{\rm cov}(n)}\big)^{T},\big(\Deltab_{\thetab;2}^{{\cal TE}(n)}\big)^{T}\big)^{T}$ or in~$\big(\big(\Deltab_{\thetab}^{{\rm cov}(n)}\big)^{T},\big(\Deltab_{\thetab;2}^{{\cal TM}(n)}\big)^{T}\big)^{T}$, depending on whether tangent elliptical or tangent vMF alternatives are considered. In the spirit of the hybrid test from Section~\ref{sec:hybrid}, detecting both types of alternatives can be achieved by considering a quadratic form in~$\big(\big(\Deltab_{\thetab}^{{\rm cov}(n)}\big)^{T},\big(\Deltab_{\thetab;2}^{{\cal TE}(n)}\big)^{T},\big(\Deltab_{\thetab;2}^{{\cal TM}(n)}\big)^{T}\big)^{T}$. The quadratic form to be used in each case naturally follows from the asymptotic covariance matrix in the (null) joint asymptotic normal distribution of these random vectors.\\

Another perspective for future research is the following. In Section~\ref{sec:distribs}, we proposed new distributions on the unit sphere~$\mathcal{S}^{p-1}$, namely tangent vMF distributions, by imposing that~$\ub_{\thetab}(\Xb)=\ub_{\thetab_1;p-2}(\Xb)$ follows its own vMF distribution over~$\mathcal{S}^{p-2}$ with location~${\pmb \mu}=\thetab_2\in\mathcal{S}^{p-2}$. In turn, one could specify that~$\ub_{\thetab_2;p-3}(\Xb)$ follows a vMF distribution over~$\mathcal{S}^{p-3}$ with location~$\thetab_3$. Iterating this construction will provide ``nested'' tangent vMF distributions that are associated with mutually orthogonal directions~$\thetab_{i}$, $i=1,\ldots,p$ (strictly speaking,~$\thetab_{i}\in\mathcal{S}^{p-i}$ but they can all be considered embedded in the original unit sphere~$\mathcal{S}^{p-1}$). Such distributions provide flexible models on the sphere that are likely to be relevant in various applications of directional statistics. 

\section*{Supplement}

The supplement details the delicate construction of Le Cam optimal location tests in the unspecified-$\thetab$ problem, provides the proofs of the main results, and presents additional simulations.

\section*{Acknowledgments}

Eduardo Garc\'ia-Portugu\'es acknowledges support from project PGC2018-097284-B-I00, IJCI-2017-32005, and MTM2016-76969-P from the Spanish Ministry of Science, Innovation and Universities, and the European Regional Development Fund. Davy Paindaveine's research is supported by a research fellowship from the Francqui Foundation and the Program of Concerted Research Actions (ARC) of the Universit\'{e} libre de Bruxelles. Thomas Verdebout's research is supported by the ARC Program of the Universit\'{e} libre de Bruxelles, the Cr\'{e}dit de Recherche  J.0134.18 of the FNRS (Communaut\'{e} Fran\c{c}aise de Belgique), and the National Bank of Belgium. The authors would like to thank the Associate Editor and the two anonymous referees for their insightful comments and suggestions that led to a substantial improvement of this work.


\appendix


\newpage
\title{Supplement to ``On optimal tests for rotational symmetry against new classes of hyperspherical distributions''}
\setlength{\droptitle}{-1cm}
\predate{}%
\postdate{}%
\date{}

\author{Eduardo Garc\'ia-Portugu\'es$^{1,2,6}$, Davy Paindaveine$^{3,4,5}$, and Thomas Verdebout$^{3,4}$}

\footnotetext[1]{
	Department of Statistics, Carlos III University of Madrid (Spain).}
\footnotetext[2]{
	UC3M-Santander Big Data Institute, Carlos III University of Madrid (Spain).}
\footnotetext[3]{
	D\'{e}partement de Math\'{e}matique, Universit\'{e} libre de Bruxelles (Belgium).}
\footnotetext[4]{
	ECARES, Universit\'{e} libre de Bruxelles (Belgium).}
\footnotetext[5]{
	Toulouse School of Economics, Universit\'{e} Toulouse Capitole (France).}
\footnotetext[6]{Corresponding author. e-mail: \href{mailto:edgarcia@est-econ.uc3m.es}{edgarcia@est-econ.uc3m.es}.}

\maketitle

\begin{abstract}
	In Section~\ref{sec:app:BB} of this supplement, we detail the delicate construction of Le Cam optimal location tests in the unspecified-$\thetab$ problem, hence in particular explain how the test~$\phi^{\rm loc}_{\rm vMF}$ from Section~\ref{sec:locationtestunsp} is obtained. In Section~\ref{sec:app:A}, we then provide the proofs of the main results, with the required lemmas given in Section~\ref{sec:app:B}. Finally, Section~\ref{sec:app:moresimus} presents additional simulations.
\end{abstract}
\begin{flushleft}
\small\textbf{Keywords:} Directional data; Local asymptotic normality; Locally asymptotically maximin tests; Rotational symmetry. 
\end{flushleft}


\section{\texorpdfstring{Construction of optimal unspecified-$\thetab$ location tests}{Construction of optimal unspecified-theta location tests}}
\label{sec:app:BB}

As discussed in Section~\ref{sec:unspecloc}, while the replacement of~$\thetab$ with a suitable estimator~$\hat{\thetab}$ does not affect the asymptotic behavior of the scatter test statistic~$Q_{\thetab}^{\rm sc}$ under the null or contiguous alternatives, this replacement has a non-negligible asymptotic impact for the location test statistic~$Q_{\thetab}^{\rm loc}$. Here, we examine this impact and construct (optimal) unspecified-$\thetab$ location tests. We first focus on the parametric case where the angular function~$g$ is specified (Section~\ref{sec:parametric}), then we will turn to the more realistic semiparametric situation where~$g$ remains unspecified (Section~\ref{sec:loctestunsp}).

\subsection{The parametric case}
\label{sec:parametric}

When the Fisher information matrix is not block-diagonal, it is well-known that inference on~$\deltab$ (we consider the model and parametrization from Section~\ref{sec:optim2}) under unspecified~$\thetab$ is to be based on the \emph{efficient central sequence}
\begin{align}
\label{effcent}
\Deltab_{\thetab,g;2*}^{{\cal TM}(n)}
:=&\,
\Deltab_{\thetab;2}^{{\cal TM}(n)}
-
\Gamb_{\thetab,g;21}^{\cal TM}
\Gamb_{\thetab,g;11}^{-}
\Deltab_{\thetab,g;1}\n\nonumber\\
=&\,
\frac{1}{\sqrt{n}} \sum_{i=1}^n
\bigg(
1-\frac{\mathcal{I}_p(g)}{{\cal J}_p(g)}
\,
\varphi_g(V_{i,\thetab}) (1-V_{i,\thetab}^2)^{1/2}
\bigg)
\Ub_{i, \thetab}
\end{align}
(throughout, ${\bf A}^{-}$ stands for the Moore--Penrose inverse of~${\bf A}$).
Under ${\rm P}\n_{\thetab,g}$, 
\begin{align*}
\Deltab_{\thetab,g;2*}^{{\cal TM}(n)}\inlaw\mathcal{N}\big({\bf0},\Gamb_{g,22*}^{\cal TM}\big),
\textrm{ with }
\
\Gamb_{g,22*}^{\cal TM}
:=
\frac{1}{p-1}
\bigg(
1
-
\frac{\mathcal{I}^2_p(g)}{\mathcal{J}_p(g)}
\bigg)
{\bf I}_{p-1},
\end{align*}
and the corresponding test,~$\phi^{\rm loc}_{\thetab,g*}$ say, consists in rejecting the null hypothesis of rotational symmetry ($\mathcal{H}_0:\deltab=\mathbf{0}$, with unspecified~$\thetab$) at asymptotic level~$\alpha$ whenever
\begin{align*}
Q_{\thetab,g*}^{\rm loc}
:=
\big(\Deltab_{\thetab,g;2*}^{{\cal TM}(n)}\big)^{T}
\big(\Gamb_{g,22*}^{\cal TM}\big)^{-1}
\Deltab_{\thetab,g;2*}^{{\cal TM}(n)}
>\chi^2_{p-1,1-\alpha}.
\end{align*}
This test has asymptotic level~$\alpha$ under~${\rm P}\n_{\thetab,g}$ and is locally asymptotically maximin, under angular function~$g\in\mathcal{G}_a$, in the unspecified-$\thetab$ problem. A direct application of Le Cam's third lemma yields that, under the same sequence of alternatives as the one considered in Corollary~\ref{powerF},
$$
\Deltab_{\thetab,g;2*}^{{\cal TM}(n)}\inlaw\mathcal{N}\big({\bf m}_{g},\Gamb_{g,22*}^{\cal TM}\big), 
$$
with
$$
{\bf m}_{g}
:=
\lim_{n\to \infty}
{\rm E}_{\thetab,g}\big[
\Deltab_{\thetab,g;2*}^{{\cal TM}(n)}
\big(\Deltab_{\thetab;2}^{{\cal TM}(n)}\big)^{T}\big]
{\bf d}_n
=
\frac{1}{p-1}
\bigg(
1
-
\frac{\mathcal{I}^2_p(g)}{\mathcal{J}_p(g)}
\bigg)
k {\pmb \mu},
$$
so that~$Q_{\thetab,g*}^{\rm loc}\inlaw\chi^2_{p-1}(\lambda)$ with non-centrality parameter $\lambda={\bf m}_g^{T} \big(\Gamb_{g,22*}^{\cal TM}\big)^{-1} {\bf m}_g
=
k^2
\big(
1
-
\mathcal{I}^2_p(g)/\mathcal{J}_p(g)
\big)/\allowbreak(p-1)$.
Note that this non-centrality parameter is smaller than or equal to the one in Corollary~\ref{powerF}. The comments below Theorem~\ref{LANvM} imply that the non-centrality parameter is larger than or equal to zero, with equality if and only if~$g$ is of the form~$g(t)=C \exp(\kappa \arcsin(t))$. In other words, %
it is only for angular densities of the previous form that the $g$-optimal unspecified-$\thetab$ test has asymptotic power~$\alpha$.\\

Now, even if we are after the construction of a parametric ($g$-fixed) test, the  test~$\phi^{\rm loc}_{\thetab,g*}$ is unfortunately infeasible because~$\thetab$ is unknown. We have the following result.
\begin{prop} \label{alignparamm} Let~$\hat{\thetab}$ satisfy \ref{assump:2}. Then, for any~$\thetab\in\mathcal{S}^{p-1}$ and any~$g\in \mathcal{G}_a\cap \mathcal{G}'$, 
	\begin{align}
	\label{gtestloc}
	Q_{\hat{\thetab},g*}^{\rm loc}
	=
	Q_{\thetab,g*}^{\rm loc}
	+
	o_{\rm P}(1)
	,
	\end{align}
	as $\ny$ under ${\rm P}\n_{\thetab,g}$.
\end{prop}

It directly follows from Proposition \ref{alignparamm} that the $g$-parametric test~$\phi^{\rm loc}_{g*}$ that rejects the null hypothesis at asymptotic level~$\alpha$ whenever~$Q_{\hat{\thetab},g*}^{\rm loc}>\chi^2_{p-1,1-\alpha}$ has the exact same asymptotic properties as the infeasible test~$\phi^{\rm loc}_{\thetab,g*}$ above. In particular, like~$\phi^{\rm loc}_{\thetab,g*}$, the test~$\phi^{\rm loc}_{g*}$ is locally asymptotically maximin at asymptotic level~$\alpha$ when testing
$
\bigcup_{\thetab\in\mathcal{S}^{p-1}} \big\{ {\rm P}\n_{\thetab,g} \big\}
$
against 
$\bigcup_{\thetab\in\mathcal{S}^{p-1}}\bigcup_{{\pmb\mu}\in\mathcal{S}^{p-2}}\bigcup_{\kappa> 0}  \big\{{\rm P}^{{\cal TM}(n)}_{\thetab,g,{\pmb\mu},\kappa}\big\}.
$
From contiguity, \eqref{gtestloc} also holds under the sequences of local tangent elliptical alternatives considered in Corollary~\ref{powerE}, which implies that~$\phi^{\rm loc}_{g*}$ has asymptotic power~$\alpha$ under such alternatives. 

\subsection{The semiparametric case}
\label{sec:loctestunsp}

The test~$\phi^{\rm loc}_{g*}$ constructed above is a purely parametric test: it requires the knowledge of the underlying angular function~$g$. In practice, of course, $g$ may hardly be assumed to be known and it is therefore desirable to define a location test that would be valid (in the sense that it meets asymptotically the nominal level constraint) under a broad range of angular functions~$g$. Two options are possible here. The first one aims at uniform optimality in~$g$ by reconstructing, at any~$g$, the test statistic~$Q_{\hat{\thetab},g*}^{\rm loc}$ above. The form of the $g$-efficient central sequence in~\eqref{effcent} makes it clear that this requires estimating nonparametrically the optimal score function~$\varphi_g$, which typically requires large sample sizes and which makes it hard to control the replacement of~$\thetab$ with~$\hat{\thetab}$. We therefore favor the second approach, that consists in robustifying the parametric test~$\phi^{\rm loc}_{g*}$ in such a way that it remains \textit{valid} away from the target angular function at which power optimality is to be achieved (of course, in general, the resulting test will not be \textit{optimal} away from the selected target density).\\

To be more specific, assume that we target optimality at the fixed angular function~$f$. Our goal is to define a test statistic that:~(\textit{i}) is asymptotically equivalent to~$Q_{\hat{\thetab},f*}^{\rm loc}$ whenever~$f$ is the true angular function (which will ensure asymptotic optimality of the resulting test at angular function~$f$);~(\textit{ii}) remains~$\chi^2_{p-1}$ under the null with angular function~$g\neq f$ (which will guarantee validity away from angular function~$f$). With these objectives in mind, consider the alternative efficient central sequence
\begin{align}
\Deltab_{\thetab,f;g;2*}^{{\cal TM}(n)}
:=&\,
\Deltab_{\thetab;2}^{{\cal TM}(n)}
-
\Gamb_{\thetab,g;21}^{\cal TM}
\Gamb_{\thetab,f;g;11}^{-}
\Deltab_{\thetab,f;1}\n\nonumber\\
=&\,
\frac{1}{\sqrt{n}} \sum_{i=1}^n
\bigg(
1-\frac{\mathcal{I}_p(g)}{{\cal J}_p(f;g)}
\,
\varphi_f(V_{i,\thetab}) (1-V_{i,\thetab}^2)^{1/2}
\bigg)
\Ub_{i, \thetab},\label{effcentalt}
\end{align}
where
$
\Gamb_{\thetab,f;g;11}
:=
({\cal J}_p(f;g)/(p-1))({\bf I}_p- \thetab\thetab^T)
$
involves the ``cross-information'' quantity
\begin{align*}
{\cal J}_p(f;g)
:=
\int_{-1}^{1} \varphi_{f}(t) \varphi_{g}(t) (1-t^2) \tilde{g}_p(t) \, dt.
\end{align*}
First note that, for~$g=f$, this alternative efficient central sequence $\Deltab_{\thetab,f;g;2*}^{{\cal TM}(n)}$ coincides with the $f$-version of the efficient central sequence in~\eqref{effcent}, so that a test based on~\eqref{effcentalt} will meet the objective~(\textit{i}) above. As for the objective~(\textit{ii}), we have the following result.

\begin{prop} \label{alignsemi} Let~$\hat{\thetab}$ satisfy \ref{assump:2}. Then, for any~$\thetab\in\mathcal{S}^{p-1}$ and any~$g\in \mathcal{G}_a\cap \mathcal{G}'$, 
	\begin{align}
	\label{gtest}
	\Deltab_{\hat\thetab,f;g;2*}^{{\cal TM}(n)}
	=
	\Deltab_{\thetab,f;g;2*}^{{\cal TM}(n)}
	+
	o_{\rm P}(1),
	\end{align}
	as $\ny$ under ${\rm P}\n_{\thetab,g}$.
\end{prop}

This confirms that the alternative efficient central sequence above is defined so that the replacement of~$\thetab$ with~$\hat{\thetab}$ has no asymptotic impact also under~$g\neq f$. Since, under~${\rm P}\n_{\thetab,g}$,
$$
\Deltab_{\thetab,f;g;2*}^{{\cal TM}(n)}\inlaw\mathcal{N}\big({\bf 0},\Gamb_{f;g;22*}^{\cal TM}\big), 
$$
with
$$
\Gamb_{f;g;22*}^{\cal TM}
:=
\frac{1}{p-1}
\bigg(
1- \frac{2\mathcal{I}_p(g){\cal H}_p(f;g)}{{\cal J}_p(f;g)}
+ \frac{\mathcal{I}_p^2(g) {\cal K}_p(f;g)}{{\cal J}_p^2(f;g)}
\bigg)
\mathbf{I}_{p-1},
$$
where we let
\begin{align*}
{\cal H}_p(f;g)
:=
\int_{-1}^{1} \varphi_{f}(t) (1-t^2)^{1/2} \tilde{g}_p(t) \, dt \quad\textrm{and}\quad
{\cal K}_p(f;g)
:=
\int_{-1}^{1} \varphi_{f}^2(t) (1-t^2) \tilde{g}_p(t)\, dt
,
\end{align*}
the resulting test rejects the null hypothesis at asymptotic level~$\alpha$ whenever
\begin{align}
\label{ftest}
Q_{\hat\thetab,f;g*}^{\rm loc}
:=
\big(\Deltab_{\hat{\thetab},f;g;2*}^{{\cal TM}(n)}\big)^{T}
\big(\Gamb_{f;g;22*}^{\cal TM}\big)^{-1}
\Deltab_{\hat{\thetab},f;g;2*}^{{\cal TM}(n)}
>\chi^2_{p-1,1-\alpha}.
\end{align}
Le Cam's third lemma allows showing that, under the sequence of alternatives considered in Corollary~\ref{powerF}, $\Deltab_{\thetab,f;g;2*}^{{\cal TM}(n)}$ is asymptotically normal with covariance $\Gamb_{f;g;22*}^{\cal TM}$ and mean
$$
{\bf m}_{f;g}
:=
\lim_{n\to \infty}
{\rm E}_{\thetab,g}\big[
\Deltab_{\thetab,f;g;2*}^{{\cal TM}(n)}
\big(\Deltab_{\thetab;2}^{{\cal TM}(n)}\big)^{T}\big]
k_n {\pmb \mu}
=
\frac{1}{p-1}
\bigg(
1
-
\frac{\mathcal{I}_p(f) \mathcal{H}_p(f;g)}{\mathcal{J}_p(f;g)}
\bigg)
k {\pmb \mu},
$$
so that~$Q_{\thetab;f;g*}^{\rm loc}$ (hence also,~$Q_{\hat{\thetab};f;g*}^{\rm loc}$) is asymptotically $\chi^2_{p-1}(\lambda)$ with non-centrality parameter $\lambda$ given by
\begin{align}
{\bf m}_{f;g}^{T} &\big(\Gamb_{f;g;22*}^{\cal TM}\big)^{-1} {\bf m}_{f;g}\nonumber\\
&=
\frac{k^2}{p-1}
\bigg(
1-\frac{\mathcal{I}_p(g) \mathcal{H}_p(f;g)}{\mathcal{J}_p(f;g)}
\bigg)^2
\,
\bigg/
\,
\bigg(
1-\frac{2\mathcal{I}_p(g){\cal H}_p(f;g)}{{\cal J}_p(f;g)}
+ \frac{\mathcal{I}_p^2(g) {\cal K}_p(f;g)}{{\cal J}_p^2(f;g)}
\bigg)
.\label{ncpsemi}
\end{align}
Now, since the test statistic~\eqref{ftest} still depends on the unknown underlying angular function~$g$, turning this pseudo-test into a genuine test requires estimating consistently the quantities~$\mathcal{I}_p(g)$, ${\cal J}_p(f;g)$, ${\cal H}_p(f;g)$, and~${\cal K}_p(f;g)$. To that aim, we express them as
\begin{align*}
\mathcal{I}_p(g)=&\,(p-2)\,{\rm E}_{\thetab,g}\bigg[ \frac{V_{1,\thetab}}{(1-V_{1,\thetab}^2)^{1/2}} \bigg],\\
{\cal J}_p(f;g)
=&\,
(p-1)\, {\rm E}_{\thetab,g}[\varphi_f(V_{1,\thetab}) V_{1,\thetab}]-{\rm E}_{\thetab,g}[\varphi_f'(V_{1,\thetab}) (1-V_{1,\thetab}^2)],\\
{\cal H}_p(f;g)
:=&\,
{\rm E}_{\thetab,g}[\varphi_f(V_{1,\thetab}) (1-V_{1,\thetab}^2)^{1/2}],
\qquad{\cal K}_p(f;g)
:=
{\rm E}_{\thetab,g}[\varphi_f^2(V_{1,\thetab}) (1-V_{1,\thetab}^2)]
\end{align*}
(the fist two identities are obtained from integration by parts, assuming that~$\varphi_f$ is differentiable). Natural estimators of these quantities are
\begin{align*}
\hat{\mathcal{I}}_p(g)
&:=
\frac{p-2}{n} \sum_{i=1}^n
\frac{V_{i,\hat{\thetab}}}{(1-V_{i,\hat{\thetab}}^2)^{1/2}}
,\\
\hat{\cal J}_p(f;g)
&:=
\frac{p-1}{n}
\sum_{i=1}^n
\varphi_f(V_{i,\hat{\thetab}}) V_{i,\hat{\thetab}}
-
\frac{1}{n}
\sum_{i=1}
\varphi_f'(V_{i,\hat{\thetab}})
(1-V_{i,\hat{\thetab}}^2)
,\\
\hat{{\cal H}}_p(f;g)
&:= 
\frac{1}{n} \sum_{i=1}^n
\varphi_f(V_{i,\hat{\thetab}}) (1-V_{i,\hat{\thetab}}^2)^{1/2},
\qquad\hat{\cal K}_p(f;g):= \frac{1}{n}
\sum_{i=1}^n
\varphi_f^2(V_{i,\hat{\thetab}}) (1-V_{i,\hat{\thetab}}^2),
\end{align*}
at any~$g$ for which~$\mathcal{I}_p(g)$, ${\cal J}_p(f;g)$, ${\cal H}_p(f;g)$, and~${\cal K}_p(f;g)$ are finite. Consistency follows by successively applying the weak law of large numbers, under~${\rm P}_{\thetab+{n^{-1/2}}\tb_n,g}\n$, with~$\thetab+{n^{-1/2}}\tb_n\in\mathcal{S}^{p-1}$, to random variables of the form~$n^{-1}\sum_{i=1}^n H_f(V_{i,\thetab+{n^{-1/2}}\tb_n})$ (with $H_f$ a suitable function), the general version of the Le Cam's third lemma  (see, e.g., Theorem~6.6 in \citealp{van1998}), and then Lemma~4.4 from \cite{Kreiss87}. \\

We consider now the important particular case~$f_\eta(r)=\exp(\eta r)$, which will provide the test~$\phi^{\rm loc}_{\rm vMF}$ described in Section~\ref{sec:locationtestunsp}. Since $f_\eta\in\mathcal{G}_a$ is the vMF angular function with concentration parameter~$\eta$ (we avoid using the standard notation~$\kappa$, as this notation was used to denote the skewness intensity in the tangent vMF model), we have~$\varphi_{f_\eta}(r)=\eta$. Letting
$$D_{p,g}:=\frac{(p-2)\,{\rm E}_{\thetab,g}[V_{1,\thetab} (1-V_{1,\thetab}^2)^{-1/2}]}{(p-1)\, {\rm E}_{\thetab,g}[V_{1,\thetab}]},$$
we have
$$
\Deltab_{\thetab,{\rm vMF},g;2*}^{{\cal TM}(n)}
:=
\Deltab_{\thetab,f_\eta;g;2*}^{{\cal TM}(n)}
=
\frac{1}{\sqrt{n}} \sum_{i=1}^n
\Big(
1-D_{p,g}\,
(1-V_{i,\thetab}^2)^{1/2}
\Big)
\Ub_{i, \thetab}
$$
and
\begin{align*}
\Gamb_{{\rm vMF},g;22*}^{{\cal TM}(n)}
:=
\Gamb_{f_\eta;g;22*}^{\cal TM}
=\frac{1}{p-1}
\,
\bigg(
1-2 D_{p,g} {\rm E}_{\thetab,g}[(1-V_{1,\thetab}^2)^{1/2}]+ D_{p,g}^2  (1-{\rm E}_{\thetab,g}[V_{1,\thetab}^2]) \bigg)
\mathbf{I}_{p-1},
\end{align*}
where the notation is justified by the fact that, quite nicely, the $f_\eta$-efficient central sequence and corresponding Fisher information matrix do not depend on~$\eta$. In the present case, the quantities to be estimated consistently are therefore
\begin{align}
\label{tofinestim}
{\rm E}_{\thetab,g}[V_{1,\thetab}(1-V_{1,\thetab}^2)^{-1/2}],
\qquad {\rm E}_{\thetab,g}[(1-V_{1,\thetab}^2)^{1/2}], \qquad 
{\rm E}_{\thetab,g}[V_{1,\thetab}],
\quad \textrm{and}\quad
{\rm E}_{\thetab,g}[V_{1,\thetab}^2],
\end{align}
and the corresponding estimators are
\begin{align}
\label{finestim}
\frac{1}{n} \sum_{i=1}^n
V_{i,\hat{\thetab}} (1-V_{i,\hat{\thetab}}^2)^{-1/2},
\qquad \frac{1}{n} \sum_{i=1}^n
\, (1-V_{i,\hat{\thetab}}^2)^{1/2}, \qquad 
\frac{1}{n} \sum_{i=1}^n
V_{i,\hat{\thetab}},
\quad \textrm{and}\quad
\frac{1}{n} \sum_{i=1}^n
V_{i,\hat{\thetab}}^2,
\end{align}
respectively. The same argument as above proves consistency of these estimators at any~$g$ in the collection~$\mathcal{G}_b$ that was introduced in Section~\ref{sec:locationtestunsp} (note that the last three expectations in~\eqref{tofinestim} are trivially finite for any~$g$).
The resulting test rejects the null of rotational symmetry about an unspecified~$\thetab$ whenever
$$
Q^{\rm loc}_{\rm vMF}
:=
\big(\widehat{\Deltab}_{\hat{\thetab},{\rm vMF};g;2*}^{{\cal TM}(n)}\big)^{T}
\big(
\widehat{\Gamb}_{{\rm vMF};g;22*}^{{\cal TM}(n)}
\big)^{-1}
\widehat{\Deltab}_{\hat{\thetab},{\rm vMF};g;2*}^{{\cal TM}(n)}
>\chi^2_{p-1,1-\alpha},
$$
where~$\widehat{\Deltab}_{\thetab,{\rm vMF};g;2*}^{{\cal TM}(n)}$ and $\widehat{\Gamb}_{\thetab,{\rm vMF};g;22*}^{{\cal TM}(n)}$ result from~$\Deltab_{\thetab,{\rm vMF};g;2*}^{{\cal TM}(n)}$ and~$\Gamb_{{\rm vMF};g;22*}^{{\cal TM}(n)}$, respectively, by replacing~$\thetab$ with~$\hat{\thetab}$ and the quantities in~\eqref{tofinestim} with their consistent estimators in~\eqref{finestim}. Clearly, this test is the test~$\phi^{\rm loc}_{\rm vMF}$ from Section~\ref{sec:locationtestunsp}. While this test was built to be locally asymptotically maximin at asymptotic level~$\alpha$ when testing
$$
\bigcup_{\thetab\in\mathcal{S}^{p-1}} \bigcup_{g\in \mathcal{G}_a\cap\mathcal{G}_b\cap\mathcal{G}'}\big\{ {\rm P}\n_{\thetab,g} \big\}
\quad\textrm{ against } \quad
\bigcup_{\thetab\in\mathcal{S}^{p-1}}\bigcup_{{\pmb\mu}\in\mathcal{S}^{p-2}}\bigcup_{\kappa> 0}  \big\{{\rm P}^{{\cal TM}(n)}_{\thetab,f_\eta,{\pmb\mu},\kappa}\big\},
$$
it is also optimal for the problem of testing 
$$
\bigcup_{\thetab\in\mathcal{S}^{p-1}} \bigcup_{g\in \mathcal{G}_a\cap\mathcal{G}_b\cap\mathcal{G}'}\big\{ {\rm P}\n_{\thetab,g} \big\}
\quad\textrm{ against } \quad
\bigcup_{\thetab\in\mathcal{S}^{p-1}} \bigcup_{\eta>0} \bigcup_{{\pmb\mu}\in\mathcal{S}^{p-2}}\bigcup_{\kappa> 0}  \big\{{\rm P}^{{\cal TM}(n)}_{\thetab,f_\eta,{\pmb\mu},\kappa}\big\}
$$
since~$Q^{\rm loc}_{\rm vMF}$ does not depend on~$\eta$. 

\section{Proofs of the main results}
\label{sec:app:A}

The lemmas given in Appendix \ref{sec:app:B} are used to prove the main results.

\begin{proof}[Proof of Theorem~\ref{pdfellipt}]
	Consider first the case~$\Xb\sim \mathcal{TE}_p(\thetab_0,g,\Lamb)$, with~${\thetab}_0:=(1,0,\ldots,0)^{T}\in \R^p$. Clearly,~$\Xb= (V, (1-V^2)^{1/2} \Ub^{T})^{T}$, with~$V:=v_{\thetab_0}(\Xb)=X_1$ and~$\Ub:=\ub_{\thetab_0}(\Xb)=(X_2,\ldots,X_p)^{T}/\allowbreak\sqrt{1-X_1^2}$, where we used the notation introduced in~\eqref{vandsignU}. By definition,~$\Ub$ takes its values in~$\mathcal{S}^{p-2}$, with density~$
	{\bf u}
	\mapsto
	c^{\cal A}_{p-1,\Lamb}
	({\bf u}^{T} \Lamb^{-1} {\bf u})^{-(p-1)/2}
	$ with respect to~$\sigma_{p-2}$. Therefore, conditional on~$V=v$, $(1-V^2)^{1/2}\Ub$ takes its values on the hypersphere~$\mathcal{S}^{p-2}(r_v)\subset\R^{p-1}$ with radius $r_v:=(1-v^2)^{1/2}$. Its density with respect to the surface area measure~$\sigma_{p-2,r}$ on~$\mathcal{S}^{p-2}(r_v)$ is (recall that~$V$ and~$\Ub$ are mutually independent)
	$$
	{\bf w}
	\mapsto
	c^{\cal A}_{p-1,\Lamb} \bigg(\frac{{\bf w}^{T}\Lamb^{-1} {\bf w}}{r_v^2} \bigg)^{-(p-1)/2} r_v^{-(p-2)},
	$$
	where~$r^{-(p-2)}$ is the Jacobian of the radial projection of~$\mathcal{S}^{p-2}(r)$ onto~$\mathcal{S}^{p-2}$.
	Since $d\sigma_{p-2,r}=r^{p-2} d\sigma_{p-2}$, the density of~$\Xb$
	with respect to the product measure~$\mu\times\sigma_{p-2}$ (where~$\mu$ stands for the Lebesgue measure on~$[-1,1]$) is
	\begin{align*}
	\xb
	\mapsto&\,
	c^{\cal A}_{p-1,\Lamb}
	\big(
	{\ub_{\thetab_0}^{T}(\xb)\Lamb^{-1} \ub_{\thetab_0}(\xb)}
	\big)^{-(p-1)/2}
	\,
	\frac{d{\rm P}^V_{\thetab_0,g,\Lamb}}{d\mu}(v_{\thetab_0}(\xb))
	\\
	&=
	c^{\cal A}_{p-1,\Lamb}
	\big(
	{\ub_{\thetab_0}^{T}(\xb)\Lamb^{-1} \ub_{\thetab_0}(\xb)}
	\big)^{-(p-1)/2}
	\,
	\omega_{p-1} c_{p,g}
	(1-v^2_{\thetab_0}(\xb))^{(p-3)/2} g(v_{\thetab_0}(\xb))
	.
	\end{align*}
	The result for~${\thetab}=\thetab_0$ then follows from the fact that (see, e.g., page~44 of \citealp{Wat1983})
	$$
	\frac{d(\mu\times\sigma_{p-2})}{d\sigma_{p-1}}(\xb)=(1-v^2_{\thetab_0}(\xb))^{(p-3)/2}.
	$$
	To obtain the result for an arbitrary value of~$\thetab$, let~$\Xb\sim \mathcal{TE}_p(\thetab, g,\Lamb)$ and pick a $p\times p$ orthogonal matrix~${\bf O}$ such that ${\bf O} \thetab=\thetab_0$. Since~${\bf O}\Gamb_{{\thetab}}=\Gamb_{\thetab_0}$, we have that~$\Ob\Xb\sim \mathcal{TE}_p(\thetab_0,g,\Lamb)$. Therefore, the result for~$\thetab=\thetab_0$ implies that the density of~$\Xb$ with respect to~$\sigma_{p-1}$ is
	\begin{align*}
	\xb
	\mapsto&\,
	|\!\det\,{\bf O}| \,
	\omega_{p-1}
	c^{\cal A}_{p-1,\Lamb}
	c_{p,g}
	g(v_{\thetab_0}({\bf O}\xb))
	\big(
	{\ub_{\thetab_0}^{T}({\bf O}\xb)\Lamb^{-1} \ub_{\thetab_0}({\bf O}\xb)}
	\big)^{-(p-1)/2}
	\\
	&=
	\omega_{p-1}
	c^{\cal A}_{p-1,\Lamb}
	c_{p,g}
	g(v_{\thetab}(\xb))
	\big(
	{\ub_{\thetab}^{T}(\xb)\Lamb^{-1} \ub_{\thetab}(\xb)}
	\big)^{-(p-1)/2}
	,
	\end{align*}
	as was to be proved. 
\end{proof}

\begin{proof}[Proof of Theorem~\ref{LANell}]
	Lemma~\ref{lem:3} readily entails that
	\begin{align*}
	\log \frac{ d{\rm P}^{{\cal TE}(n)}_{\thetab_n,g,\Lamb_n}}{ d{\rm P}\n_{\thetab,g}}
	=
	\log\frac{ d{\rm P}^{{\cal TE}(n)}_{\thetab_n,g,\Lamb_n}}{ d{\rm P}_{\thetab_n,g}\n}
	+
	\log\frac{ d{\rm P}_{\thetab_n,g}\n}{ d{\rm P}_{\thetab,g}\n} =
	\log\frac{ d{\rm P}^{{\cal TE}(n)}_{\thetab_n,g,\Lamb_n}}{ d{\rm P}_{\thetab_n,g}\n}
	+
	{\bf t}_n^{T}\Deltab_{\thetab,g;1}^{(n)}- \frac{1}{2} {\bf t}_n^{T} \Gamb_{\thetab,g;11} {\bf t}_n+o_{\rm P}(1)
	\end{align*}
	as $\ny$ under $ {\rm P}_{\thetab,g}\n$.
	Therefore, we only need to show that
	\begin{align*}
	L_n:=
	\log\frac{ d{\rm P}^{{\cal TE}(n)}_{\thetab_n,g,\Lamb_n}}{ d{\rm P}_{\thetab_n,g}\n}
	=
	(\vechrond{\mathbf{L}}_n)^{T}\Deltab_{\thetab;2}^{{\cal TE}(n)}- \frac{1}{2} (\vechrond{\mathbf{L}}_n)^{T}\, \Gamb_{\thetab;22}^{\cal TE}(\vechrond{\mathbf{L}}_n)+o_{\rm P}(1)
	\end{align*}
	as $\ny$ under $ {\rm P}_{\thetab,g}\n$.
	First note that Theorem~\ref{pdfellipt} gives
	\begin{align}
	\label{Vnn}
	L_n
	=
	-\frac{n}{2} \log\big({\rm det}\, \Lamb_n\big)
	-
	\frac{p-1}{2} \sum_{i=1}^n \log\big({\bf U}_{i,\thetab_n}^{T}\Lamb_n^{-1} {\bf U}_{i,\thetab_n}\big)
	=:L_{n,1}+L_{n,2}
	,
	\end{align}
	say. Since $\log({\rm det}({\bf I}_{p-1}+ {\bf A}))={\rm tr}[{\bf A}]- \frac{1}{2}{\rm tr}[{\bf A}^2]+o(\| {\bf A}\|^2)$ as $\| {\bf A}\| \to 0$, we have that
	\begin{align}
	\label{constterm}
	L_{n,1}
	=
	-\frac{n}{2} \log\big({\rm det}({\bf I}_{p-1}+n^{-1/2}{\mathbf{L}}_n)\big)
	=
	\frac{1}{4} {\rm tr}[{\mathbf{L}}_n^2]+o(1)
	\end{align}
	as $\ny$ (recall that~${\rm tr}[{\mathbf{L}}_n]=0$).
	Now,
	write
	\begin{align*}
	L_{n,2}
	&=
	-
	\frac{p-1}{2}
	\sum_{i=1}^n \log\big(1+{\bf U}_{i,\thetab_n}^{T}(\Lamb_n^{-1}- {\bf I}_{p-1}) {\bf U}_{i,\thetab_n}\big)
	\\
	&=
	-
	\frac{p-1}{2}
	\sum_{i=1}^n \log\big(1+{\rm tr}[{\bf U}_{i,\thetab_n}{\bf U}_{i,\thetab_n}^{T} (\Lamb_n^{-1}- {\bf I}_{p-1})]\big) \\
	&=:
	-
	\frac{p-1}{2}
	\sum_{i=1}^n \log\big(1+T_{i,n}\big)
	.
	\end{align*}
	Using (9)--(10) in pages 218--219 from \cite{MagNeu},
	we obtain that 
	$
	T_{i,n}
	=\break
	-n^{-1/2} {\bf U}_{i,\thetab_n}^{T} {\mathbf{L}}_n{\bf U}_{i,\thetab_n}
	+n^{-1} {\bf U}_{i,\thetab_n}^{T} {\mathbf{L}}_n^2 {\bf U}_{i,\thetab_n}
	+R_{i,n}
	$,
	where (due to the uniform boundedness of the ${\bf U}_{i,\thetab_n}$'s) $\max_{i=1,\ldots,n} R_{i,n}=O_{\rm P}(n^{-3/2})$ as~$\ny$ under~$ {\rm P}_{\thetab,g}\n$. Using the fact that $\log(1+x)=x- \frac{1}{2}x^2+o(x^2)$ as $x \to 0$, it follows that
	\begin{align*}
	L_{n,2}
	&=
	-
	\frac{p-1}{2}
	\sum_{i=1}^n
	\log\bigg(1
	- \frac{1}{\sqrt{n}} {\bf U}_{i,\thetab_n}^{T} {\mathbf{L}}_n{\bf U}_{i,\thetab_n}
	+\frac{1}{n} {\bf U}_{i,\thetab_n}^{T}{\mathbf{L}}_n^2 {\bf U}_{i,\thetab_n}
	+R_{i,n}\bigg)
	\\
	&=
	-
	\frac{p-1}{2}
	\sum_{i=1}^n
	\bigg\{
	- \frac{1}{\sqrt{n}} {\bf U}_{i,\thetab_n}^{T}{\mathbf{L}}_n{\bf U}_{i,\thetab_n}
	+\frac{1}{n} {\bf U}_{i,\thetab_n}^{T} {\mathbf{L}}_n^2 {\bf U}_{i,\thetab_n}
	-\frac{1}{2n}({\bf U}_{i,\thetab_n}^{T}{\mathbf{L}}_n{\bf U}_{i,\thetab_n})^2
	\bigg\}
	+o_{\rm P}(1)
	\end{align*}
	as~$\ny$ under~$ {\rm P}_{\thetab,g}\n$. Using Lemma \ref{lem:2}, the law of large numbers for triangular arrays then yields
	\begin{align*}
	L_{n,2}
	=&\,
	\bigg(\frac{p-1}{2\sqrt{n}} \sum_{i=1}^n {\bf U}_{i,\thetab_n}^{T}{\mathbf{L}}_n{\bf U}_{i,\thetab_n}\bigg)
	-\frac{p-1}{2}
	\,
	{\rm E}\bigg[{\bf U}_{1,\thetab_n}^{T} {\mathbf{L}}_n^2{\bf U}_{1,\thetab_n}- \frac{1}{2}({\bf U}_{1,\thetab_n}^{T} {\mathbf{L}}_n {\bf U}_{1,\thetab_n})^2\bigg]+o_{\rm P}(1)
	\nonumber \\
	=&\,
	\frac{p-1}{2\sqrt{n}}
	(\veco{\mathbf{L}}_n)^{T} \sum_{i=1}^n \veco({\bf U}_{i,\thetab_n}{\bf U}_{i,\thetab_n}^{T})
	\nonumber  \\
	&-\frac{p-1}{2} (\veco{\mathbf{L}}_n)^{T}\left[ \frac{1}{p-1}\,{\bf I}_{(p-1)^2} - \frac{1}{2(p^2-1)} ({\bf I}_{(p-1)^2}+ {\bf K}_{p-1}+{\bf J}_{p-1}) \right]
	(\veco {\mathbf{L}}_n) +o_{\rm P}(1)
	\end{align*}
	as $\ny$ under $ {\rm P}_{\thetab,g}\n$. Applying Part~\ref{lem:4:3:te} of Lemma~\ref{lem:4}, and using the identities~${\bf K}_{p-1}(\veco \Ab)=\veco (\Ab^{T})$ and $(\veco \Ab)^{T} (\veco \Bb)={\rm tr}[\Ab^{T}\Bb]$ (this second identity provides~$(\veco {\bf I}_{p-1})^{T}(\veco{\mathbf{L}}_n) ={\rm tr}[{\mathbf{L}}_n]=0$, hence~${\bf J}_{p-1} (\veco{\mathbf{L}}_n)={\bf 0}$), we obtain
	\begin{align}
	L_{n,2}
	=
	\frac{p-1}{2\sqrt{n}}
	(\veco{\mathbf{L}}_n)^{T} \sum_{i=1}^n \veco\bigg({\bf U}_{i,\thetab}{\bf U}_{i,\thetab}^{T} - \frac{1}{p-1}\,{\bf I}_{p-1} \bigg)
	-\frac{p}{2(p+1)} \,{\rm tr}[{\mathbf{L}}_n^2] +o_{\rm P}(1)
	\label{Snnbis}
	\end{align}
	as $\ny$ under $ {\rm P}_{\thetab,g}\n$.
	Plugging~\eqref{constterm}--\eqref{Snnbis} in~\eqref{Vnn} then provides
	\begin{align*}
	L_n
	&=
	\frac{p-1}{2\sqrt{n}}
	(\veco{\mathbf{L}}_n)^{T}\sum_{i=1}^n \veco\bigg({\bf U}_{i,\thetab}{\bf U}_{i,\thetab}^{T} - \frac{1}{p-1}\,{\bf I}_{p-1}\bigg)
	-\frac{p-1}{4(p+1)} \,{\rm tr}[{\mathbf{L}}_n^2] +o_{\rm P}(1)
	,
	\end{align*}
	as $\ny$ under $ {\rm P}_{\thetab,g}\n$, which,  by using the definition of~$\Mb_p$ and the matrix identities above, yields~\eqref{LAQell}. Finally, the CLT ensures that,  under $ {\rm P}_{\thetab,g}\n$, $
	\Deltab^{{\cal TE}(n)}_{\thetab;2}\inlaw\mathcal{N}\big({\bf 0}, \Gamb_{\thetab;22}^{\cal TE}\big)$ with
	$$
	{\bf M}_p \bigg( \frac{p-1}{4(p+1)} ({\bf I}_{(p-1)^2}+ {\bf K}_{p-1}+{\bf J}_{p-1}) - \frac{1}{4}\,{\bf J}_{p-1}\bigg)  {\bf M}_p^{T}
	=
	\frac{p-1}{4(p+1)} {\bf M}_p \big({\bf I}_{(p-1)^2}+ {\bf K}_{p-1}\big)  {\bf M}_p^{T}
	=
	\Gamb_{\thetab;22}^{\cal TE}
	,
	$$
	where we used the fact that~$\Mb_p(\veco {\bf I}_{p-1})={\bf 0}$ (see (\textit{v}) in Lemma~4.2 of \cite{Pai2008}).
\end{proof}

\begin{proof}[Proof of Theorem~\ref{LANvM}]
	First note that
	\begin{align*}
	\log \frac{ d{\rm P}^{{\cal TM}(n)}_{\thetab_n,g,\deltab_n}}{ d{\rm P}\n_{\thetab,g}} =
	\log\frac{ d{\rm P}^{{\cal TM}(n)}_{\thetab_n,g,\deltab_n}}{ d{\rm P}_{\thetab_n,g}\n}
	+
	\log\frac{ d{\rm P}_{\thetab_n,g}\n}{ d{\rm P}_{\thetab,g}\n} =
	\log\frac{ d{\rm P}^{{\cal TM}(n)}_{\thetab_n,g,\deltab_n}}{ d{\rm P}_{\thetab_n,g}\n}
	+
	{\bf t}_n^{T}\Deltab_{\thetab,g;1}^{(n)}- \frac{1}{2} {\bf t}_n^{T} \Gamb_{\thetab,g;11} {\bf t}_n+o_{\rm P}(1).
	\end{align*}
	In the parametrization adopted in Theorem~\ref{LANvM}, recall that~$\deltab_n$ corresponds to a skewness direction~${\pmb \mu}_n:=\deltab_n/\|\deltab_n\|={\bf d}_n/\|{\bf d}_n\|$ and a skewness intensity~$\kappa_n:=\|\deltab_n\|=n^{-1/2}\|{\bf d}_n\|$.
	From Theorem~\ref{pdf2}, we then readily obtain
	$$
	G_n
	:=
	\log\frac{ d{\rm P}^{{\cal TM}(n)}_{\thetab_n,g,\deltab_n}}{ d{\rm P}_{\thetab_n,g}\n}
	=
	n \big(\log(c_{p-1, n^{-1/2}\| {\bf d}\n\|})-\log (c_{p-1, 0})\big)
	+
	\big({\bf d}\n\big)^{T} \frac{1}{\sqrt{n}} \sum_{i=1}^n {\bf U}_{i,\thetab_n}
	.
	$$
	Lemma A.1 in \cite{Cut2017} implies that
	$$
	n \big(\log (c_{p-1, n^{-1/2}\| {\bf d}\n\|}) - \log (c_{p-1, 0})\big)
	=
	-\frac{1}{2(p-1)}\,\| {\bf d}\n\|^2
	+o(1)
	$$
	as $\ny$, which, by using \ref{lem:4:3:tm} in Lemma~\ref{lem:4}, yields
	\begin{align*}
	G_n
	&=
	({\bf d}\n)^T \frac{1}{\sqrt{n}} \sum_{i=1}^n {\bf U}_{i,\thetab_n}
	- \frac{1}{2(p-1)}\,\| {\bf d}\n\|^2
	+o_{\rm P}(1)
	\\
	&=
	({\bf d}\n)^T
	\bigg[
	\frac{1}{\sqrt{n}} \sum_{i=1}^n {\bf U}_{i,\thetab}
	-\frac{\mathcal{I}_p(g)}{p-1} \Gamb_{\thetab}^{T} {\bf t}_n
	\bigg]
	- \frac{1}{2(p-1)}\,\| {\bf d}\n\|^2
	+o_{\rm P}(1)
	\end{align*}
	as $\ny$ under ${\rm P}_{\thetab,g}\n$. Therefore,
	\begin{align*}
	\log \frac{ d{\rm P}^{{\cal TM}(n)}_{\thetab_n,g,\deltab_n}}{ d{\rm P}\n_{\thetab,g}}
	=\,&
	{\bf t}_n^{T}\Deltab_{1, \thetab}\n
	+
	({\bf d}\n)^{T}
	\frac{1}{\sqrt{n}} \sum_{i=1}^n {\bf U}_{i,\thetab}
	\\
	&-\frac{1}{2}
	\bigg(
	{\bf t}_n^{T}\Gamb_{\thetab,g;11} {\bf t}_n
	+
	\frac{2\mathcal{I}_p(g)}{p-1} ({\bf d}\n)^{T} \Gamb_{\thetab}^{T} {\bf t}_n
	+ \frac{1}{p-1}\,\| {\bf d}\n\|^2
	\bigg)
	+o_{\rm P}(1)
	,
	\end{align*}
	as $\ny$ under ${\rm P}_{\thetab,g}\n$, which establishes the result.
\end{proof}

\begin{proof}[Proof of Proposition~\ref{alignparamm}]
	Proposition~3.1 of \cite{Leyetal2013} directly implies that if~$\hat{\thetab}$ satisfies \ref{assump:2}, then
	\begin{align}
	\label{glinLeyetal2013}
	\Deltab_{\hat{\thetab},g;1}\n
	=
	\Deltab_{\thetab,g;1}\n
	-
	\Gamb_{\thetab,g;11}
	\sqrt{n}(\hat{\thetab}- \thetab)
	+
	o_{\rm P}(1)
	\end{align}
	as $\ny$ under ${\rm P}\n_{\thetab,g}$.
	Using this and~\eqref{linlin}, we obtain that, again as $\ny$ under ${\rm P}\n_{\thetab,g}$,
	\begin{align*}
	\Deltab_{\hat{\thetab},g;2*}^{{\cal TM}(n)}
	&=
	\Deltab_{\hat{\thetab};2}^{{\cal TM}(n)}
	-
	\Gamb_{\hat{\thetab},g;21}^{\cal TM}
	\Gamb_{\hat{\thetab},g;11}^{-}
	\Deltab_{\hat{\thetab},g;1}\n
	\\
	&=
	\big(\Deltab_{\thetab;2}^{{\cal TM}(n)}
	-
	\Gamb_{\thetab,g;21}^{\cal TM}
	\sqrt{n}(\hat{\thetab}- \thetab)\big)
	-
	\Gamb_{\hat{\thetab},g;21}^{\cal TM}
	\Gamb_{\hat{\thetab},g;11}^{-}
	(\Deltab_{\thetab,g;1}\n
	-
	\Gamb_{\thetab,g;11}
	\sqrt{n}(\hat{\thetab}- \thetab))
	+
	o_{\rm P}(1)
	\\
	&=
	\Deltab_{\thetab,g;2*}^{{\cal TM}(n)}
	-
	\Gamb_{\thetab,g;21}^{\cal TM}
	\sqrt{n}(\hat{\thetab}- \thetab)
	+
	\Gamb_{\hat{\thetab},g;21}^{\cal TM}
	\Gamb_{\hat{\thetab},g;11}^{-}
	\Gamb_{\thetab,g;11}
	\sqrt{n}(\hat{\thetab}- \thetab)
	+
	o_{\rm P}(1)
	\\
	&=
	\Deltab_{\thetab,g;2*}^{{\cal TM}(n)}
	-
	\Gamb_{\thetab,g;21}^{\cal TM}
	\sqrt{n}(\hat{\thetab}- \thetab)
	+
	\Gamb_{\thetab,g;21}^{\cal TM}
	\Gamb_{\thetab,g;11}^{-}
	\Gamb_{\thetab,g;11}
	\sqrt{n}(\hat{\thetab}- \thetab)
	+
	o_{\rm P}(1)
	\\
	&=
	\Deltab_{\thetab,g;2*}^{{\cal TM}(n)}
	+
	o_{\rm P}(1)
	,
	\end{align*}
	where the last equality follows from the fact that~$
	(\mathbf{I}_{p}-\thetab\thetab^{T})
	\sqrt{n} (\hat{\thetab}- \thetab)
	=
	\sqrt{n} (\hat{\thetab}- \thetab)
	+
	o_{\rm P}(1)$
	as $\ny$ under ${\rm P}\n_{\thetab,g}$. The result follows.
\end{proof}

\begin{proof}[Proof of Proposition~\ref{alignsemi}]
	Proposition~3.1 of \cite{Leyetal2013} actually shows that the asymptotic linearity property in \eqref{glinLeyetal2013} generalizes into
	$$
	\Deltab_{\hat{\thetab},f;1}\n
	=
	\Deltab_{\thetab,f;1}\n
	-
	\Gamb_{\thetab,f;g;11}
	\sqrt{n}(\hat{\thetab}- \thetab)
	+
	o_{\rm P}(1)
	$$
	as $\ny$ under ${\rm P}\n_{\thetab,g}$,
	which, jointly with~\eqref{linlin}, provides
	\begin{align*}
	\Deltab_{\hat{\thetab},f;g;2*}^{{\cal TM}(n)}
	&=
	\Deltab_{\hat{\thetab};2}^{{\cal TM}(n)}
	-
	\Gamb_{\hat{\thetab},g;21}^{\cal TM}
	\Gamb_{\hat{\thetab},f;g;11}^{-}
	\Deltab_{\hat{\thetab},f;1}\n
	\\
	&=
	\big(\Deltab_{\thetab;2}^{{\cal TM}(n)}
	-
	\Gamb_{\thetab,g;21}^{\cal TM}
	\sqrt{n}(\hat{\thetab}- \thetab)\big)
	-
	\Gamb_{\hat{\thetab},g;21}^{\cal TM}
	\Gamb_{\hat{\thetab},f;g;11}^{-}
	(\Deltab_{\thetab,f;1}\n
	-
	\Gamb_{\thetab,f;g;11}
	\sqrt{n}(\hat{\thetab}- \thetab))
	+
	o_{\rm P}(1)
	\\
	&=\Deltab_{\thetab,f;g;2*}^{{\cal TM}(n)}
	-
	\Gamb_{\thetab,g;21}^{\cal TM}
	\sqrt{n}(\hat{\thetab}- \thetab)
	+
	\Gamb_{\hat{\thetab},g;21}^{\cal TM}
	\Gamb_{\hat{\thetab},f;g;11}^{-}
	\Gamb_{\thetab,f;g;11}
	\sqrt{n}(\hat{\thetab}- \thetab)
	+
	o_{\rm P}(1)
	\\
	&=\Deltab_{\thetab,f;g;2*}^{{\cal TM}(n)}
	+
	o_{\rm P}(1)
	\end{align*}
	as $\ny$ under ${\rm P}\n_{\thetab,g}$, which is the desired result.
\end{proof}

\section{Required lemmas}
\label{sec:app:B}

\begin{lem}
	\label{lem:1}
	For any~$\thetab\in\mathcal{S}^{p-1}$ and~$g\in\mathcal{G}$, under $ {\rm P}_{\thetab,g}\n$:
	\begin{enumerate}[label=\textit{(\roman{*})},ref=(\textit{\roman{*}})]
		\item ${\rm E}[{\bf U}_{1,\thetab}] = \mathbf{0}$, \label{lem:1:1}
		\item ${\rm E}[{\bf U}_{1,\thetab}{\bf U}_{1,\thetab}^{T}]
		=
		\frac{1}{p-1} {\bf I}_{p-1}$, \label{lem:1:2}
		\item ${\rm E}[\veco({\bf U}_{1,\thetab}{\bf U}_{1,\thetab}^{T}) \veco ({\bf U}_{1,\thetab}{\bf U}_{1,\thetab}^{T})^{T}]
		=
		\frac{1}{p^2-1}
		\big({\bf I}_{(p-1)^2}+ {\bf K}_{p-1}+{\bf J}_{p-1}\big)$. \label{lem:1:3}
	\end{enumerate}
\end{lem}

\begin{proof}[Proof of Lemma \ref{lem:1}]
	The result is a direct consequence of Lemma~A.2 in \cite{PV16}.
\end{proof}

\begin{lem}
	\label{lem:2}
	For any~$\thetab\in\mathcal{S}^{p-1}$, $g\in\mathcal{G}$, and any bounded sequence~$({\bf t}_n)$ in~$\R^p$ such that~$\thetab_n=\thetab+n^{-1/2} {\bf t}_n\in\mathcal{S}^{p-1}$ for any~$n$, we have that, as $\ny$ under ${\rm P}_{\thetab,g}\n$:
	\begin{enumerate}[label=\textit{(\roman{*})},ref=(\textit{\roman{*}})]
		\item ${\rm E}[{\bf U}_{1,\thetab_n}] = o(1)$, \label{lem:2:1}
		\item ${\rm E}[{\bf U}_{1,\thetab_n}{\bf U}_{1,\thetab_n}^{T}] = \frac{1}{p-1} {\bf I}_{p-1} +o(1)$, \label{lem:2:2}
		\item ${\rm E}[\veco({\bf U}_{1,\thetab_n}{\bf U}_{1,\thetab_n}^{T})\veco({\bf U}_{1,\thetab_n}{\bf U}_{1,\thetab_n}^{T})^{T}]
		=
		\frac{1}{p^2-1}
		\,
		({\bf I}_{(p-1)^2}+ {\bf K}_{p-1}+{\bf J}_{p-1})
		+o(1)$.\label{lem:2:3}
	\end{enumerate}
\end{lem}

\begin{proof}[Proof of Lemma \ref{lem:2}]
	All expectations in this proof are under ${\rm P}_{\thetab,g}\n$ and all convergences are as $\ny$.
	For \ref{lem:2:1} first note that, letting~${\bf Z}_{1, \thetab}:= \Gamb_{\thetab}^{T}\Xb_1$ and $d_{1, \thetab}:=\|{\bf Z}_{1, \thetab}\|$, we have
	\begin{align*}
	\|{\bf U}_{1,\thetab_n}-{\bf U}_{1,\thetab} \|
	&\leq
	\bigg\| \frac{{\bf Z}_{1, {\thetab_n}}}{d_{1, {\thetab_n}}}-\frac{{\bf Z}_{1, {\thetab_n}}}{d_{1, {\thetab}}}\bigg\|
	+
	\bigg\|\frac{{\bf Z}_{1, {\thetab_n}}}{d_{1, {\thetab}}}- \frac{{\bf Z}_{1, {\thetab}}}{d_{1, {\thetab}}} \bigg\| \\
	&\leq
	\bigg|\frac{1}{d_{1, {\thetab_n}}}-\frac{1}{d_{1, {\thetab}}}\bigg|
	\,
	\| {\bf Z}_{1, {\thetab_n}}\|
	+
	\frac{1}{d_{1, {\thetab}}} \,
	\| {\bf Z}_{1, {\thetab_n}}- {\bf Z}_{1, {\thetab}}\|\\
	&\leq
	\frac{|d_{1, {\thetab_n}}-d_{1, {\thetab}}|}{d_{1, {\thetab}}}
	+
	\frac{1}{d_{1, {\thetab}}} \,
	\| {\bf Z}_{1, {\thetab_n}}- {\bf Z}_{1, {\thetab}}\|\\
	&\leq
	\frac{2\| {\bf Z}_{1, {\thetab_n}}- {\bf Z}_{1, {\thetab}}\|}{d_{1, {\thetab}}}
	,
	\end{align*}
	which implies that~$\|{\bf U}_{1,\thetab_n}-{\bf U}_{1,\thetab} \|=o_{\rm P}(1)$. Uniform integrability follows because~$\|{\bf U}_{1,\thetab_n}-{\bf U}_{1,\thetab} \|\leq 2$ almost surely, hence
	\begin{align}
	\label{tos}
	{\rm E}[\|{\bf U}_{1,\thetab_n} - {\bf U}_{1,\thetab}\|^2]=o(1)
	.
	\end{align}
	Since
	$
	\|{\rm E}[{\bf U}_{1,\thetab_n}]\|^2
	=
	\|{\rm E}[{\bf U}_{1,\thetab_n} - {\bf U}_{1,\thetab}]\|^2
	\leq
	\big({\rm E}[\|{\bf U}_{1,\thetab_n} - {\bf U}_{1,\thetab}\|]\big)^2
	\leq
	{\rm E}[\|{\bf U}_{1,\thetab_n} - {\bf U}_{1,\thetab}\|^2],
	$
	the result then follows from \ref{lem:1:1} in Lemma \ref{lem:1}.
	For proving \ref{lem:2:2}, we have that, since \ref{lem:1:2} in Lemma~\ref{lem:1} entails that
	$$
	\bigg\|
	\veco
	\bigg(
	{\rm E}[{\bf U}_{1,\thetab_n} {\bf U}_{1,\thetab_n}^{T}]- \frac{1}{p-1}\, {\bf I}_{p-1}
	\bigg)
	\bigg\|^2
	=
	\big\|
	{\rm E}[\veco({\bf U}_{1,\thetab_n}{\bf U}_{1,\thetab_n}^{T})-\veco({\bf U}_{1,\thetab}{\bf U}_{1,\thetab}^{T})]
	\big\|^2,
	$$
	it is enough to show that
	\begin{align}
	\label{tos2}
	{\rm E}[\|\veco({\bf U}_{1,\thetab_n}{\bf U}_{1,\thetab_n}^{T})-\veco({\bf U}_{1,\thetab}{\bf U}_{1,\thetab}^{T})\|^2]=o(1).
	\end{align}
	This follows from~\eqref{tos}, the fact that~$
	\|{\rm vec}({\bf U}_{1,\thetab_n}{\bf U}_{1,\thetab_n}^{T})- {\rm vec}({\bf U}_{1,\thetab}{\bf U}_{1,\thetab}^{T})\|^2
	=
	{\rm tr}[({\bf U}_{1,\thetab_n}{\bf U}_{1,\thetab_n}^{T}-{\bf U}_{1,\thetab}{\bf U}_{1,\thetab}^{T})^2]
	=
	2(1-({\bf U}_{1,\thetab_n}^{T}{\bf U}_{1,\thetab})^2)
	=
	\|{\bf U}_{1,\thetab_n} - {\bf U}_{1,\thetab}\|^2
	$, and the arguments in the proof of \ref{lem:2:1}.
	For \ref{lem:2:3}, we proceed as above and use that \ref{lem:1:3} in Lemma \ref{lem:1} entails that it is sufficient to show that
	$$
	w_n:=
	{\rm E}[
	\|
	\veco({\bf U}_{1,\thetab_n}{\bf U}_{1,\thetab_n}^{T})\veco ({\bf U}_{1,\thetab_n}{\bf U}_{1,\thetab_n}^{T})^{T}-\veco({\bf U}_{1,\thetab}{\bf U}_{1,\thetab}^{T})\veco({\bf U}_{1,\thetab}{\bf U}_{1,\thetab}^{T})^{T}\|^2
	]=o(1).
	$$
	Since~$w_n\leq 2(w_{1n}+w_{2n})$, with
	\begin{align*}
	w_{1n}&:=
	{\rm E}[\|(\veco({\bf U}_{1,\thetab_n}{\bf U}_{1,\thetab_n}^{T})-\veco({\bf U}_{1,\thetab}{\bf U}_{1,\thetab}^{T}))\veco({\bf U}_{1,\thetab_n}{\bf U}_{1,\thetab_n}^{T})^{T}\|^2],\\
	w_{2n}&:=
	{\rm E}[\|\veco({\bf U}_{1,\thetab}{\bf U}_{1,\thetab}^{T})(\veco({\bf U}_{1,\thetab_n}{\bf U}_{1,\thetab_n}^{T})-\veco({\bf U}_{1,\thetab}{\bf U}_{1,\thetab}^{T})^{T})\|^2],
	\end{align*}
	and since~${\bf U}_{1,\thetab_n}$ and~${\bf U}_{1,\thetab}$ are bounded almost surely, the result follows from~\eqref{tos2}.
\end{proof}

\begin{lem}
	\label{lem:3}
	Fix~$\thetab\in\mathcal{S}^{p-1}$, $g\in\mathcal{G}_a$, and let~$({\bf t}_n)$ be a bounded sequence in~$\R^p$ such that~$\thetab_n:=\thetab+n^{-1/2} {\bf t}_n\in\mathcal{S}^{p-1}$ for any~$n$. Then,
	\begin{align*} %
	\log
	\frac{ d{\rm P}^{(n)}_{\thetab_n,g}}{ d{\rm P}\n_{\thetab,g}}
	=
	{\bf t}_n^{T} \Deltab_{\thetab,g;1}^{(n)}
	- \frac{1}{2} {\bf t}_n^{T} \, \Gamb_{\thetab, g;11} {{\bf t}_n}+o_{\rm P}(1),
	\end{align*}
	as $\ny$ under $ {\rm P}_{\thetab,g}\n$, where~$\Deltab_{\thetab,g;1}^{(n)}$ and~$\Gamb_{\thetab, g;11}$ are as in Theorems~\ref{LANell} and~\ref{LANvM}.
\end{lem}

\begin{proof}[Proof of Lemma~\ref{lem:3}]
	This follows from Proposition~2.2 in \cite{Leyetal2013}.
\end{proof}

\begin{lem}
	\label{lem:4}
	For any~$\thetab\in\mathcal{S}^{p-1}$, $g\in\mathcal{G}_a$, and any bounded sequence~$({\bf t}_n)$ in~$\R^p$ such that~$\thetab_n:=\thetab+n^{-1/2} {\bf t}_n\in\mathcal{S}^{p-1}$ for any~$n$, we have that, as $\ny$ under ${\rm P}_{\thetab,g}\n$:
	\begin{enumerate}[label=$(\roman{*}_{\cal TM})$,ref=$(\roman{*}_{\cal TM})$]
		\item $\frac{1}{\sqrt{n}}
		\sum_{i=1}^n \Big(
		{\bf U}_{i,\thetab_n}-{\bf U}_{i,\thetab}-{\rm E}[{\bf U}_{i,\thetab_n}]
		\Big) = o_{\rm P}(1)$,\label{lem:4:1:tm}
		\item $\frac{1}{\sqrt{n}}
		\sum_{i=1}^n
		{\rm E}[{\bf U}_{i,\thetab_n}]
		=
		-\frac{\mathcal{I}_p(g)}{p-1} \Gamb_{\thetab}^{T} {\bf t}_n
		+o(1)
		$,\label{lem:4:2:tm}
		\item $\frac{1}{\sqrt{n}}
		\sum_{i=1}^n ({\bf U}_{i,\thetab_n} -{\bf U}_{i,\thetab})
		=
		-\frac{\mathcal{I}_p(g)}{p-1} \Gamb_{\thetab}^{T} {\bf t}_n
		+o_{\rm P}(1)
		$,\label{lem:4:3:tm}
	\end{enumerate}%
	\begin{enumerate}[label=$(\roman{*}_{\cal TE})$,ref=$(\roman{*}_{\cal TE})$]
		\item $\frac{1}{\sqrt{n}}
		\sum_{i=1}^n \Big(
		{\bf U}_{i,\thetab_n}{\bf U}_{i,\thetab_n}^{T}-{\bf U}_{i,\thetab}{\bf U}_{i,\thetab}^{T}-{\rm E}[{\bf U}_{i,\thetab_n}{\bf U}_{i,\thetab_n}^{T}]+\frac{1}{p-1}\,{\bf I}_{p-1}
		\Big)
		=
		o_{\rm P}(1)$,\label{lem:4:1:te}
		\item $\frac{1}{\sqrt{n}}
		\sum_{i=1}^n \Big(
		{\rm E}[{\bf U}_{i,\thetab_n}{\bf U}_{i,\thetab_n}^{T}]-\frac{1}{p-1}\,{\bf I}_{p-1}\Big)
		=
		o(1)$,\label{lem:4:2:te}
		\item $\frac{1}{\sqrt{n}}
		\sum_{i=1}^n \big({\bf U}_{i,\thetab_n}{\bf U}_{i,\thetab_n}^{T}-{\bf U}_{i,\thetab}{\bf U}_{i,\thetab}^{T}\big)=o_{\rm P}(1)
		$. \label{lem:4:3:te}
	\end{enumerate}
\end{lem}

\begin{proof}[Proof of Lemma \ref{lem:4}]
	Throughout this proof, expectations are under ${\rm P}_{\thetab,g}\n$, convergences are as $\ny$, and superscript~${\cal T}$ stands for~``${\cal TM}$ (respectively, ${\cal TE}$)''.
	For \ref{lem:4:1:tm}--\ref{lem:4:1:te}, let~${\bf N}^{\cal TM}_{i,n}:={\bf U}_{i,\thetab_n} - {\bf U}_{i,\thetab}$ and~${\bf N}^{\cal TE}_{i,n}:={\rm vec}({\bf U}_{i,\thetab_n}{\bf U}_{i,\thetab_n}^{T}) - {\rm vec}({\bf U}_{i,\thetab}{\bf U}_{i,\thetab}^{T})$. We have to show that
	$$
	{\bf T}_{n}^{\cal T}= n^{-1/2} \sum_{i=1}^n ({\bf N}^{\cal T}_{i,n}- {\rm E}[{\bf N}^{\cal T}_{i,n}])=o_{\rm P}(1).
	$$
	Since
	${\rm E}[ \| {\bf T}^{{\cal T}}_{n} \|^2]
	=
	n^{-1} \sum_{i,j=1}^n
	{\rm E}[ ({\bf N}_{i,n}^{{\cal T}}- {\rm E}[{\bf N}_{i,n}^{{\cal T}}])^{T}({\bf N}_{jn}^{{\cal T}}- {\rm E}[{\bf N}_{jn}^{{\cal T}}]) ]
	=
	{\rm E}[ \|{\bf N}_{1n}^{{\cal T}}- {\rm E}[{\bf N}_{1n}^{{\cal T}}]\|^2 ]
	\leq
	{\rm E}[ \|{\bf N}_{1n}^{{\cal T}}\|^2 ],
	$
	the result follows from~\eqref{tos} for ${\cal TM}$, and from \eqref{tos2} for ${\cal TE}$. For \ref{lem:4:2:tm}--\ref{lem:4:2:te}, we consider
	$$
	{\Sb}_{\thetab}^{{\cal TM}(n)}
	:=
	\frac{1}{\sqrt{n}}
	\sum_{i=1}^n
	{\bf U}_{i,\thetab}
	\quad\textrm{and}
	\quad
	{\Sb}_{\thetab}^{{\cal TE}(n)}
	:=
	\frac{1}{\sqrt{n}}
	\sum_{i=1}^n {\rm vec}\left({\bf U}_{i,\thetab}{\bf U}_{i,\thetab}^{T}-\frac{1}{p-1}\,{\bf I}_{p-1} \right).
	$$
	By using Lemma~\ref{lem:1}, the %
	CLT for triangular arrays implies that, under ${\rm P}_{\thetab_n,g}\n$,
	\begin{align}
	\label{jointas}
	\bigg(
	\begin{array}{cc}
	{\Sb}_{\thetab}^{{\cal T}(n)} \\
	\Deltab_{\thetab_n,g;1}\n
	\end{array}
	\bigg)
	\inlaw
	\mathcal{N}\bigg({\bf 0},
	\bigg(
	\begin{array}{cc}
	\Sigb^{{\cal T}} & (\mathbf{C}_{\thetab}^{{\cal T}})^T \\
	\mathbf{C}_{\thetab}^{{\cal T}} & \Gamb_{\thetab,g;11}
	\end{array}
	\bigg)
	\bigg),
	\end{align}
	where~$\mathbf{C}_{\thetab}^{\cal TM}=\frac{\mathcal{I}_p(g)}{p-1}\Gamb_\thetab$, $\mathbf{C}_{\thetab}^{\cal TE}=\mathbf{0}$, $
	\Sigb^{\cal TM}
	:= \frac{1}{p-1}\,{\bf I}_{p-1}$, and
	$$
	\Sigb^{\cal TE}
	:=
	\frac{1}{p^2-1}
	\,
	({\bf I}_{(p-1)^2}+ {\bf K}_{p-1}+{\bf J}_{p-1})
	-
	\frac{1}{(p-1)^2}\,
	{\bf J}_{p-1}.
	$$
	By using Lemma~\ref{lem:3}, Le Cam's first lemma implies that ${\rm P}_{\thetab_n,g}\n$ and ${\rm P}_{\thetab,g}\n$ are mutually contiguous. Therefore, one can apply Le Cam's third lemma to the joint asymptotic normality results in~\eqref{jointas}, which yields that, under ${\rm P}_{\thetab,g}\n$,
	\begin{align}
	\label{Lem3aF}
	\frac{1}{\sqrt{n}} \sum_{i=1}^n {\bf U}_{i,\thetab_n}
	+ \frac{\mathcal{I}_p(g)}{p-1} \Gamb_{\thetab}^{T}{\bf t}_n
	=
	{\bf S}_{\thetab_n}^{{\cal TM}(n)}
	+ (\mathbf{C}_{\thetab}^{{\cal TM}})^{T} {\bf t}_n
	\inlaw
	\mathcal{N}\big({\bf 0}, \Sigb^{\cal TM}\big)
	\end{align}
	and
	\begin{align}
	\label{Lem3aE}
	\frac{1}{\sqrt{n}} \sum_{i=1}^n {\rm vec}\left({\bf U}_{i,\thetab_n}{\bf U}_{i,\thetab_n}^{T}-\frac{1}{p-1}\,{\bf I}_{p-1} \right)
	=
	{\bf S}_{\thetab_n}^{{\cal TE}(n)}
	+ (\mathbf{C}_{\thetab}^{{\cal TE}})^{T} {\bf t}_n
	\inlaw
	\mathcal{N}\big({\bf 0}, \Sigb^{\cal TE}\big).
	\end{align}
	Now, by using Lemma \ref{lem:2}, the CLT for triangular arrays shows that, still under ${\rm P}_{\thetab,g}\n$,
	\begin{align}
	\label{Lem3bF}
	\frac{1}{\sqrt{n}} \sum_{i=1}^n ({\bf U}_{i,\thetab_n} - {\rm E}[{\bf U}_{i,\thetab_n}] )
	\inlaw
	\mathcal{N}\big({\bf 0}, \Sigb^{\cal TM}\big)
	,
	\end{align}
	and
	\begin{align}
	\label{Lem3bE}
	\frac{1}{\sqrt{n}} \sum_{i=1}^n {\rm vec}\left({\bf U}_{i,\thetab_n}{\bf U}_{i,\thetab_n}^T- {\rm E}[{\bf U}_{i,\thetab_n}{\bf U}_{i,\thetab_n}^{T}] \right)
	\inlaw
	\mathcal{N}\big({\bf 0}, \Sigb^{\cal TE}\big)
	,
	\end{align}
	where the expectations are under~${\rm P}_{\thetab,g}\n$.
	The result~\ref{lem:4:2:tm} then follows from~\eqref{Lem3aF} and~\eqref{Lem3bF}, whereas \ref{lem:4:2:te} similarly follows from~\eqref{Lem3aE} and~\eqref{Lem3bE}.
	Finally, \ref{lem:4:3:tm}--\ref{lem:4:3:te} are a direct consequence of~\ref{lem:4:1:tm}--\ref{lem:4:1:te} and~\ref{lem:4:2:tm}--\ref{lem:4:2:te}.
\end{proof}

\section{Additional simulations}
\label{sec:app:moresimus}

\subsection{\texorpdfstring{The specified-$\thetab$ problem on~$\mathcal{S}^3$}{The specified-theta problem on S3}}
\label{sec:simuspecp4}

\begin{figure}[!htpb]
	\centering
	\includegraphics[width=.9\textwidth]{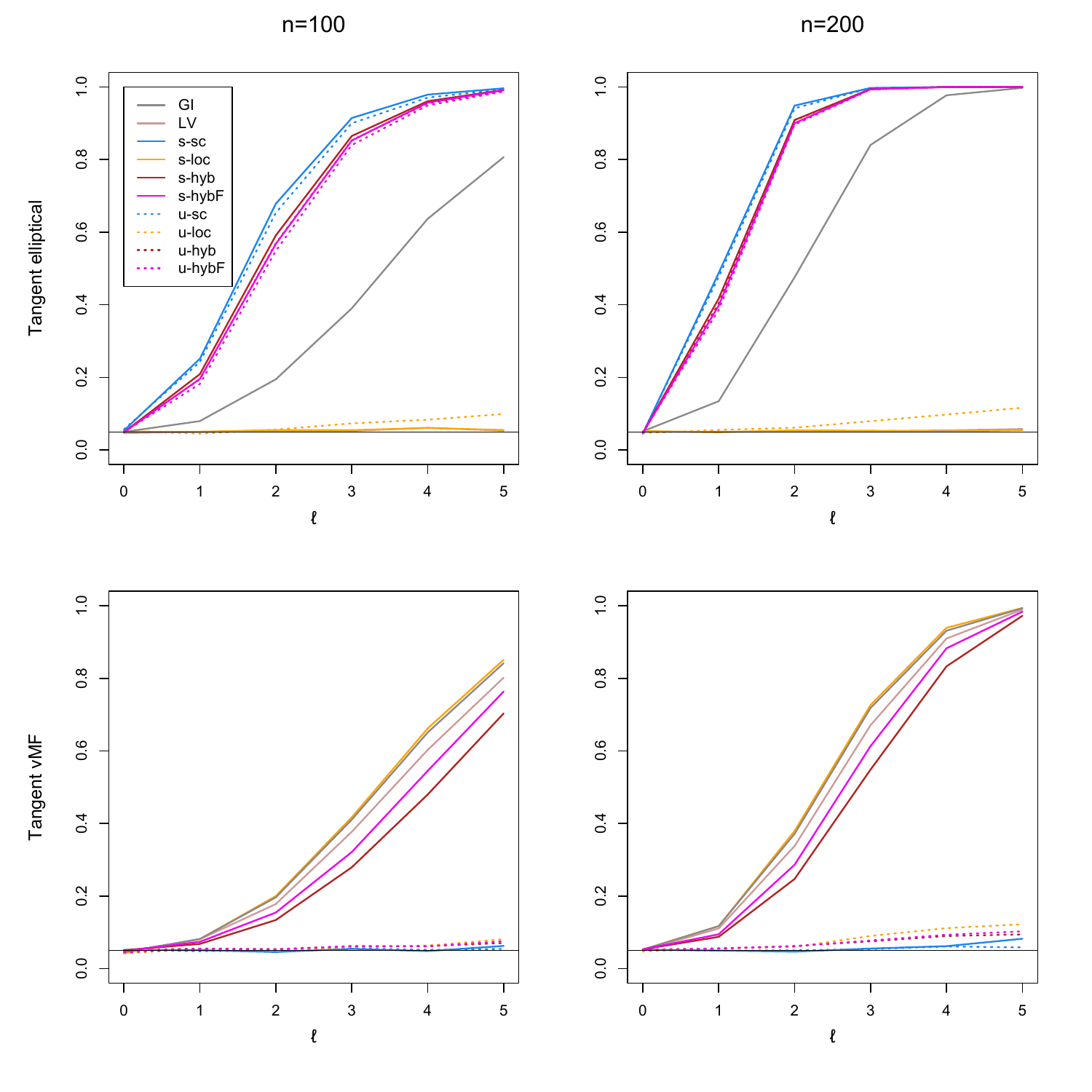}\vspace*{-0.75cm}
	\caption{\small Rejection frequencies, under tangent elliptical alternatives (top row) and tangent vMF alternatives (bottom row), of the specified-$\thetab$ tests $\phi_\thetab^{\rm sc}$ (s-sc), $\phi_\thetab^{\rm loc}$ (s-loc), $\phi_\thetab^{\rm hyb}$ (s-hyb), $\tilde{\phi}_\thetab^{\rm hyb}$ (s-hybF), $\phi_\thetab^{\rm LV}$ (LV), and $\phi_\thetab^{\rm GI}$ (GI), as well as the unspecified-$\thetab$ tests~$\phi_\dagger^{\rm sc}$ (u-sc), $\phi_{\rm vMF}^{\rm loc}$ (u-loc), $\phi_{\rm vMF}^{\rm hyb}$ (u-hyb), and $\tilde{\phi}_{\rm vMF}^{\rm hyb}$ (u-hybF). Sample sizes are $n=100$ (left column) and $n=200$ (right column). All tests are performed at asymptotic level~$5\%$; see Section~\ref{sec:simuspecp4} for details.
		\label{power4d}}
\end{figure}

The third simulation exercise essentially replicates the one in Section \ref{sec:simuspecp3} on~$\mathcal{S}^{3}$. Since the Kuiper test~$\phi_\thetab^{\rm KU}$ only applies for data on~$\mathcal{S}^2$, we replaced it with the Gin\'{e} test $\phi_\thetab^{\rm GI}$, that, as the Kuiper test, is an omnibus test addressing the specified-$\thetab$ problem. For sample sizes~$n=100$ and~$n=200$ and for two types of alternatives to rotational symmetry~($r=1,2$), we generated
$N = 5,\!000$ mutually independent random samples of the form
${\Xb}_{i;\ell}^{(r)}$, 
$i=1,\ldots, n$,  
$\ell=0, \ldots, 5$,  
$r=1, 2$,
with values in~$\mathcal{S}^3$. The ${\Xb}_{i;\ell}^{(1)}$'s follow a $\mathcal{TE}_4(\thetab,g_2,{\Lamb}_{\ell})$ with location~$\thetab:=(1, 0, 0, 0)^T$ and shape~${\Lamb}_{\ell}:=3{\rm diag}(1+\ell/2,1, 1)/(3+\ell/2)$. The~${\Xb}_{i;\ell}^{(2)}$'s follow a $\mathcal{TM}_4(\thetab,g_2,{\pmb \mu},\kappa_{\ell})$ with skewness direction~${\pmb \mu}=(1,0,0)^T$ and skewness intensity $\kappa_{\ell}:=\ell/8$. As in the previous simulation exercises, $\ell=0$ corresponds to the null of rotational symmetry about~$\thetab$ and~$\ell=1, \ldots, 5$ provide increasingly severe alternatives. For each replication, we performed, at asymptotic level~$5\%$, the specified-$\thetab$ tests~$\phi_\thetab^{\rm loc}$, $\phi_\thetab^{\rm sc}$,  $\phi_\thetab^{\rm hyb}$, $\phi_\thetab^{\rm LV}$, and the Gin\'{e} test~$\phi_\thetab^{\rm GI}$, as well as the unspecified-$\thetab$ tests~$\phi_\dagger^{\rm sc}$, $\phi_{\rm vMF}^{\rm loc}$, and~$\phi_{\rm vMF}^{\rm hyb}$ (still based on the spherical mean). In addition, we also considered the Fisher-aggregated hybrid tests $\tilde{\phi}_\thetab^{\rm hyb}$ and $\tilde{\phi}_{\rm vMF}^{\rm hyb}$, as in $\mathcal{S}^3$ they differ from $\phi_\thetab^{\rm hyb}$ and $\phi_{\rm vMF}^{\rm hyb}$. As expected due to their construction, the Fisher-aggregated tests dominate their hybrid counterparts under tangent vMF alternatives, while they are dominated by the latter under tangent elliptical alternatives. The rest of the power curves, that are provided in Figure~\ref{power4d}, lead to conclusions that are very similar to those reported in the simulation exercise conducted in Section~\ref{sec:simuspecp3}.

\subsection{\texorpdfstring{Mixtures in the specified-$\thetab$ case on~$\mathcal{S}^2$}{Mixtures}}
\label{sec:simusmix}

\begin{figure}[!htpb]
	\centering
	\includegraphics[width=.9\textwidth]{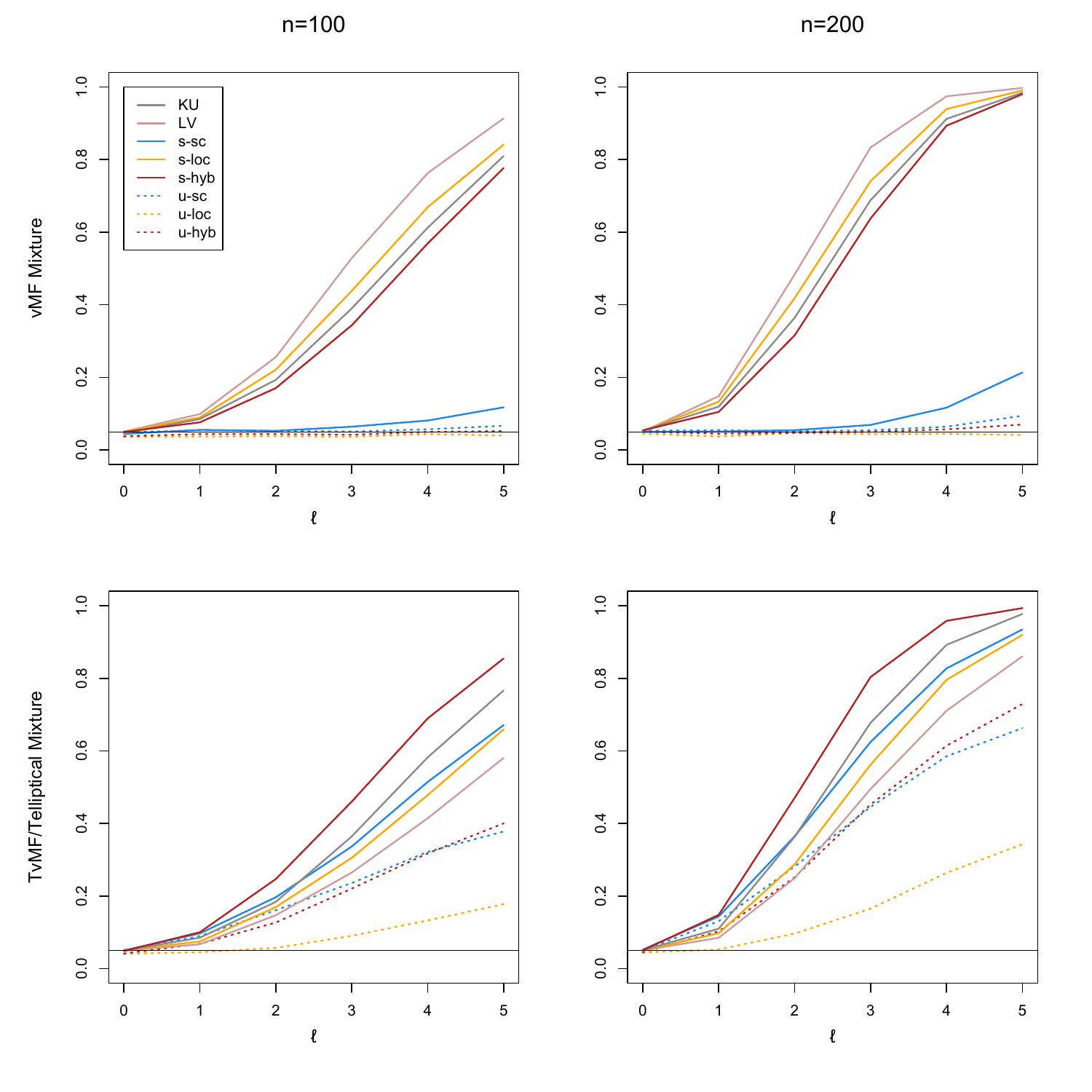}\vspace*{-0.75cm}
	\caption{\small Rejection frequencies, under mixtures of vMF's and mixtures of tangent vMF and tangent elliptical, of the specified-$\thetab$ tests $\phi_\thetab^{\rm sc}$ (s-sc), $\phi_\thetab^{\rm loc}$ (s-loc), $\phi_\thetab^{\rm hyb}$ (s-hyb), $\phi_\thetab^{\rm LV}$ (LV), and $\phi_\thetab^{\rm GI}$ (GI), as well as the unspecified-$\thetab$ tests~$\phi_\dagger^{\rm sc}$ (u-sc), $\phi_{\rm vMF}^{\rm loc}$ (u-loc), and $\phi_{\rm vMF}^{\rm hyb}$ (u-hyb). Sample sizes are $n=100$ (left column) and $n=200$ (right column). All tests are performed at asymptotic level~$5\%$; see Section~\ref{sec:simusmix} for details.
		\label{powermix}}
\end{figure}

For the last simulation exercise, we consider two types of mixture distributions on~${\cal S}^2$: mixtures of vMF distributions ($r=1$) and mixtures of tangent vMF and tangent elliptical distributions ($r=2$). For sample sizes~$n=100$ and~$n=200$ and for both types of mixtures, we generated $N=5,\!000$ mutually independent random samples of the form
${\Xb}_{i;\ell}^{(r)}$, 
$i=1,\ldots, n$,  
$\ell=0, \ldots, 5$, 
$r=1,2$,
with values on~$\mathcal{S}^2$. The ${\Xb}_{i;\ell}^{(1)}$'s are distributed as the mixture $\frac{1}{2} \Yb+ \frac{1}{2} \Zb_{\ell}$, where~$\Yb$ and~$\Zb_{\ell}$ are independent vMF random vectors with respective locations~$\thetab=(1, 0, 0)^T$ and~$\thetab_\ell=(\cos(\ell/40), \sin(\ell/40), 0)^T$ and common concentration~$\kappa=5$. The ${\Xb}_{i;\ell}^{(2)}$'s are distributed as the mixture $\frac{1}{2} \Zb_{1, \ell}+ \frac{1}{2} \Zb_{2, \ell},$ where~$\Zb_{1, \ell}\sim \mathcal{TM}_3(\thetab, \exp(5u), {\pmb \mu}, \kappa_\ell)$ and $\Zb_{2, \ell}\sim \mathcal{TE}_3(\thetab, \exp(5u), \Lamb_\ell)$ are independent, with $\thetab=(1, 0, 0)^T$, ${\pmb \mu}=(1, 0)^T$, $\kappa_\ell= \ell/6$, and~$\Lamb_\ell=2{\rm diag}(1+\ell/2,1)/(2+\ell/2)$, $\ell=0, \ldots, 5$. For~$r=1,2$, the value $\ell=0$ corresponds to the null hypothesis of rotational symmetry about~$\thetab$, whereas~$\ell=1, \ldots, 5$ provide increasingly severe alternatives. For each replication, we performed, at asymptotic level~$\alpha=5\%$, the specified-$\thetab$ tests~$\phi_\thetab^{\rm loc}$, $\phi_\thetab^{\rm sc}$, $\phi_\thetab^{\rm hyb}$, $\phi_\thetab^{\rm LV}$, and~$\phi_\thetab^{\rm KU}$ (based on the true value of~$\thetab$). For the sake of comparison, we also considered the unspecified-$\thetab$ tests~$\phi_\dagger^{\rm sc}$, $\phi_{\rm vMF}^{\rm loc}$, and~$\phi_{\rm vMF}^{\rm hyb}$, based on the spherical mean. Figure~\ref{powermix} plots the resulting empirical power curves for sample sizes $n=100$ and $n=200$ and for both types of mixtures. Inspection of Figure~\ref{powermix} reveals that the \cite{LVD16} test performs well against mixtures of vMF distributions, while, as we might have guessed, the (specified-$\thetab$) hybrid test dominates the other tests for mixtures of tangent vMF and tangent elliptical distributions. The location and hybrid tests perform well overall.

\end{document}